\documentclass[twocolumn,aps,pre,superscriptaddress,10pt]{revtex4-2} 
\usepackage{graphicx}  
\usepackage{dcolumn}   
\usepackage{bm}        
\usepackage{amssymb}  
\usepackage{amsmath}   
\UseRawInputEncoding
\usepackage{hyperref}
\usepackage[version=4]{mhchem}
\usepackage{siunitx}
\usepackage{natbib}
\usepackage{braket}
\usepackage{appendix}
\usepackage{soul}

\begin{document}

\title{A projected complex Langevin sampling method for bosons in the canonical and microcanonical ensembles} 
\author{Ethan C.~McGarrigle} 
\affiliation{Department of Chemical Engineering, University of California, Santa Barbara, 93106, California, USA}

\author{Hector D.~Ceniceros}
\affiliation{Department of Mathematics, University of California, Santa Barbara, 93106, California, USA}
 
\author{Glenn H.~Fredrickson}
\email{ghf@ucsb.edu}
\affiliation{Department of Chemical Engineering, University of California, Santa Barbara, 93106, California, USA}
\affiliation{Materials Research Laboratory, University of California, Santa Barbara, 93106, California, USA}
\affiliation{Materials Department, University of California, Santa Barbara, 93106, California, USA}


\begin{abstract}
We introduce a projected complex Langevin (CL) numerical sampling method -- a fictitious Langevin dynamics scheme that uses numerical projection to sample a constrained stationary distribution with highly oscillatory character. Despite the complex-valued degrees of freedom and associated sign-problem, the projected CL method succeeds as a natural extension of real-valued projected Langevin processes. In the new proposed method, complex-valued Lagrange multipliers are determined to enforce constraints to machine precision at each iteration. To illustrate the efficacy of this approach, we adapt the projected CL method to sample coherent state quantum field theories describing interacting Bose gases, which are realized in modern cold-atom experiments. We apply projected CL to two scenarios with holomorphic constraints, the canonical and microcanonical ensembles, and show that projected CL reproduces the correct thermodynamic observables. We further observe improved numerical stability and accuracy at larger timesteps when compared to the previous state-of-the-art method for performing constrained CL sampling. 
\end{abstract}

\maketitle
\section{Introduction}
Computational methods in statistical mechanics offer a systematic and powerful framework for uncovering the behavior of a panoply of systems, ranging from proteins and polymers to magnetic materials, metals, and quantum fluids. Methods such as Markov Chain Monte Carlo, molecular dynamics, Brownian dynamics, and density functional theory provide estimates of thermodynamic or transport properties via equilibrium or non-equilibrium formalisms \cite{shollDensityFunctionalTheory2011, hessGROMACSAlgorithmsHighly2008, andradeInqModernGPUAccelerated2021, berendsenGROMACSMessagepassingParallel1995, binderMonteCarloMethods1986, gubernatisQuantumMonteCarlo2016, boninsegniWormAlgorithmDiagrammatic2006}. However, many physical scenarios require a restriction of the phase space to embed constraints such as conservation laws, presenting a significant challenge for stochastic sampling methods in particular.  

Numerical methods in statistical physics have successfully enforced constraints via projection. In Markov Chain Monte Carlo, a projection step is layered on top of the usual Metropolis procedure such that only projection steps which obey a detailed balance condition are accepted \cite{zappaMonteCarloManifolds2018}. In computational fluid mechanics, time-dependent solutions to the Navier Stokes equations incorporate projection steps via a splitting scheme that enforces the continuity equation at each time point \cite{chorinNumericalSolutionNavierStokes1967, temamApproximationSolutionEquations1969}. For protein and molecular simulations that neglect vibrational degrees of freedom, distances between neighboring atoms are constrained to fixed lengths via projection \cite{gronbech-jensenLongtimeOverdampedLangevin1994, pellegriniNewConstrainedLangevin1998}. In general Brownian dynamics and Langevin schemes \cite{gronbech-jensenLongtimeOverdampedLangevin1994, alfonsiConstrainedOverdampedLangevin2022, kallemovSecondOrderStrongMethod2011}, a projecting force is included with a Lagrange multiplier, which is then determined at each iteration to enforce the constraints. 

Among Brownian dynamics and Langevin approaches, complex Langevin (CL) sampling \cite{parisiComplexProbabilities1983, klauderLangevinApproachFermion1983, bergerComplexLangevinOther2021a} has emerged as a scalable computational method for tackling systems with a ``sign problem'', where the statistical weight $e^{-S}$ is not positive definite and has highly oscillatory character that worsens with system size \cite{lohSignProblemNumerical1990, dornheimFermionSignProblem2019}. Sign problems arise typically in fermionic systems, where negative signs alter the effective probability weights in the path-integral representation due to antisymmetric exchange. Sign problems additionally arise in cases where the effective action $S$ is expressed in a complex-valued basis, such as the bosonic coherent state path integral \cite{negeleQuantumManyparticleSystems1988}. Systems that suffer from a sign problem cannot be tackled with Monte Carlo methods without reweighting or approximations, where reweighting may fail depending on the severity of the sign problem \cite{troyerComputationalComplexityFundamental2005}. CL directly confronts the sign problem by complexifying all degrees of freedom and then prescribing their evolution by an overdamped Langevin dynamics. Dynamical trajectories of the stochastic process are used to sample a Markov chain of states, which implicitly importance sample the stationary distribution of interest \cite{bergerComplexLangevinOther2021a}. 

CL sampling has enjoyed routine application in disciplines where sign problems arise naturally, such as polymer physics \cite{fredricksonEquilibriumTheoryInhomogeneous2005, beardsleyComputationallyEfficientFieldTheoretic2019}, quantum chromodynamics \cite{aartsAdaptiveStepsizeInstabilities2010, aartsComplexLangevinDynamics2012, blochComplexLangevinSimulation2018, nagataComplexLangevinCalculations2018, nagataJustificationComplexLangevin2016}, lattice gauge theories \cite{makinoComplexLangevinMethod2015, mollgaardComplexLangevinDynamics2013, hirasawaComplexLangevinAnalysis2020a, greensiteComparisonComplexLangevin2014}, and ultracold atomic physics \cite{fredricksonFieldTheoreticSimulationsSoft2023, simmonsThermodynamicEngineQuantum2023, attanasioThermodynamicsSpinorbitcoupledBosons2020, heinenComplexLangevinApproach2022}. In the case of ultracold atomic gases, a grand canonical formulation is typically employed and serves as a natural starting point for CL methods \cite{negeleQuantumManyparticleSystems1988, fredricksonFieldTheoreticSimulationsSoft2023, heinenComplexLangevinApproach2022}. However, most ultracold atom experiments occur at fixed particle number and in isolation \cite{grossQuantumGasMicroscopy2021, simmonsThermodynamicEngineQuantum2023, hallerSingleatomImagingFermions2015, wenzFewManyObserving2013}, which is better represented by a canonical or microcanonical statistical treatment. Furthermore, grand canonical treatments may fail to describe the fluctuation spectrum of systems with fixed particle number or internal energy at very low temperature or atom numbers \cite{bedinghamBoseEinsteinCondensationCanonical2003, barghathiTheoryNoninteractingFermions2020}. While path integral Monte Carlo (PIMC) is one natural choice for simulating bosons in the grand canonical or canonical ensemble \cite{boninsegniWormAlgorithmContinuousSpace2006, boninsegniWormAlgorithmDiagrammatic2006}, CL sampling of an equivalent coherent states field theory enables simulations of boson models with a sign problem, such as spin-orbit coupled or rotating Bose-Einstein condensates (BEC) \cite{mcgarrigleEmergenceSpinMicroemulsion2023, fredricksonFieldTheoreticSimulationsSoft2023}.   

The previous state-of-the-art method for CL sampling cold atomic BECs in the canonical or microcanonical ensemble involved first employing a coherent states field theory and then inserting delta functions for each constraint into the partition function \cite{delaneyNumericalSimulationFiniteTemperature2020, fredricksonFieldTheoreticSimulationsSoft2023}. Invoking the Fourier integral representation of the delta function introduces an additional fluctuating Lagrange multiplier degree of freedom in the partition function. For each Lagrange multiplier, an additional stochastic differential equation (SDE) is prescribed and numerically solved in tandem with the coherent state fields' Langevin equations. In this approach, the coherent state dynamical trajectories enforce the constraints on average. Nevertheless, the method suffers from numerical instabilities that practically restrict the range of operable timesteps and accessible systems. 

In this paper, we detail a projected complex Langevin method as a more robust choice for sampling statistical field theories and Bose quantum field theories with constraints. Instead of prescribing a Langevin equation of motion for the Lagrange multipliers, the multipliers are determined at every step of the stochastic process in order to enforce the constraints rigorously to machine precision. We demonstrate projected CL's efficacy for simulating large ensembles of bosons in the canonical (NVT) and microcanonical ensembles (UVN) using a coherent states field-theoretic representation. Notably, we find that the projected CL method shows improved numerical stability and accuracy at higher timestep discretizations when compared to the previous Lagrange multiplier SDE method. The projected CL formalism and demonstration provide a blueprint for sampling other constrained statistical theories with a sign problem. 

We organize the paper as follows. In Sec.~\ref{background}, we introduce and define a model for systems of interacting bosons in the continuum. Then we review the canonical and microcanonical statistical ensembles for considering these systems at finite temperature, employing a coherent states field-theoretic representation. Section \ref{background} also reviews the prior Lagrange multiplier SDE approach and corresponding numerical methods for sampling the constrained coherent state field theory. In Sec.~\ref{projectedCL}, we introduce a framework and numerical approach for projected CL and then detail their application to the canonical and microcanonical ensembles. Section \ref{results} provides numerical demonstrations of the projected CL method and a direct comparison to the Lagrange multiplier SDE method. The results in Sec.~\ref{results} compare the numerical properties and accuracy of each method. Finally, we conclude and discuss implications of this work in Sec.~\ref{conclusion}.  

\section{Background} \label{background}

\subsection{Model System: Interacting Bosons in the Continuum}
We begin by considering a many-body assembly of interacting bosons -- particles obeying Bose-Einstein statistics in the quantum regime -- and their quantum field-theoretic representation. Such systems are broadly described by a second-quantized Hamiltonian in the continuum \cite{fetterQuantumTheoryManyParticle2012}: 
\begin{equation}
  \begin{split}
 \hat{H} &= \int d\mathbf{r}\ \hat{\psi}^{\dagger}(\mathbf{r}) \left [ -\frac{\hbar^2}{2m} \nabla^2  \right ] \hat{\psi}^{\vphantom{\dagger}} (\mathbf{r}) \\
  &+ \frac{1}{2} \int d\mathbf{r} \int d\mathbf{r'}\ \hat{\psi}^{\dagger}(\mathbf{r}) \hat{\psi}^{\dagger}(\mathbf{r'}) u(|\mathbf{r} - \mathbf{r}'|) \hat{\psi}^{\vphantom{\dagger}} (\mathbf{r}) \hat{\psi}^{\vphantom{\dagger}} (\mathbf{r'}) ,
  \label{eq: H}
\end{split}
\end{equation}
\noindent{where} $\hat{\psi}$ ($\hat{\psi}^{\dagger}$) represents a second-quantized destruction (creation) operator obeying Bose commutation relations $[\hat{\psi}(\mathbf{r}), \hat{\psi}^{\dagger} (\mathbf{r'})] = \delta (\mathbf{r} - \mathbf{r'}) $ \cite{fetterQuantumTheoryManyParticle2012}, $\hbar$ is the reduced Planck's constant, $m$ is the atomic mass, and $u(|\mathbf{r} - \mathbf{r'}|)$ is a pair-potential describing the interactions between two bosons. Spatial integrals are taken over a domain in $d$ dimensions, $D_{r} \subset \mathbb{R}^{d}$.  In this work, we assume hypercubic geometries such that the system volume is $V = L^{d}$ and impose periodic boundary conditions on the domain. 

This Hamiltonian readily describes interacting gaseous Bose-Einstein condensates such as $^{87}$Rb or $^{23}$Na \cite{pitaevskiiBoseEinsteinCondensationSuperfluidity2016} as well as superfluid $^{4}$He \cite{ceperleyPathIntegralsTheory1995a}, which are routinely accessed in modern experiments. In this work, we restrict ourselves to the short-ranged pseudopotential $u(|\mathbf{r} - \mathbf{r'}|) = u_{0} \delta (\mathbf{r} - \mathbf{r'})$ with contact interaction strength parameter $u_{0}$. This choice of the pair-potential is appropriate for dilute ultracold Bose-Einstein condensates \cite{pitaevskiiBoseEinsteinCondensationSuperfluidity2016}, where $u_{0}$ is well-characterized by the atomic s-wave scattering length $a_{s}$ via $u_{0} = 4\pi a_{s} \hbar^2 / m$. 

\subsection{Canonical Ensemble}
An interacting assembly of $N$ identical bosons in a volume $V$ in thermal equilibrium with a heat reservoir at temperature $T$ is characterized by the canonical partition function \cite{barghathiTheoryNoninteractingFermions2020, shenStableRecursiveAuxiliary2023, bedinghamBoseEinsteinCondensationCanonical2003} 
\begin{equation}
    \mathcal{Z}_{c} (N, V, T) = \text{Tr} \left [ e^{-\beta \hat{H}} \delta_{\hat{N}, N}  \right]
 \label{eq: Zcanonical},
\end{equation}
\noindent{where} $\beta = 1 / k_{B} T$ with Boltzmann constant $k_{B}$, $\hat{N} = \int d\mathbf{r}\ \hat{\psi}^{\dagger} (\mathbf{r}) \hat{\psi}^{\vphantom{\dagger}} (\mathbf{r})$ is the particle number operator, and the Kronecker delta ensures that only states containing the desired particle number $N$ are counted. The Kronecker delta is then expressed via an integral representation:   
\begin{equation}
 \delta_{\hat{N}, N} = \frac{1}{2\pi} \int_{-\pi}^{\pi} d\psi_{N}\ e^{i \psi_N (\hat{N} - N)} .
\end{equation}
Next, we follow the Feynman path integral approach \cite{negeleQuantumManyparticleSystems1988, prokofevWormAlgorithmQuantum1998, ceperleyPathIntegralsTheory1995a} by discretizing the inverse temperature $\beta$ and $\psi_{N}$ into $N_{\tau}$ imaginary time points and factorizing: 
\begin{equation} 
  e^{-\beta \hat{H} + i \psi_{N} \hat{N}} = \prod_{j=0}^{N_{\tau}-1} e^{-\Delta_{\tau} \hat{H} + i \psi_{N} \hat{N}/N_{\tau}} ,
\end{equation}
\noindent{where} we have defined the imaginary time discretization $\Delta_{\tau} \equiv \beta / N_{\tau}$ and utilized the fact that $[\hat{H}, \hat{N} ] = 0$ for the class of Hamiltonians defined in equation\ (\ref{eq: H}).

Next, we invoke the basis of many-body coherent states $\ket{\phi} \equiv \prod_{\mathbf{k}} e^{\phi^{\vphantom{\dagger}}_{\mathbf{k}} \hat{b}^{\dagger}_{\mathbf{k}}} \ket{0}$, where $\mathbf{k}$ denotes a plane-wave mode from a single-particle basis and $\hat{b}^{\dagger}_{\mathbf{k}}$ is a boson creation operator for state $\mathbf{k}$ \cite{negeleQuantumManyparticleSystems1988}. In the basis of coherent states, we express the trace of any operator $\hat{A}$ as $\text{Tr}[\hat{A}] \equiv \int \mathcal{D} (\phi^{*}_{0} , \phi^{\vphantom{*}}_{0}) e^{-\int d\mathbf{r} \phi^*_0 (\mathbf{r}) \phi^{\vphantom{*}}_0 (\mathbf{r}) } \bra{\phi_0} \hat{A} \ket{\phi_0} $ where $\mathcal{D}(\phi^{*}_{0} , \phi^{\vphantom{*}}_{0})$ denotes a functional integral that consists of integration over the real and imaginary parts of $\phi_{\mathbf{k}}$ for each $\mathbf{k}$. We insert a copy of the identity operator $\hat{1} = \int \mathcal{D} (\phi^*_{j}, \phi^{\vphantom{*}}_{j})\ e^{-\int d\mathbf{r} \phi^*_{j} (\mathbf{r}) \phi^{\vphantom{*}}_{j} (\mathbf{r}) } \ket{\phi_{j}} \bra{\phi_{j}} $ at each imaginary time slice $j$ separating the exponential factors. A field theory that is first-order accurate in imaginary time emerges from evaluating matrix elements for the $j$th slice: 
\begin{equation}
\begin{split}
&\bra{\phi_{j}} e^{-\Delta_{\tau} \hat{H} + \frac{i \psi_{N} \hat{N}}{N_{\tau}}} \ket{\phi_{j-1}} = \\
&\bra{\phi_{j}} 1 - \Delta_{\tau} \hat{H} + \frac{i \psi_{N}}{N_{\tau}} \hat{N} \ket{\phi_{j-1}} + \mathcal{O} (N^{-2}_{\tau} ) \\ 
&= \langle \phi_{j} | \phi_{j-1} \rangle e^{-\beta \tilde{U}[\phi^*_{j}, \phi_{j-1}] + \frac{i\psi_{N}}{N_{\tau}} \tilde{N}[\phi^*_{j}, \phi_{j-1}] } ,
\end{split}
 \label{eq: 1st_order_expansion}
\end{equation}
\noindent{where} the overlap of coherent states is $\langle \phi_{j} | \phi_{j-1} \rangle = e^{\int d\mathbf{r} \phi^*_{j} (\mathbf{r}) \phi^{\vphantom{*}}_{j-1} (\mathbf{r})}$. After collecting all the exponential factors, we define global particle number and internal energy functionals: 
\begin{equation}
\tilde{N}[\phi^* , \phi^{\vphantom{*}}] = \frac{1}{N_{\tau}} \sum_{j=0}^{N_\tau -1} \int d\mathbf{r}\ \phi^*_{j} (\mathbf{r}) \phi^{\vphantom{*}}_{j-1} (\mathbf{r}) ,
 \label{eq: N_CSfunctional}
\end{equation} 
and 
\begin{equation}
  \begin{split}
\tilde{U}[\phi^* , \phi^{\vphantom{*}}] &=  \frac{1}{N_{\tau}} \sum_{j=0}^{N_{\tau} -1} \int d\mathbf{r}\ \phi^*_{j} (\mathbf{r}) \hspace{1.5px} \left [ -\frac{\hbar^2 \nabla^2}{2m} \right] \phi^{\vphantom{*}}_{j-1} (\mathbf{r}) \\
 &+ \frac{1}{2 N_{\tau}} \sum_{j=0}^{N_{\tau} -1} \int d\mathbf{r} \int d\mathbf{r}\ \phi^*_{j} (\mathbf{r}) \phi^*_{j} (\mathbf{r'}) u(|\mathbf{r} - \mathbf{r'}|)^{\vphantom{*}}  \\ 
    & \hspace{125px} \times \phi^{\vphantom{*}}_{j-1} (\mathbf{r}) \phi^{\vphantom{*}}_{j-1} (\mathbf{r'}) . \label{eq: U_tilde_functional} \\
  \end{split}
\end{equation} 

\noindent{Now}, the canonical partition function can be expressed succinctly: 
\begin{equation}
 \mathcal{Z}_{c} = \frac{1}{2\pi} \int d\psi_{N} \int \mathcal{D} (\phi^* , \phi^{\vphantom{*}} )\ e^{-S_{\text{CE}}[\phi^* , \phi^{\vphantom{*}} ; \psi_{N}]} , 
  \label{eq: Z_CE}
\end{equation} weighted by the canonical action $S_{\text{CE}}[\phi^* , \phi^{\vphantom{*}} ; \psi_{N}] \equiv S_0 [\phi^* , \phi^{\vphantom{*}}] + \beta \tilde{U}[\phi^* , \phi^{\vphantom{*}}] + S_{N}[\phi^* , \phi^{\vphantom{*}}; \psi_{N}]$: 
\begin{equation}
  S_{0}[\phi^* , \phi^{\vphantom{*}}] = \sum_{j=0}^{N_{\tau} -1} \int d\mathbf{r}\ \phi^*_{j} (\mathbf{r}) [ \phi^{\vphantom{*}}_{j} (\mathbf{r}) -  \phi^{\vphantom{*}}_{j-1} (\mathbf{r}) ] ,
    \label{eq: S_0}
\end{equation}
\begin{equation}
 S_{N} [\phi^* , \phi^{\vphantom{*}}; \psi_{N}] = -i \psi_{N} \left[\tilde{N}[\phi^* , \phi^{\vphantom{*}}] - N \right ] . 
  \label{eq: S_N}
\end{equation}
Besides the constraint variable $\psi_N$, the degrees of freedom in the partition function are a complex-conjugate pair of $d+1$-dimensional fields $\phi$ and $\phi^*$. As such, equation\ (\ref{eq: Z_CE}) consists of functional integrals over the coherent state (CS) fields, which live on a space-imaginary time domain $ D = D_{r} \cup D_{\tau}$, where $D_{\tau}$ is a discrete domain of imaginary time points $D_{\tau} = \{ \tau_{j} \in \mathbb{R} : \tau_{j} = j \Delta_{\tau} ,  j \in \{0, 1, 2, \dots, N_{\tau} - 1 \} \}$. The effective field theory described by equation\ (\ref{eq: Z_CE}) represents an ensemble of bosons with fixed particle number. 

\subsection{Microcanonical Ensemble}
An isolated system of $N$ bosons in a volume $V$ with internal energy $U$ can be described by a microcanonical partition function \cite{fredricksonFieldTheoreticSimulationsSoft2023}: 
\begin{equation}
   \Omega (U, V, N) = \text{Tr} \left [\delta_{\hat{N}, N} \delta_{\hat{H}, U}  \right] .
    \label{eq: Microcanonical_partition}
\end{equation}
In the same vein as the previous section, we call upon the Kronecker delta integral representations and combine exponential terms since $[\hat{H}, \hat{N}]=0$:
\begin{equation}
\delta_{\hat{N}, N} \delta_{\hat{H}, U} = \frac{1}{(2\pi)^2} \int d\psi_{N} \int d\psi_{U}\ e^{i\psi_{N}(\hat{N} - N) - i\psi_{U}(\hat{H} - U)} ,
\end{equation}
\noindent{where} we have performed a variable substitution $\psi_{U} \to -\psi_{U}$ for later convenience. 
As before, we factorize all exponentiated operators into $N_{\tau}$ parts and take the trace using the coherent states basis, inserting identities between each exponential factor and subsequently evaluating the matrix elements to first-order in $N^{-1}_{\tau}$ (as in equation \ref{eq: 1st_order_expansion}). After simplifying, we arrive at a field-theoretic microcanonical partition function: 
\begin{equation}
  \Omega = \frac{1}{(2\pi)^2} \int d\psi_{N} \int d\psi_{U} \int \mathcal{D} (\phi^* , \phi^{\vphantom{*}})\ e^{-S_{\text{MCE}}[\phi^* , \phi^{\vphantom{*}} ; \psi_{N}, \psi_{U}]},
   \label{eq: Z_MCE}
\end{equation}
with the microcanonical action $S_{\text{MCE}}[\phi^* , \phi^{\vphantom{*}} ; \psi_{N}, \psi_{U}] \equiv S_{0} [\phi^* , \phi^{\vphantom{*}}] + S_{U}[\phi^* , \phi^{\vphantom{*}} ; \lambda_{U}] + S_{N} [\phi^* , \phi^{\vphantom{*}} ; \lambda_{N}]$ defined using equations\ (\ref{eq: S_0}) and (\ref{eq: S_N}) as in the canonical case. In contrast to the canonical ensemble, we embed the energy constraint via this additional contribution to the action:
\begin{equation}
 S_{U} [\phi^* , \phi^{\vphantom{*}}; \psi_{N}] = i \psi_{U} \left( \tilde{U}[\phi^* , \phi^{\vphantom{*}}] - U \right ) .
  \label{eq: S_U}
\end{equation}
We emphasize that $U$ and $N$ are real-valued scalars, while $\tilde{U} [\phi^* , \phi^{\vphantom{*}}]$ and $\tilde{N}[\phi^* , \phi^{\vphantom{*}}]$ are estimators of the internal energy and particle number from a coherent state field configuration and thus may be complex-valued at an instant.

Contributions from equations\ (\ref{eq: S_U}) and (\ref{eq: S_N}) to the partition function are highly oscillatory and present an explicit sign-problem. As such, complex Langevin is a judicious choice for sampling the effective field theories in equations\ (\ref{eq: Z_CE}) or (\ref{eq: Z_MCE}).

\subsection{Previous Method: Lagrange Multiplier Stochastic Differential Equation}
Previously, Delaney and Fredrickson \cite{fredricksonFieldTheoreticSimulationsSoft2023} applied the complex Langevin (CL) sampling method to the canonical and microcanonical partition functions, equations\ (\ref{eq: Z_CE}) and (\ref{eq: Z_MCE}). Overdamped Langevin equations of motion are prescribed for $\psi_{N}$ and $\psi_{U}$ in addition to the coherent states fields $\phi$ and $\phi^*$ in order to enforce the particle number and energy constraints while sampling quantum and thermal fluctuations:
\begin{subequations}
  \begin{align}
  \frac{\partial}{\partial t}  \phi^{\vphantom{*}}_{j} (\mathbf{r}, t) &= - \frac{\delta S[\phi^* , \phi^{\vphantom{*}}; \lambda_{N}, \lambda_{U}]}{\delta \phi^*_{j} (\mathbf{r}, t) }  + \eta^{\vphantom{*}}_{j} (\mathbf{r}, t) , \label{eq: CL_phi} \\
  \frac{\partial}{\partial t} \phi^*_{j} (\mathbf{r}, t) &= - \frac{\delta S[\phi^* , \phi^{\vphantom{*}}; \lambda_{N}, \lambda_{U}]}{\delta \phi^{\vphantom{*}}_{j} (\mathbf{r}, t) } + \eta^*_{j} (\mathbf{r}, t) ,
  \label{eq: CL_phistar}
    \end{align}
\end{subequations}
\begin{subequations}
 \begin{align}
  \frac{\partial}{\partial t}  \psi_{N} (t) &= - \frac{\delta S[\phi^* , \phi^{\vphantom{*}}; \lambda_{N}, \lambda_{U}]}{\delta \psi_{N} (t)}  + \eta_{\psi_{N}} (t) , \label{eq: CL_psiN} \\
  \frac{\partial}{\partial t}  \psi_{U} (t) &= - \frac{\delta S[\phi^* , \phi^{\vphantom{*}}; \lambda_{N}, \lambda_{U}]}{\delta \psi_{U} (t)}  + \eta_{\psi_{U}} (t) ,
  \label{eq: CL_psiU}
  \end{align}
\end{subequations}
\noindent{where} $t$ represents a fictitious time, $\eta_{\psi_{X}}$ are real-valued white noises with zero mean and second moment $\langle \eta_{\psi_{X}}(t) \eta_{\psi_{X'}}(t') \rangle = 2 \delta_{X, X'}\delta(t - t')$, and $S$ is either the canonical or microcanonical action appearing in equations\ (\ref{eq: Z_CE}) and (\ref{eq: Z_MCE}). For the coherent states CL equations, $\eta^{\vphantom{*}}_{j} = \eta^{(1)}_{j} + i \eta^{(2)}_{j}$ and $\eta^*_{j} = \eta^{(1)}_{j} - i \eta^{(2)}_{j}$ are complex conjugate white noise sources that are built from real noise fields $\eta^{(1)}_{j}$ and $\eta^{(2)}_{j}$ with $\langle \eta^{(\mu)}_{j} (\mathbf{r}, t) \rangle = 0 $ and $\langle \eta^{(\mu)}_{j} (\mathbf{r}, t)\ \eta^{(\nu)}_{k } (\mathbf{r'}, t') \rangle = \delta_{\mu, \nu} \delta_{j, k}  \delta (\mathbf{r} - \mathbf{r'}) \delta (t - t') $. 

Taking a forward difference for the fictitious time derivative with discretization $\Delta t$, Euler-Maruyama updates are readily applicable to propagate the constraint variables to the next timepoint $t^{(\ell+1)} = t^{(\ell)} + \Delta t$:  
\begin{subequations}
  \begin{align}
  \psi^{\ell + 1}_{N} = \psi^{\ell}_{N} - i \Delta t ( \tilde{N}[\phi^* , \phi^{\vphantom{*}}]^{(\ell)} - N) + \sqrt{2 \Delta t}\ \eta^{(\ell)}_{\psi_{N}} , \label{eq: EM_1} \\
 \psi^{\ell+1}_{U} = \psi^{\ell}_{U} - i \Delta t ( \tilde{U}[\phi^* , \phi^{\vphantom{*}}]^{(\ell)} - U) + \sqrt{2 \Delta t}\ \eta^{(\ell)}_{\psi_{U}} ,
    \label{eq: EM_2}
 \end{align}
\end{subequations}
\noindent{where} $\eta^{(\ell)}$ is shorthand notation for real-valued noise generated with a normal distribution with zero mean and unit variance at CL iteration $(\ell)$. The Euler-Maruyama update scheme is weak first-order accurate and generally suffers from numerical instability. 

Fortunately, there is freedom to rescale the timestep $\Delta t \to \alpha \Delta t$ in each Langevin equation of motion, where $\alpha$ is a mobility parameter that controls the relative relaxation rate of the degree of freedom. For the coherent state Langevin equations, we find that fixing the mobility to $N_{\tau}$ is an optimal choice, and we denote the mobilities of $\psi_{N}$ and $\psi_{U}$ in equations \ (\ref{eq: EM_1}) and (\ref{eq: EM_2}) as $\alpha_{N}$ and $\alpha_{U}$, respectively. The choice of mobility bears no thermodynamic consequence (i.e., bias to the stationary state) \cite{aartsComplexLangevinDynamics2012} but is observed to improve numerical stability for $0 < \alpha_{N} < 1$ and $ 0 < \alpha_{U}  < 1$. In some cases, separate mobility values $\alpha_{N}$ and $\alpha_{U}$ were used for equations \ (\ref{eq: EM_1}) and (\ref{eq: EM_2}), respectively, to ensure numerical stability in the microcanonical ensemble. For the canonical ensemble, dynamical equation (\ref{eq: CL_psiU}) and its propagation algorithm (eq. \ref{eq: EM_2}) are omitted and the canonical action from equation\ (\ref{eq: Z_CE}) is used to drive the updates. When sampling for long fictitious times, equations\ (\ref{eq: EM_1}) and (\ref{eq: EM_2}) ensure that the constraints are enforced after Langevin time-averaging. In the limit of large $\alpha \gg 1$, the variance of the constraint residuals will decrease; however, numerical stability suffers.

In tandem with the Euler-Maruyama updates of the constraint variables, a pseudospectral approach is used \cite{delaneyNumericalSimulationFiniteTemperature2020} to integrate the coherent state fields forward in fictitious time. We perform $(d+1)$-dimensional Fourier transforms on the coherent state fields, where functions of space are transformed via $f_{\mathbf{k}} = 1/ V \int d\mathbf{r}\ e^{- i \mathbf{k} \cdot \mathbf{r}} f(\mathbf{r})$ using a set of discrete wavevectors $\mathbf{k}_{n} = 2\pi L^{-1} (n_x , n_y , n_z)$, where $n_{\nu} \in \mathbb{Z}$. Furthermore, the coherent state fields are transformed in the imaginary time coordinate $\tau$ with a discrete set of Matsubara frequencies $\omega_{n} = 2\pi i n/ \beta $ via $\phi_{n, \mathbf{k}} = 1 / N_{\tau} \sum_{j=0}^{N_\tau -1} e^{-i \omega_{n} \tau_{j}} \phi_{j, \mathbf{k}}$, where $\tau_{j} = \beta j / N_{\tau} $, with $j, n \in \mathbb{Z}$. Both transforms leverage Fast-Fourier transform (FFT) algorithms that keep the cost of a forward or reverse transform to $\mathcal{O} (N_{\tau} N^{d}_{x} \log (N_{\tau} N^{d}_{x} ))$ complex operations. This approach uses spectral collocation with $N_{x}$ plane waves in each spatial dimension and $N_{\tau}$ points in imaginary time, assuming periodic boundary conditions in all $d+1$ dimensions. Finally, an exponential-time-differencing (ETD) integration algorithm with weak first-order accuracy \cite{villetEfficientFieldtheoreticSimulation2014a, delaneyNumericalSimulationFiniteTemperature2020} is utilized
\begin{equation}
  \begin{split}  
 \phi_{\mathbf{k}, n}^{(\ell + 1)} &= e^{-A_{\mathbf{k}, n} \Delta t} \phi^{(\ell)}_{\mathbf{k},n} \\
  &- \frac{1 - e^{-A_{\mathbf{k}, n} \Delta t} }{A_{\mathbf{k}, n}} \mathcal{N}_{\mathbf{k}, n}[\phi^* , \phi^{\vphantom{*}} ; \psi_{N}, \psi_{U}]^{(\ell)} \\
  &+ \sqrt{\frac{1 - e^{-2A_{\mathbf{k}, n} \Delta t}}{2 A_{\mathbf{k},n}}} R_{\mathbf{k}, n} ,
   \label{eq: ETD_phi}
  \end{split}
\end{equation}
\begin{equation}
 \begin{split}  
( \phi^*_{\mathbf{k}, n})^{(\ell+1)} &= e^{-A^*_{\mathbf{k}, n} \Delta t} (\phi^*_{\mathbf{k},n})^{(\ell)} \\
&- \frac{1 - e^{-A^*_{\mathbf{k}, n} \Delta t} }{A^*_{\mathbf{k}, n}} \mathcal{N}^*_{\mathbf{k}, n} [\phi^* , \phi^{\vphantom{*}} ; \psi_{N}, \psi_{U}]^{(\ell)} \\
  &+ \sqrt{\frac{1 - e^{-2A^*_{\mathbf{k}, n} \Delta t}}{2 A^*_{\mathbf{k},n}}} R^*_{\mathbf{k}, n} ,
   \label{eq: ETD_phistar}
  \end{split}
\end{equation}
\noindent{where} the linear force contributions $A_{\mathbf{k}, n}$ and $A^*_{\mathbf{k},n}$ as well as the non-linear force contributions $\mathcal{N}$ and $\mathcal{N}^*$ are detailed in Appendix \ref{appendixA}. $R_{\mathbf{k}, n}$ and $R^*_{\mathbf{k}, n}$ are complex-conjugate Gaussian white noise sources that are first generated in real space and imaginary time and then numerically Fourier transformed. 

When numerical trajectories converge to a steady state, the solutions sample an effective stationary distribution proptional to $e^{-S}$ \cite{bergerComplexLangevinOther2021a, fredricksonFieldTheoreticSimulationsSoft2023}. Expectation values of thermodynamic quantities $\langle \mathcal{O} \rangle$ are computed by a sample average of the corresponding holomorphic ``field operator'' $\tilde{O}[\phi^* , \phi^{\vphantom{*}}; \psi_{N}, \psi_{U}]$, a functional calculated in the coherent states basis: 
\begin{equation}
  \begin{split}
   \langle O \rangle &= \frac{\int \mathcal{D}(\phi^* , \phi^{\vphantom{*}}) \int d\psi_{N} \int d\psi_{U}\ e^{-S} \tilde{O}[\phi, \phi^*; \psi_{N}, \psi_{U}]}{\mathcal{Z}} \\
   &\approx \frac{1}{N_{t}} \sum_{\ell=0}^{N_{t} - 1}\tilde{O}[\phi^* , \phi^{\vphantom{*}} ; \psi_{N}, \psi_{U}]^{(\ell)} ,
   \end{split}
\end{equation}
\noindent{where} $N_{t}$ is the number of timesteps taken. Although physical observables may be complex-valued at each time $t^{\ell}$, their imaginary parts will vanish upon sample averaging for long Langevin times.

The $\psi_{N}$ and $\psi_{U}$ degrees of freedom are conjugate to the atom number and internal energy constraint, respectively, and are thus deeply connected to thermodynamic quantities. Considering a Wick rotation, it is clear that $i\psi_{N}$ and $i\psi_{U}$ are proportional to the chemical potential $\mu$ and inverse temperature $\beta$, respectively. Thus, in the microcanonical ensemble, the average inverse temperature is accessed via an ensemble average of the $\psi_{U}$ degree of freedom. In both the microcanonical and canonical ensembles, the chemical potential can be accessed by an expectation value of the $\psi_{N}$:
\begin{subequations}
  \begin{equation}
  \beta = i\langle \psi_{U} \rangle ,
  \label{eq: beta_noProjection}
\end{equation}
  \begin{equation}
 \mu = \frac{i\langle \psi_{N} \rangle }{\beta} .
  \label{eq: mu_noProjection}
\end{equation}
\end{subequations}
\noindent{These} identifications underscore how $\psi_{N}$ and $\psi_{U}$ function as Lagrange multiplier degrees of freedom in the effective field theory \cite{mcquarrieStatisticalMechanics2000}. Since $i \psi_{N}$ and $i \psi_{U}$ are conjugate to the real-valued constraints, we expect their thermal averages to be purely real-valued. 

These ensembles provide direct access to the relevant thermodynamic potentials,  Helmholtz free energy \cite{fredricksonDirectFreeEnergy2022} and entropy \cite{fredricksonFieldTheoreticSimulationsSoft2023}, respectively, in the canonical and microcanonical ensembles. The Helmholtz free energy $A$ and entropy $S$ are determined via the following thermodynamic relations: 
\begin{equation}
    A = - PV + \mu N ,
     \label{eq: Helmholtz}
\end{equation}
\begin{equation}
    \frac{S}{k_{B}} = \beta (U - A) .
     \label{eq: entropy}
\end{equation}

\section{Projected Complex Langevin Method} \label{projectedCL}
Alternatively, a projection approach omits equations\ (\ref{eq: CL_psiN}) and (\ref{eq: CL_psiU}) altogether and instead determines the Lagrange multipliers at every Langevin time step, to enforce the constraints strongly. In such a procedure, the multipliers are determined such that the coherent state configuration at the next time point will satisfy the constraints to machine precision. The Fourier representation of the delta function is no longer required; instead, the Lagrange multipliers are introduced by building a Lagrangian that augments the action with the constraints, as done in constrained optimization problems. We introduce and detail the general projected formalism and then apply it to the canonical and microcanonical cases. 

\subsection{Formalism}
An equivalent formulation of a constrained statistical theory is produced by defining a Lagrangian that augments the action $S[\phi^* , \phi^{\vphantom{*}}]$ with a set of $M$ constraints $\{g_{m} [\phi^* , \phi^{\vphantom{*}}] =0\}_{m=1}^{M}$, where we define $\mathbf{g} \in \mathbb{C}^{M}$ to be the vector of complex-valued constraints. Like a constrained optimization approach, the Lagrangian functional is generally 
\begin{equation}
 \mathcal{L}[\phi^* , \phi^{\vphantom{*}} ; \bm{\lambda}] = S[\phi^* , \phi^{\vphantom{*}} ] - \sum_{m=1}^{M} \lambda_{m} g_{m} [\phi^* , \phi^{\vphantom{*}}] , 
\label{eq: Lagrangian}
\end{equation}
\noindent{where} $\bm{\lambda} \in \mathbb{C}^{M}$ represents a vector of Lagrange multipliers that will enforce the constraints at each $\phi^* , \phi^{\vphantom{*}}$ configuration. The Lagrange multipliers have been complexified to accommodate the complex-valued coherent state degrees of freedom as well as potentially complex-valued constraints. This allows us to sample a generalized partition function $\mathcal{Z} = \int \mathcal{D} (\phi^* , \phi^{\vphantom{*}} ) \hspace{2px} e^{-S_{c} [\phi^* , \phi^{\vphantom{*}} ] }$, where the theory is governed by a constrained distribution $ e^{-S_{c}} \equiv e^{-S[\phi^* , \phi^{\vphantom{*}}]} \prod_{m=1}^{M} \delta ( g_{m} [\phi^* , \phi^{\vphantom{*}} ] )$. In other words, $S_{c} \in \mathbb{C}$ is sampled only when the set of constraints are rigorously satisfied and $S_{c}[\phi^* , \phi^{\vphantom{*}}] \subset S[\phi^* , \phi^{\vphantom{*}} ]$. As a result, the degrees of freedom $\phi^* , \phi^{\vphantom{*}}$ live on a manifold $\mathcal{M} = \{(\phi^* , \phi^{\vphantom{*}}) \in \mathcal{H} \hspace{4px} | \hspace{4px} \mathbf{g}[\phi^* , \phi^{\vphantom{*}}] = \mathbf{0} \} $, where $\mathcal{H}$ is a finite-dimensional Hilbert space \cite{lelievreLangevinDynamicsConstraints2012a, lelivreFreeEnergyComputations2010, lelievreHybridMonteCarlo2019}. 

A numerical sampling strategy for the constrained action is then introduced by recasting the CL equations of motion for the coherent states fields as
\begin{equation}
   \begin{split}
  \frac{\partial}{\partial t}  \phi^{\vphantom{*}}_{j} (\mathbf{r}, t) &= - \frac{\delta \mathcal{L}[\phi^* , \phi^{\vphantom{*}}; \bm{\lambda}]}{\delta \phi^*_{j} (\mathbf{r}, t) }  + \eta^{\vphantom{*}}_{j} (\mathbf{r}, t) \\
  \frac{\partial}{\partial t} \phi^*_{j} (\mathbf{r}, t) &= - \frac{\delta \mathcal{L}[\phi^* , \phi^{\vphantom{*}}; \bm{\lambda}]}{\delta \phi^{\vphantom{*}}_{j} (\mathbf{r}, t) } + \eta^*_{j} (\mathbf{r}, t)  ,
  \label{eq: CL_Lagrangian}
    \end{split}
\end{equation}
\noindent{where} the noises retain the same statistical properties as previously defined for the equations\ (\ref{eq: CL_phi}) and (\ref{eq: CL_phistar}), and we keep the off-diagonal descent scheme optimally chosen for the coherent state fields \cite{manCoherentStatesFormulation2014}. Crucially, we omit a CL equation of motion for the multipliers $\bm{\lambda}$. In the projected approach, $\bm{\lambda}$ is determined at each time $t$ in order to guarantee configurations $(\phi^* , \phi^{\vphantom{*}})_{t^{+}} \in \mathcal{M}$, where $t^{+}$ denotes a time infinitesimally larger than time $t$. 

\subsection{Numerical Method}
Here we present a general numerical strategy for propagating the constrained Langevin equations forward in fictitious time. We discretize time $t^{(\ell+1)} = t^{(\ell)} + \Delta t$ and choose $\Delta t \ll 1 $ to maintain a desired accuracy. For well-posedness, we select the initial field configuration to belong to the manifold $\mathcal{M}$, i.e. $\mathbf{g}[\phi^* , \phi^{\vphantom{*}}]^{(\ell=0)} = \mathbf{0}$. Using an initial condition on the manifold and choosing a small enough Langevin timestep $\Delta t$ will ensure that the projection step is well-defined \cite{alfonsiConstrainedOverdampedLangevin2022}. 

A projected numerical scheme with weak first-order accuracy is defined via a splitting approach. First, an unconstrained step is performed based on the unconstrained portion of the action $S[\phi^* , \phi^{\vphantom{*}}]$ to propagate the fields to an intermediate configuration, denoted $(\bar{\phi}^* , \bar{\phi}^{\vphantom{*}})$. Then, a projection step is taken so that $(\phi^* , \phi^{\vphantom{*}} ) \in \mathcal{M}$ at the next time point. Thus, a CL iteration proceeds via two intermediate steps for each field at time $t^{\ell}$: 
\begin{equation}
  (\bar{\phi}^*, \bar{\phi}) = \underset{\Delta t, S }{\text{Step}} \left [(\phi^* , \phi^{\vphantom{*}} )^{(\ell)}  \right ],
      \label{eq: step_1}
\end{equation} 

\begin{subequations}
    \begin{equation}
      \phi_{j} (\mathbf{r})^{(\ell+1)} = \bar{\phi}_{j} (\mathbf{r}) + \Delta t \sum_{m=1}^{M} \lambda^{(\ell+1)}_{m} \left (\frac{\delta g_{m}[\phi^* , \phi^{\vphantom{*}}]}{\delta \phi^*_{j}(\mathbf{r})} \right )^{(\ell)},
        \label{eq: phi_proj}
      \end{equation}
    \begin{equation}
       (\phi^*_{j} (\mathbf{r}))^{(\ell+1)} = \bar{\phi}^*_{j} (\mathbf{r}) + \Delta t \sum_{m=1}^{M} \lambda^{(\ell+1)}_{m} \left (\frac{\delta g_{m} [\phi^* , \phi^{\vphantom{*}}]}{\delta \phi_{j}(\mathbf{r})} \right )^{(\ell)} ,
       \label{eq: phistar_proj}
    \end{equation}
\end{subequations}
\begin{equation}
    s.t. \hspace{4px} \mathbf{g}[ \phi^* , \phi^{\vphantom{*}}]^{(\ell+1)} = \bm{0},
     \label{eq: g_ell}
\end{equation}
\noindent{where} $S[\phi^* , \phi^{\vphantom{*}}]$ is the unconstrained action. In equation\ (\ref{eq: step_1}), a first-order time stepping method is denoted as `Step' and propagates the CS fields to an intermediate configuration by taking an unconstrained complex Langevin step with the full noise incorporated. For example, this unconstrained step can be done using the ETD scheme in equations\ (\ref{eq: ETD_phi}) and (\ref{eq: ETD_phistar}) or another first-order time stepping method such as Euler-Maruyama. Then, the coherent state fields are projected onto the manifold $\mathcal{M}$ via the Lagrange multipliers in equations\ (\ref{eq: phi_proj}) and (\ref{eq: phistar_proj}). The $\lambda_{m}^{(\ell+1)}$ notation is used to clarify that updates\ (\ref{eq: phi_proj}) and (\ref{eq: phistar_proj}) cannot proceed until the Lagrange multipliers are determined by solving equation\ (\ref{eq: g_ell}), whose solution at time $t^{(\ell + 1)}$ is denoted by $\bm{\lambda}^{(\ell+1)}$. By weighting the Lagrange multipliers by the constraint functional derivatives evaluated using $(\phi^* , \phi^{\vphantom{*}}) \in \mathcal{M}$, the projection steps will ensure proper sampling of the constrained stationary distribution $e^{-S_{c}}$, assuming that the CL correctness criteria are additionally met \cite{bergerComplexLangevinOther2021a}. 

To proceed with the projection step in practice, the intermediate configurations $\bar{\phi}^{*} , \bar{\phi}^{\vphantom{*}}$ along with the constraint functional derivatives are computed and stored at each time step $(\ell)$, and then the constraint equations are expressed as explicit functions of $\bm{\lambda}$ by substituting the projected update equations\ (\ref{eq: phi_proj}) and (\ref{eq: phistar_proj}) into the constraints evaluated at $t^{(\ell + 1)}$, yielding $\mathbf{g} [\bm{\lambda}]^{\ell + 1} = \mathbf{0}$ to solve. 

For a general set of $M$ constraints $\mathbf{g}[\phi^* , \phi^{\vphantom{*}}] = \mathbf{0}$, nonlinear root-finding methods must be used. For nonlinear equations, spurious unphysical solutions exist and require thoughtful root selection and rejection strategies. Finding and choosing the physical root at every CL iteration is paramount to ensure robust sampling of the physical system of interest. Fortunately, there are physical models which involve linear or quadratic constraints, such that analytical solutions may exist at each CL iteration, and the physical root can be readily identified. Some examples include particle number conservation in the canonical ensemble or magnetization constraints in cold alkali atom mixtures, where the constraint(s) in both cases are bi-linear in the coherent state fields and are thus amenable to inexpensive and robust projection methods. On the other hand, the microcanonical ensemble involves a coupled particle number constraint and energy constraint that must be solved simultaneously using root-finding algorithms. 

\subsection{Canonical Ensemble}
For a continuum Bose fluid in the canonical ensemble, there is a single global constraint on the total particle number: 
\begin{equation}
 g_{N}[\phi^* , \phi^{\vphantom{*}}] = \tilde{N}[\phi^* , \phi^{\vphantom{*}} ] - N = 0
  \label{eq: N_global},
\end{equation}
\noindent{where} the definition of $\tilde{N}$ is the same as previously defined in equation\ (\ref{eq: N_CSfunctional}). Because of the constraint's bilinear form, the constraint's functional derivatives are effectively linear in the CS fields: 
\begin{equation}
 \frac{\delta \tilde{N} [\phi^* , \phi^{\vphantom{*}}]}{\delta \phi^*_{j}(\mathbf{r})} = \frac{1}{N_{\tau}} \phi^{\vphantom{*}}_{j-1} (\mathbf{r}), 
  \hspace{20px} 
  \frac{\delta \tilde{N} [\phi^* , \phi^{\vphantom{*}}]}{\delta \phi^{\vphantom{*}}_{j}(\mathbf{r})} = \frac{1}{N_{\tau}} \phi^*_{j+1} (\mathbf{r})  .
\end{equation}  
Furthermore, we define a Lagrangian $\mathcal{L}[\phi^* , \phi^{\vphantom{*}}; \lambda_{N}] = S_0 [\phi^* , \phi^{\vphantom{*}}] + \beta \tilde{U}[\phi^* , \phi^{\vphantom{*}}] - \lambda_{N} g_{N}[\phi^* , \phi^{\vphantom{*}}]$ and apply the scheme in equation\ (\ref{eq: CL_Lagrangian}).

For our numerical implementation, we easily generate initial configurations on $\mathcal{M}$ by choosing the spatially independent and $\tau$-independent configuration $\phi^{(l=0)} = (\phi^*)^{(l=0)} = \sqrt{N / V}$, which ensures that $\tilde{N}[\phi^* , \phi^{\vphantom{*}}]|_{t=0} = N$. After initialization, we propagate the coherent state fields forward in time via the two-step procedure discussed in the previous section. We take the unconstrained step defined in equation\ (\ref{eq: step_1}) based on the unconstrained canonical action $S_{0}[\phi^* , \phi^{\vphantom{*}}] + \beta \tilde{U}[\phi^* , \phi^{\vphantom{*}}]$. Then, we proceed with the following projection steps: 
\begin{subequations}
 \begin{equation}
\phi_{j} (\mathbf{r})^{(\ell+1)} = \bar{\phi}_{j} (\mathbf{r}) + \Delta t \lambda^{(\ell+1)}_{N} \left ( \frac{\delta \tilde{N} [\phi^* , \phi^{\vphantom{*}}]}{\delta \phi^*_{j} (\mathbf{r})} \right )^{(\ell)}, 
  \label{eq: phi_N_projection}
 \end{equation}
\begin{equation}
  \phi^*_{j} (\mathbf{r})^{(\ell+1)} = \bar{\phi}^{*}_{j} (\mathbf{r}) + \Delta t \lambda^{(\ell+1)}_{N} \left ( \frac{\delta \tilde{N} [\phi^* , \phi^{\vphantom{*}}]}{\delta \phi_{j} (\mathbf{r}) } \right )^{(\ell)}. \label{eq: phistar_N_projection}
\end{equation}
\end{subequations} 
\noindent{To} determine $\lambda_{N}$, we substitute the update equations\ (\ref{eq: phi_N_projection}) and (\ref{eq: phistar_N_projection}) into the particle number constraint equation\ (\ref{eq: N_global}), yielding an equation in terms of $\lambda^{(\ell + 1)}_{N}$. Because of the bilinear form of the particle number constraint, this procedures yields a complex-valued quadratic equation: 
\begin{equation}
A \lambda_{N}^2 + B\lambda_{N} + C = 0, 
\end{equation}
\noindent{where} we have denoted $\lambda^{(\ell + 1)}_{N} \to \lambda_{N}$ for convenience of notation, and the complex-valued coefficients of the quadratic are computed at each Langevin time step as a function of the constraint functional derivatives and the intermediate CS field configuration. The coefficients $A$, $B$, and $C$ are 
\begin{subequations}
\begin{align}
A &= \frac{1}{N_{\tau}} \sum_{j=0}^{N_{\tau}-1} \int d\mathbf{r} \left ( \frac{\delta \tilde{N}[\phi^* , \phi^{\vphantom{*}}]}{\delta \phi^{\vphantom{*}}_{j}(\mathbf{r})} \right )^{(\ell)}_{j} \left ( \frac{\delta \tilde{N}[\phi^* , \phi^{\vphantom{*}}]}{\delta \phi^*_{j}(\mathbf{r})} \right )^{(\ell)}_{j-1} , \label{eq: A} \\ 
 \begin{split}
 B &= \frac{1}{N_{\tau}} \sum_{j=0}^{N_{\tau}-1} \int d\mathbf{r} \left [ \bar{\phi}^*_{j} (\mathbf{r}) \left ( \frac{\delta \tilde{N}[\phi^* , \phi^{\vphantom{*}}]}{\delta \phi^{*}_{j}(\mathbf{r})} \right )^{(\ell)}_{j-1} \right. \label{eq: B}\\ 
  & \hspace{60px}  + \left. \left ( \frac{\delta \tilde{N}[\phi^* , \phi^{\vphantom{*}}]}{\delta \phi_{j}(\mathbf{r})} \right )^{(\ell)}_{j} \bar{\phi}^{\vphantom{*}}_{j-1} (\mathbf{r}) \right ]  ,
  \end{split} \\
C &= \tilde{N}[\bar{\phi}^* , \bar{\phi}^{\vphantom{*}} ] - N . \label{eq: C}
\end{align}
\end{subequations}

\noindent{Clearly}, the $C$ coefficient is the residual of the global particle number constraint and thus quantifies the deviation in the intermediate configuration from $\mathcal{M}$. Notably, $C \to 0$ as $\Delta t \to 0$ assuming $(\phi^* , \phi^{\vphantom{*}})^{\ell} \in \mathcal{M}$, so $\lambda_{N} \to 0$ is the expected solution as $\Delta t \to 0$, since no Lagrange multiplier is required to enforce the constraint. $A$ and $B$ are quantities that are expected to be real on average for the regime of ultracold atomic BECs. Complex-valued quadratic equations share the same solution as their real counterpart, which we invoke here: 
\begin{equation} 
  \lambda_{N} = \frac{-B + \sqrt{B^2 - 4AC}}{2A} .
   \label{eq: lambda} 
\end{equation}
While quadratic equations have two solutions, we have herein identified the positive solution as the physical one by requiring $\lambda \to 0$ in the limit where $C \to 0$ or equivalently $\Delta t \to 0$. Furthermore, care must be taken when working with complex-valued square-roots, which possess a branch-cut discontinuity assumed to be taken along the negative real-axis. Lapses in analyticity are speculated to lead to a breakdown in ergodicity in CL sampling \cite{bergerComplexLangevinOther2021a}. In particular, the discriminant $D \equiv B^2 - 4AC$ must be monitored throughout a CL simulation. In the regime of cold atomic gases with particle number constraints $N \sim 10^3 - 10^7$, we have found no instances of the discriminant winding around the origin.

A remarkable feature of the field-theoretic canonical ensemble method is its direct access to the Helmholtz free energy $A = -PV + \mu N$ \cite{fredricksonDirectFreeEnergy2022}, where $P$ is the thermodynamic pressure. A pressure field operator functional is readily obtained by quantifying the volume derivative of the internal energy portion of the action $P[\phi^* , \phi] \propto \partial \tilde{U} [\phi^* , \phi] / \partial V $, following a procedure in reference\ (\cite{fredricksonDirectFreeEnergy2022, fredricksonFieldTheoreticSimulationsSoft2023}). Furthermore, the chemical potential $\mu$ is directly conjugate to the particle number and thus can be directly related to $\lambda_{N}$. With the projected complex Langevin method, we recover the chemical potential with first-order accuracy in $\Delta t$: 
\begin{equation}
 \mu^{(\ell)} = \beta^{-1} \lambda^{(\ell)}_{N} + \mathcal{O} (\Delta t^2 ) . 
\label{eq: mu}
\end{equation}
As a result, we expect $\lambda_{N}$ to be real on average, but we emphasize that it need not be real at each CL step.  

This constrained complex Langevin algorithm enjoys the same $\mathcal{O} (N_{\tau} N^{d}_{x} \log (N_{\tau} N^{d}_{x}) )$ scaling as the unconstrained coherent state CL method, since the projection step does not require any additional Fast Fourier Transforms. Furthermore, the projection step is parallelized across the space and imaginary time lattice and allows the usage of parallel CPU and GPU hardware for efficient application to systems with fine spatial and imaginary time resolutions. 

\subsection{Microcanonical Ensemble}
For a continuum Bose fluid in the microcanonical ensemble, there is an internal energy constraint in addition to the particle number constraint (equation\ (\ref{eq: N_global})):
\begin{equation}
  g_{U} [\phi^* , \phi^{\vphantom{*}}] = \tilde{U} [\phi^* , \phi^{\vphantom{*}} ] - U = 0 .
   \label{eq: U_global}
\end{equation}
We define the Lagrangian $\mathcal{L}[\phi^* , \phi^{\vphantom{*}}; \lambda_{N}, \lambda_{U}] \equiv S_{0}[\phi^* , \phi^{\vphantom{*}}] - \lambda_{N} g_{N} [\phi^* , \phi^{\vphantom{*}}] - \lambda_{U} g_{U} [\phi^* , \phi^{\vphantom{*}}]$ and prescribe the CL scheme in equation\ (\ref{eq: CL_Lagrangian}). Sampling this scheme requires first functional derivatives of the energy constraint, which are detailed in Appendix \ref{appendixA} in equations\ (\ref{eq: dUdphistar}) and (\ref{eq: dUdphi}). 

After discretizing in fictitious time, we arrive at a similar two-step update algorithm at each CL time point $t_{\ell}$: 
\begin{equation}
(\bar{\phi}^*, \bar{\phi}) = \underset{\Delta t, S_{0} }{\text{Step}} \left [(\phi^* , \phi^{\vphantom{*}} )^{(\ell)}  \right ] ,
\end{equation}
\begin{equation}
  \begin{split}
    \phi_{j} (\mathbf{r})^{(\ell+1)} &= \bar{\phi}_{j} (\mathbf{r}) + \Delta t \lambda^{(\ell+1)}_{N} \left ( \frac{\delta \tilde{N} [\phi^* , \phi^{\vphantom{*}}] }{\delta \phi^*_{j} (\mathbf{r})} \right )^{(\ell)} \\ 
    &+ \Delta t \lambda^{(\ell+1)}_{U} \left ( \frac{\delta \tilde{U} [\phi^* , \phi^{\vphantom{*}}]}{\delta \phi^*_{j} (\mathbf{r})} \right )^{(\ell)} ,
    \label{eq: phi_MCE_projection}
    \end{split}
\end{equation}
\begin{equation}
  \begin{split}
    \phi^*_{j} (\mathbf{r})^{(\ell+1)} &= \bar{\phi}^{*}_{j} (\mathbf{r}) + \Delta t \lambda^{(\ell+1)}_{N} \left ( \frac{\delta \tilde{N} [\phi^* , \phi^{\vphantom{*}}]}{\delta \phi_{j} (\mathbf{r}) } \right )^{(\ell)} \\
    &+ \Delta t \lambda^{(\ell+1)}_{U} \left ( \frac{\delta \tilde{U} [\phi^* , \phi^{\vphantom{*}}]}{\delta \phi_{j} (\mathbf{r}) }\right )^{(\ell)} .
    \label{eq: phistar_MCE_projection}
    \end{split}
\end{equation}
\noindent{At} each CL iteration, we must determine the vector of multipliers $\bm{\lambda} \equiv [ \lambda_{N} , \lambda_{U} ]^{T}$ to satisfy both the particle number and energy constraints (equations\ (\ref{eq: N_global}) and (\ref{eq: U_global})) at the next time point. 

As before, the projected method requires an initial configuration that is near or on $\mathcal{M}$, such that $\mathbf{g}[(\phi^* , \phi^{\vphantom{*}})^{(\ell=0)}] \approx \mathbf{0}$. To generate such a configuration, we first run a brief canonical ensemble simulation at the particle number $N$ and temperature $T$ of interest in order to generate an equilibrated CS field configuration, making sure to adjust the temperature until the canonical ensemble simulation's average internal energy $\langle U \rangle$ is near the desired internal energy. We initialize the microcanonical ensemble's Lagrange multipliers using the $\beta$ and $\mu$ values from the reference canonical simulation. 

To determine the Lagrange multipliers at each iteration in the microcanonical ensemble, we express the constraints as a function of the multipliers $\mathbf{g}[\bm{\lambda}]$ by substituting the coherent state field updates (equations\ (\ref{eq: phi_MCE_projection}) and (\ref{eq: phistar_MCE_projection})) into the constraint equations (equations\ (\ref{eq: N_global} and (\ref{eq: U_global})). This yields two coupled constraint equations to be solved simultaneously for $\lambda^{(\ell + 1)}_{N}$ and $\lambda^{(\ell + 1)}_{U}$, which are detailed in Appendix \ref{appendixB}. While the particle number constraint is a complex-valued conic, the energy equation is a complex-valued quartic equation due to the interaction terms in the internal energy. As a result, numerical root finding methods must be used at each CL iteration. We use a modified Levenberg-Marquardt algorithm \cite{kanzowLevenbergMarquardtMethods2004} to determine $\bm{\lambda}^{(\ell + 1))}$ at each CL step, with the details provided in Appendix \ref{appendixC}. 

Care must be taken with ensuring that the solutions are physical. Nonlinear equations can host many spurious solutions which would spoil the CL simulation by contributing unphysical field configurations. To check for spurious solutions, we take a candidate solution $\bm{\lambda}_{0}$ and check that each component corresponds to the ($+$) solution of the particle number conic equation as a consistency check. If an unphysical solution is encountered, the solution candidate is discarded and the Langevin timestep is repeated with different realization of the noise. 

In the same vein as the Lagrange multiplier SDE method, the chemical potential $\mu$ and the inverse temperature $\beta$ are extracted in the microcanonical ensemble by ensemble averages of the Lagrange multipliers. An estimator for the inverse temperature is obtained by negating $\lambda_{U}$: 
\begin{equation}
 \beta = - \langle \lambda_{U} \rangle + \mathcal{O} (\Delta t^2) ,
 \label{eq: beta_operator}
\end{equation}
\noindent{where} angle brackets $\langle . \rangle$ denote the noise average over the Langevin process. Furthermore, the chemical potential can be extracted by a procedure similar to equation\ (\ref{eq: mu}): 
\begin{equation}
\mu = \frac{\langle \lambda_{N} \rangle}{\beta} + \mathcal{O} (\Delta t^2) .
 \label{eq: mu_MCE_operator}
\end{equation}

\begin{figure*}[t] 
\includegraphics[scale = 0.36]{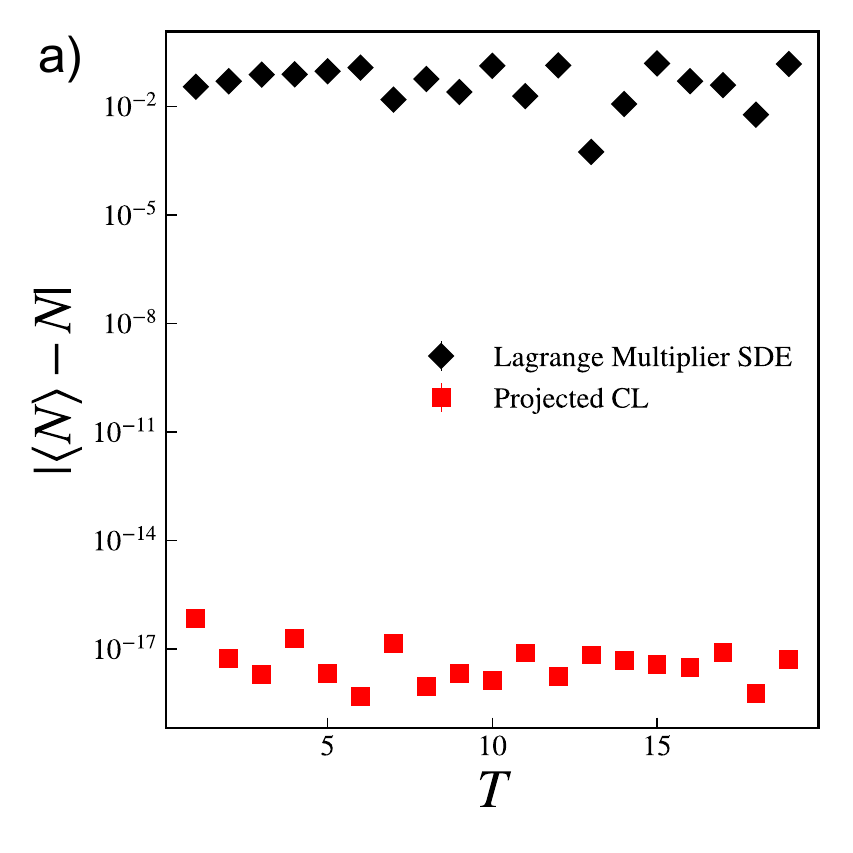}
\includegraphics[scale = 0.36]{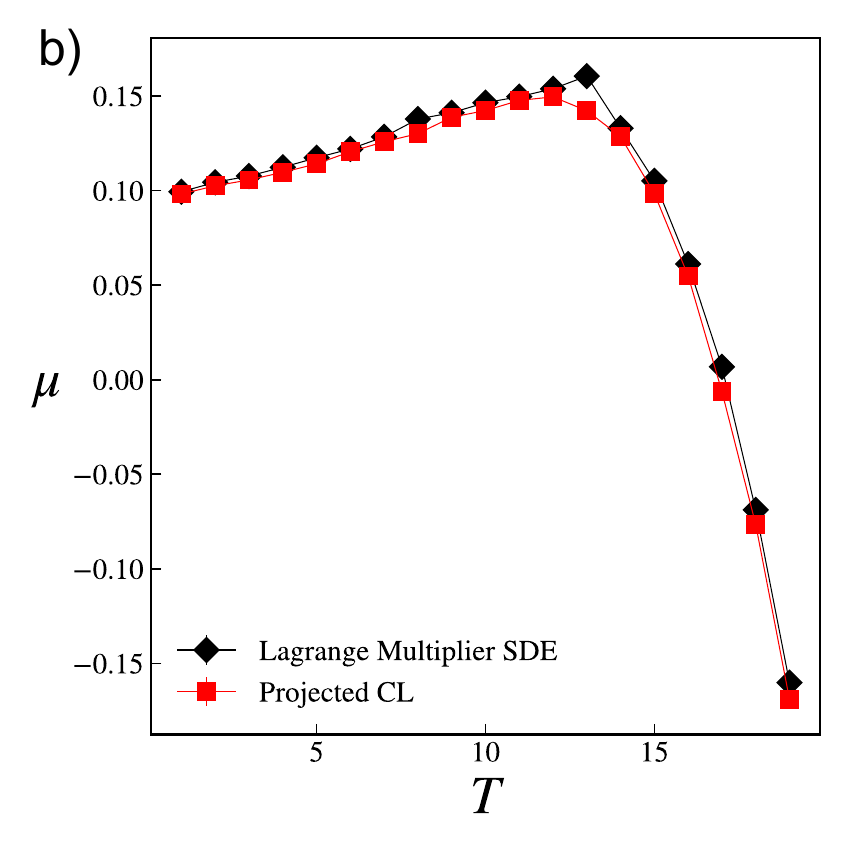}
\includegraphics[scale = 0.36]{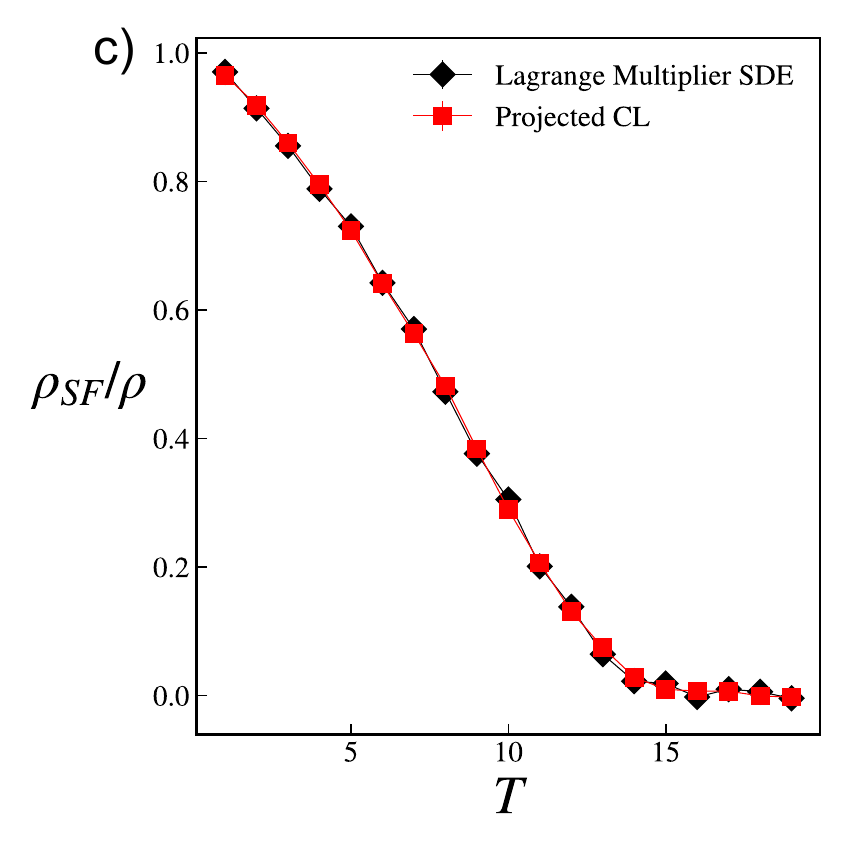}
\caption{Demonstration of the projected CL method (red squares) as compared to Lagrange Multiplier SDE method (black diamonds) for simulating a Bose fluid with $N = 1000$ particles across a range of temperatures (Kelvin units). a) Modulus of the particle number constraint residuals. b) Chemical potential and c) superfluid fraction estimates with both methods, confirming the accuracy of the projected CL method. All simulations were conducted with both methods using system parameters in two dimensions with $m = 4.0026$ Da, $u_0 = 0.1$ K\r{A}$^2$, $L = 32$ \r{A}, $\Delta x = 0.5$, and $\Delta t = 0.025$. While varying the temperature ($\beta^{-1}$), $N_{\tau}$ was simultaneously varied to maintain an approximately fixed imaginary time discretization $\Delta_{\tau} = 0.004$, where $N_{\tau}$ was rounded to the nearest integer and multiple of 4.  Error bars are calculated as the standard error of the mean using all equilibrated samples in the Langevin process. In some cases, error bars are smaller than the marker symbol.}  \label{fig: CE_Tsweep} 
\end{figure*}

\section{Numerical Demonstration and Discussion} \label{results}
It is imperative that the projected CL method reproduces the same unbiased thermodynamic observables as the Lagrange multiplier SDE method. By incorporating the constraint gradients into the projection updates, numerical solutions of the Langevin equations\ (\ref{eq: CL_Lagrangian}) will importance sample the constrained distribution of interest and reproduce the correct physics \cite{lelievreLangevinDynamicsConstraints2012a}, assuming that the CL correctness criteria are not violated \cite{bergerComplexLangevinOther2021a}. 

To demonstrate, we provide a direct comparison between the projected CL and Lagrange multiplier SDE methods in the context of cold atomic Bose-Einstein condensates. We compare the accuracy of each method for a range of temperatures and interaction strengths in the canonical and microcanonical ensembles, respectively. Furthermore, we provide a comparison of the numerical timestep convergence as well as the numerical stability of both methods.

To assess the stability of the each method, it is necessary to look at the simulation runtime for many simulation replicates with different pseudorandom number seeds and at a wide range of timesteps. To quantify the numerical stability at a given timestep, we compute the mean time to divergence \cite{vigilQuantitativeComparisonFieldUpdate2021} by calculating the harmonic mean of the $\tau_{\text{div},i}$ time it takes for the $i^{\text{th}}$ CL simulation to diverge:
\begin{equation}
    \bar{\tau}_{\text{div}} = \left ( \frac{1}{N_{\text{trials}}} \sum_{i=1}^{N_{\text{trials}}} \frac{1}{\tau_{\text{div}, i}} \right )^{-1} ,
\end{equation}
\noindent{where} $\bar{\tau}_{\text{div}}$ is the mean time until divergence, and $N_{\text{trials}}$ is the number of replicates initialized with different random number seeds. We use $N_{\text{trials}} = 10$ for all estimates of the mean time to divergence in this work and run each simulation for $2 \times 10^6$ timesteps. We consider a simulation to be divergent if any field value registers as +INF, -INF, or NaN, where it fails to be represented by a IEEE-754 double precision floating point number.

\subsection{Canonical Ensemble}
We begin by comparing the methods across a range of temperatures for a Bose fluid in the canonical ensemble. This temperature range provides a comparison in three regimes of interest: 1) the low temperature quantum regime at $T \ll T_{c}$ where the system is mostly superfluid, 2) near the critical point $T \sim T_{c}$, and 3) above the critical point in the classical, normal-fluid regime. Importantly, thermal fluctuation effects are paramount for regimes 2) and 3) away from the $T \to 0$ limit. At low temperature, thermal fluctuations are negligible, but quantum fluctuations remain.  

Figure\ (\ref{fig: CE_Tsweep}a) compares the performance of each method in terms of satisfying the particle number constraint across the temperature range of interest. The projected approach satisfies the desired constraint to machine precision throughout the temperature range, while the non-projected approach plateaus at $10^{-2}$. For the Lagrange multiplier SDE method, increasing the mobility $\alpha$ leads to tighter enforcement of the constraint, since the Lagrange multiplier is more responsive to deviations away from constraint satisfaction. All canonical ensemble simulations conducted using the Lagrange multiplier SDE approach used a mobility of $\alpha = 0.01$ to ensure numerical stability. 

Furthermore, we benchmark the projected CL method to confirm its accuracy and ability to produce unbiased thermodynamics. Figures\ (\ref{fig: CE_Tsweep}b) and (\ref{fig: CE_Tsweep}c) provide a comparison of the accuracy of both canonical ensemble methods when estimating thermodynamic quantities of interest, namely the chemical potential $\mu$ and the superfluid fraction $\rho_{\text{SF}} / \rho$. The chemical potential is determined via equations\ (\ref{eq: mu}) and (\ref{eq: mu_noProjection}) for each method, while the superfluid fraction is calculated by subtracting off the normal fluid density, which is proportional to the population variance of the fluid's momentum \cite{mcgarrigleEmergenceSpinMicroemulsion2023}:
\begin{equation}
   \rho_{\text{SF}} = \langle \rho \rangle - \frac{\beta}{2 m V} \sum_{\nu} \left [ \langle P^2_{\nu} \rangle - \langle P_{\nu} \rangle^2 \right] ,
    \label{eq: rhoSF}
\end{equation}
\noindent{where} we have utilized the bulk particle number density, which is obtained by $\langle \rho \rangle = \langle \tilde{N}[\phi^* , \phi^{\vphantom{*}}] \rangle / V$. Furthermore, $P_{\nu}$ is the momentum of the bulk fluid in the $\nu$ direction, which is obtained by averaging a functional of the coherent state fields $\tilde{P}_{\nu} [\phi^* , \phi^{\vphantom{*}}] = 1/N_{\tau} \sum_{j=0}^{N_{\tau}-1} \int d\mathbf{r}\ \phi^*_{j} (\mathbf{r}) [ -i \hbar \partial_{\nu} ] \phi^{\vphantom{*}}_{j-1} (\mathbf{r})$. 
We observe excellent agreement between both methods, highlighting the projected method's ability to capture the system's thermodynamics without approximation. The tight agreement between both methods' superfluid fraction estimates emphasizes the projected method's ability to capture the fluctuation spectrum and excitations with full accuracy.   

\begin{figure}[ht] 
\includegraphics[scale = 0.28]{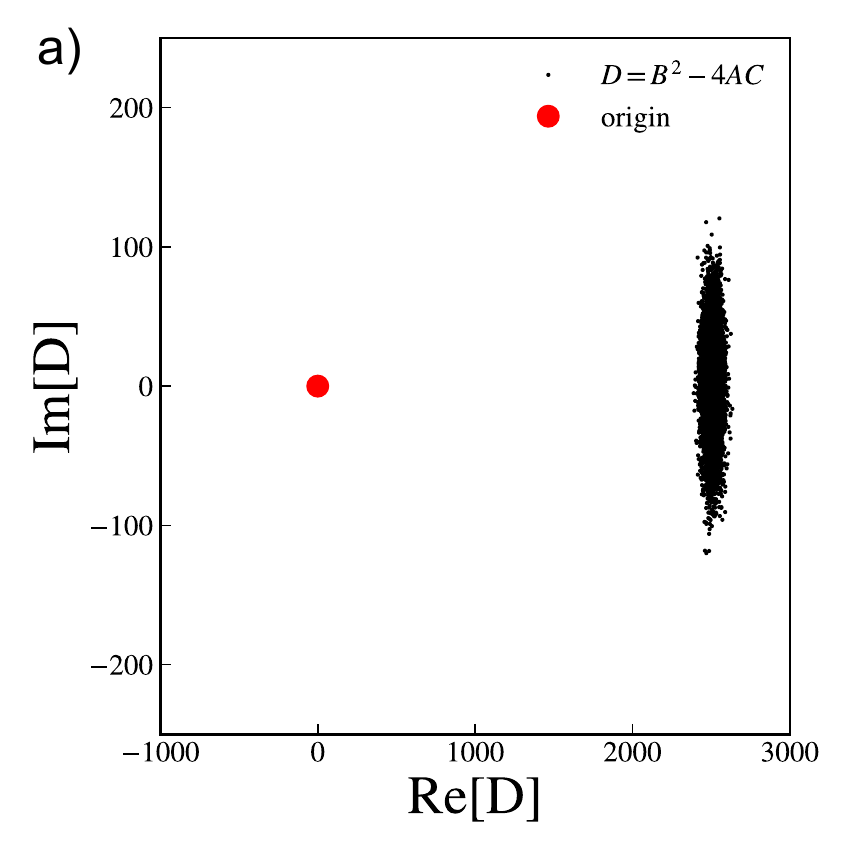}
\includegraphics[scale = 0.28]{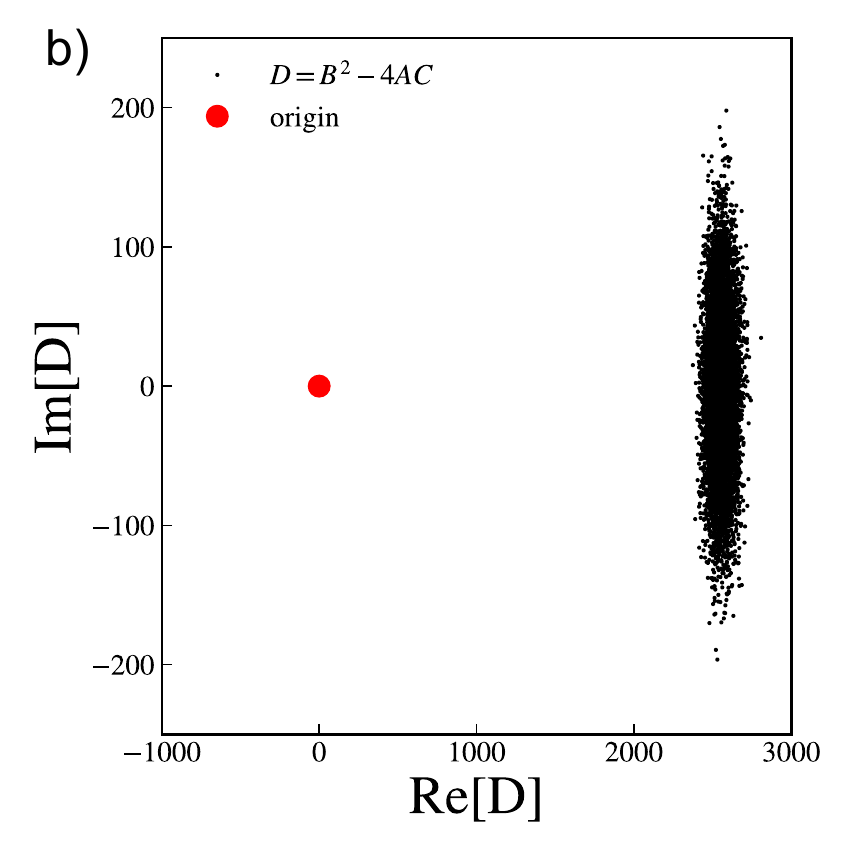}
\caption{Discriminant behavior in the complex plane for two canonical ensemble simulations conducted with projected CL, showing 10000 samples each. a) $T = 4$ K with $N_{\tau} = 72$, and b) $T = 15$ K with $N_{\tau} = 32$. Both simulations used the ETD algorithm with $\Delta t = 0.025$ and consisted of $N=1000$ particles in two dimensions with system parameters $m = 4.0026$ Da, $u_0 = 0.1$ K\r{A}$^2$, $L = 32$ \r{A}, and $\Delta x = 0.5$. The origin of the complex plane (D = 0) is depicted with a filled red circle.}  \label{fig: CE_discriminant} 
\end{figure}

As discussed previously, CL may experience failures in accuracy if system configurations are allowed to explore non-analytic portions of the complex plane. In the canonical ensemble, there is a danger present when calculating the Lagrange multiplier $\lambda_{N}$ using the square root function, which contains a branch cut discontinuity along the negative real axis. Figure\ (\ref{fig: CE_discriminant}) shows samples of the discriminant $D = B^4 - 4AC$ calculated via the quadratic coefficients in equations\ (\ref{eq: A}), (\ref{eq: B}), and (\ref{eq: C}) for both low temperature and high temperature examples. In both cases, we observe a localized behavior of the discriminant on the positive real axis and far away from the origin $(\text{Re}[D] \gg 0)$. Although there are significant fluctuations in the discriminant's imaginary part during CL simulations, the imaginary part vanishes upon sample averaging. We see a greater variance in the discriminant's imaginary part in Figure\ (\ref{fig: CE_discriminant}b), due to the stronger thermal fluctuations at higher temperature. Importantly, the discriminant does not wind around the origin and cross the branch cut discontinuity on the negative real axis in either case. Therefore, Figure\ (\ref{fig: CE_discriminant}) supports our conclusions from our method comparison in Figures (\ref{fig: CE_Tsweep}a) and (\ref{fig: CE_Tsweep}b), where the projected CL results are expected to be accurate because the discriminant avoids the branch cut for all time steps. 

\begin{figure}[ht] 
\includegraphics[scale = 0.28]{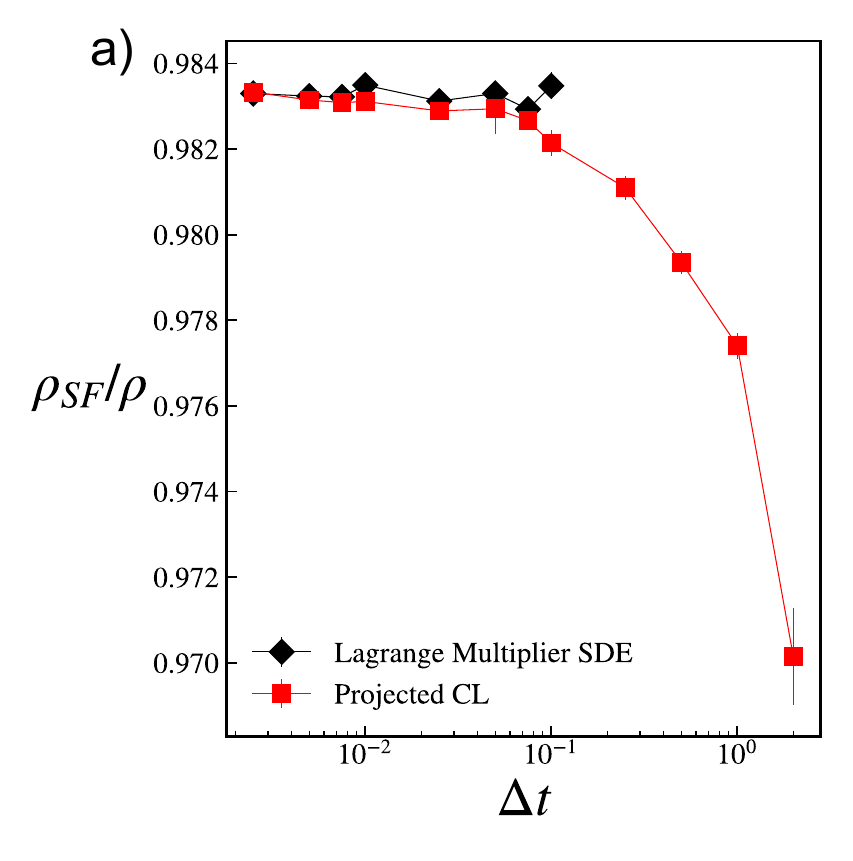}
\includegraphics[scale = 0.28]{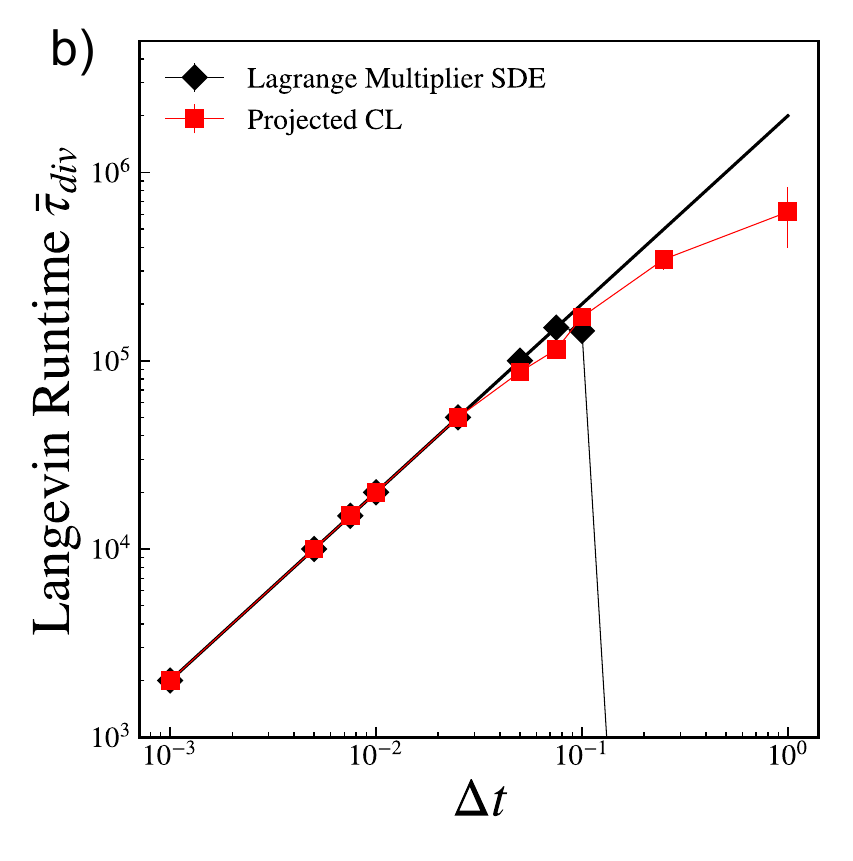}
\caption{Numerical timestep convergence and stability comparison of the projected CL (red squares) and Lagrange multiplier SDE methods (black diamonds) for a $N = 3000$ particle system in the canonical ensemble. a) Timestep convergence for the superfluid fraction, and b) mean time-to-divergence, where data points lying on the solid line signal robust numerical stability. Each simulation in b) was run for 2 million CL steps, where the solid line depicts the maximum possible Langevin time. All simulations were conducted in two dimensions using the ETD algorithm and system parameters $m = 4.0026$ Da, $u_0 = 0.15$ K\r{A}$^2$, $L = 32$ \r{A}, $\Delta x = 0.8$, and $N_{\tau} = 64$ with $\beta = 0.5$ K$^{-1}$.}  \label{fig: CE_numerical} 
\end{figure}

Next, we compare the numerical properties of each method. Figure\ (\ref{fig: CE_numerical}a) shows the numerical timestep bias for both methods for estimating the superfluid fraction. The Lagrange multiplier SDE method was unstable for larger timesteps $\Delta t > 0.1$, so no data was collected in that range. The projected CL method shows the expected behavior as a weak-first order method, and our comparison highlights projected CL's wider range of numerical stability. 

Figure\ (\ref{fig: CE_numerical}b) shows a comparison of the mean time-to-divergence between the projected CL and Lagrange multiplier SDE methods across a range of timestep discretizations. The solid line depicts the maximum amount of Langevin time possible for a simulation conducted at a timestep $\Delta t$ because each simulation ran for $2 \times 10^{6}$ timesteps. Both methods are quite stable until timesteps $\Delta t > 0.075$, where the Lagrange multiplier SDE method shows a sharp decline in numerical stability. In contrast, the projected CL method is quite stable even with timesteps $\Delta t \sim \mathcal{O}(1)$, where the mean time-to-divergence decreases modestly. 

The projected CL method shows a clear stability improvement over the Lagrange multiplier SDE method. The stability limitation of the previous method is not surprising given the poor stability of Euler-Maruyama schemes at larger timesteps. Furthermore, the projected CL method limits the exploration of the enlarged complexifed phase space by precisely keeping the coherent state fields on the physical manifold $\mathcal{M}$. As a result, escapes of the physical basin and divergence events are less likely to occur compared to schemes where the coherent state fields are allowed to deviate from $\mathcal{M}$, as in the Lagrange multiplier SDE approach.  

The boson canonical ensemble problem is a case where projected CL is particularly efficient due to the bi-linear form of the particle number constraint. Because iterations are not required to compute the physical Lagrange multiplier solution, more sophisticated numerical methods are easily conceivable. For example, an implicit projection numerical scheme involves modifying equations\ (\ref{eq: phi_N_projection}) and (\ref{eq: phistar_N_projection}) by replacing the constraint derivatives with their values at the $\emph{next}$ time point $(\ell + 1)$. In practice, such a scheme can be achieved by iterating the projection step with updated constraint gradients at each iteration until convergence within a desired tolerance. We anticipate this implicit scheme would further improve the numerical stability of the projected CL method. Furthermore, higher-order schemes such as predictor-corrector methods are conceivable, where projection steps are incorporated into both predictor and corrector stages. Such a scheme is anticipated to achieve weak second-order accuracy.  

\subsection{Microcanonical Ensemble}
Having demonstrated the efficacy and accuracy of projected CL in the canonical ensemble, we now turn to the microcanonical ensemble. The microcanonical ensemble is significantly more challenging, since two coupled nonlinear equations must be solved iteratively for the multipliers $\lambda_{N}$ and $\lambda_{U}$ at each CL step. While the root finding procedure reduces projected CL's efficiency, the number of iterations required is not prohibitive for simulating large systems in up to 3 spatial dimensions. Starting from a well-posed initial CS field configuration with appropriate initial guesses for $\bm{\lambda}$, we find that our modified Levenberg-Marquardt algorithm converges in under $10$ iterations per CL timestep when choosing a timestep of $\Delta t < 0.05$. For the Lagrange multiplier SDE method, we found that a mobility ratio $\alpha_{U} / \alpha_{N} = 0.2$ worked well to ensure numerical stability during equilibration and production sampling.

\begin{figure*}[t] 
\includegraphics[scale = 0.36]{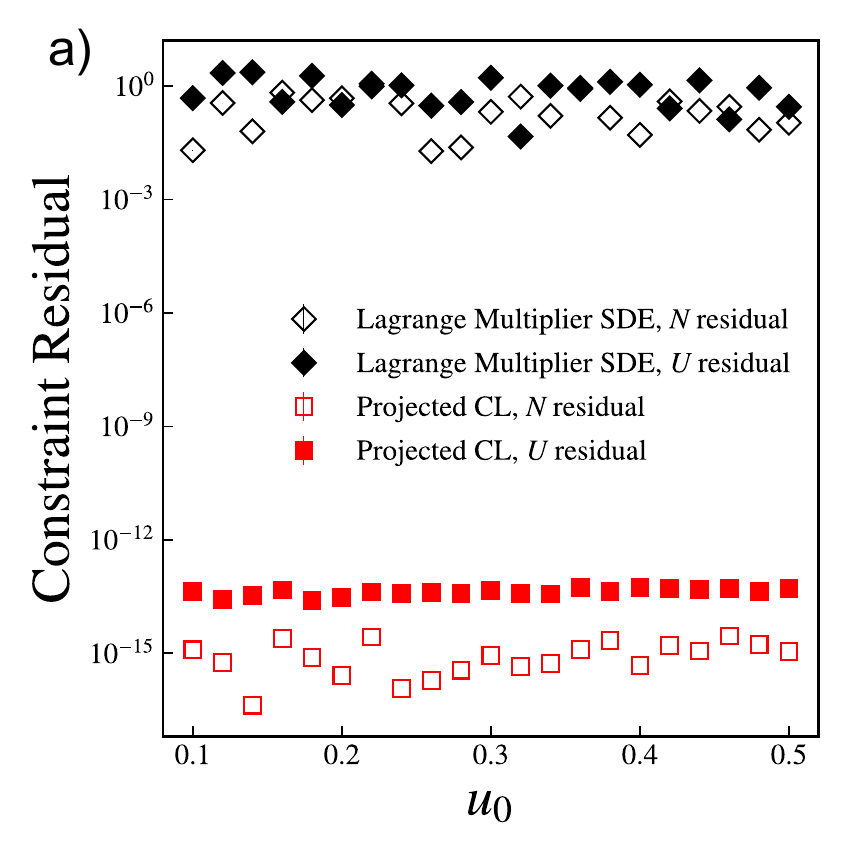}
\includegraphics[scale = 0.36]{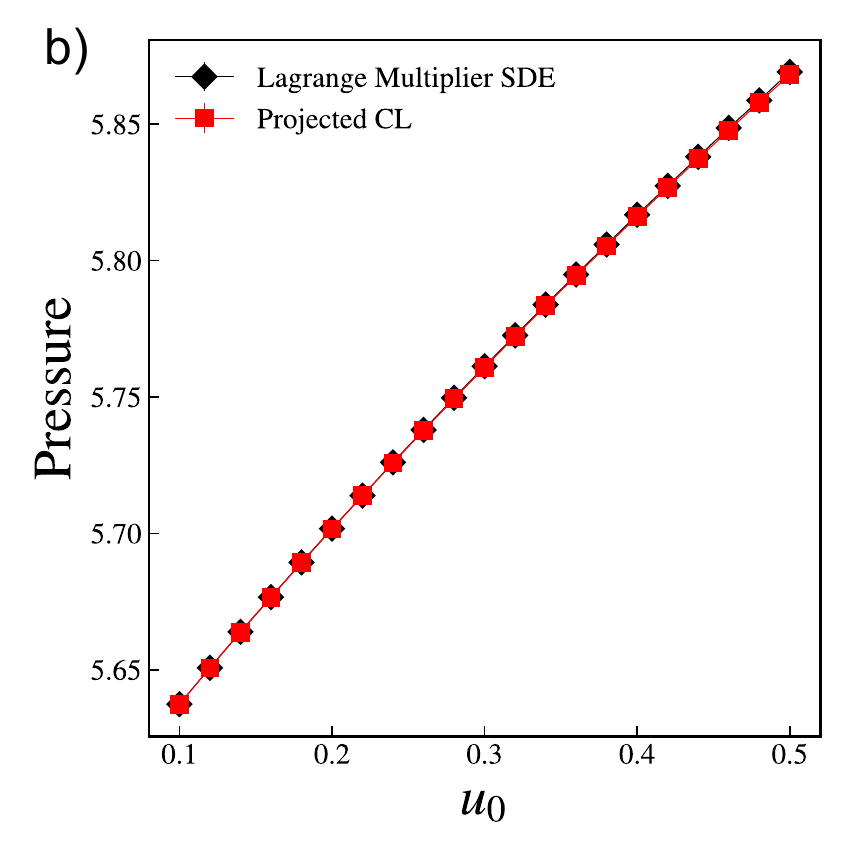}
\includegraphics[scale = 0.36]{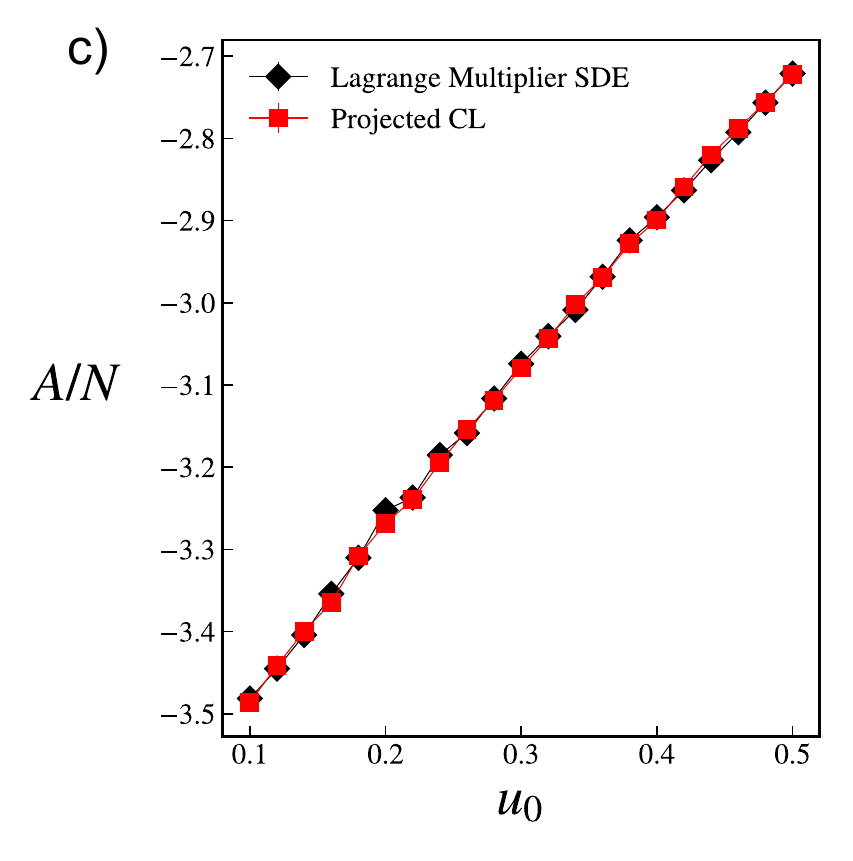}
\caption{Comparison of the projected CL method (red squares) with the Lagrange Multiplier SDE method (black diamonds), for simulating a Bose fluid with $N = 12000$ particles with internal energy $U = 66800$ Kelvin in the microcanonical ensemble for varying interaction strength $u_{0}$ (K\r{A}$^3$ units). a) Modulus of the particle number (open markers) and energy (filled markers) constraint residuals. b) Pressure and c) Helmholtz free energy per particle estimates with both methods to confirm the accuracy of the projected CL method in the microcanonical ensemble. All simulations were conducted with both methods via the ETD algorithm using system parameters in three dimensions with $m = 23$ Da, $L = 20$ \r{A}, $N_{\tau} = 32$, $N_{x} = 24$ plane waves in each direction, and $\Delta t = 0.005$. All Lagrange multiplier SDE simulations were performed with a mobility parameter $\alpha_{N} = 0.0025$ and the mobility ratio $\alpha_{U} = 0.2 \alpha_{N}$. For both methods, error bars are standard errors of the mean from the Langevin sampling process.}  \label{fig: MCE_u0sweep} 
\end{figure*}

Figure\ (\ref{fig: MCE_u0sweep}a) compares each method in terms of its accuracy and ability to satisfy the particle number and energy constraints simultaneously, considering a range of interaction strengths. As anticipated, projected CL ensures both constraints are satisfied to near machine precision. The numerical root-finding procedure is iterated to a specified tolerance, which governs how well the constraints are satisfied. We anticipate that an even stricter tolerance than $10^{-14}$ would lead to a decrease in the $U$ residual for projected CL shown in Figure\ (\ref{fig: MCE_u0sweep}a).

We assess the accuracy of both methods by studying two thermodynamic quantities of interest, namely the pressure $P$ and the Helmholtz free energy per particle $A/N$. The pressure can be determined by ensemble averaging a pressure field operator functional \cite{fredricksonDirectFreeEnergy2022}:
\begin{equation}
 \begin{split}
\tilde{P}[\phi^* , \phi^{\vphantom{*}}] &= \frac{1}{N_{\tau}V} \sum_{j=0}^{N_{\tau}-1} \int d\mathbf{r}\ \left [ \phi^*_{j} (\mathbf{r}) \left (\frac{-\hbar^2 \nabla^2}{m d}  \right ) \phi^{\vphantom{*}}_{j-1} (\mathbf{r}) \right. \\
 &+ \left. \frac{ u_{0} }{2} ( \phi^*_{j} (\mathbf{r}) \phi^{\vphantom{*}}_{j-1} (\mathbf{r}) )^2  \right ]  , 
   \end{split} \label{eq: Pressure}
\end{equation}
\noindent{where} $P = \langle \tilde{P}[\phi^* , \phi] \rangle$ and $P = -\langle \lambda_{U} \tilde{P} \rangle / \beta $ in the canonical and microcanonical ensembles, respectively. Using the pressure and chemical potential estimates, the Helmholtz-free energy is determined via equation\ (\ref{eq: Helmholtz}). Despite the difference in how the methods enforce the constraints, Figures\ (\ref{fig: MCE_u0sweep}b) and (\ref{fig: MCE_u0sweep}c) demonstrate how both methods produce unbiased thermodynamic results, as we see excellent agreement between the two methods. 

\begin{figure}[h] 
\includegraphics[scale = 0.28]{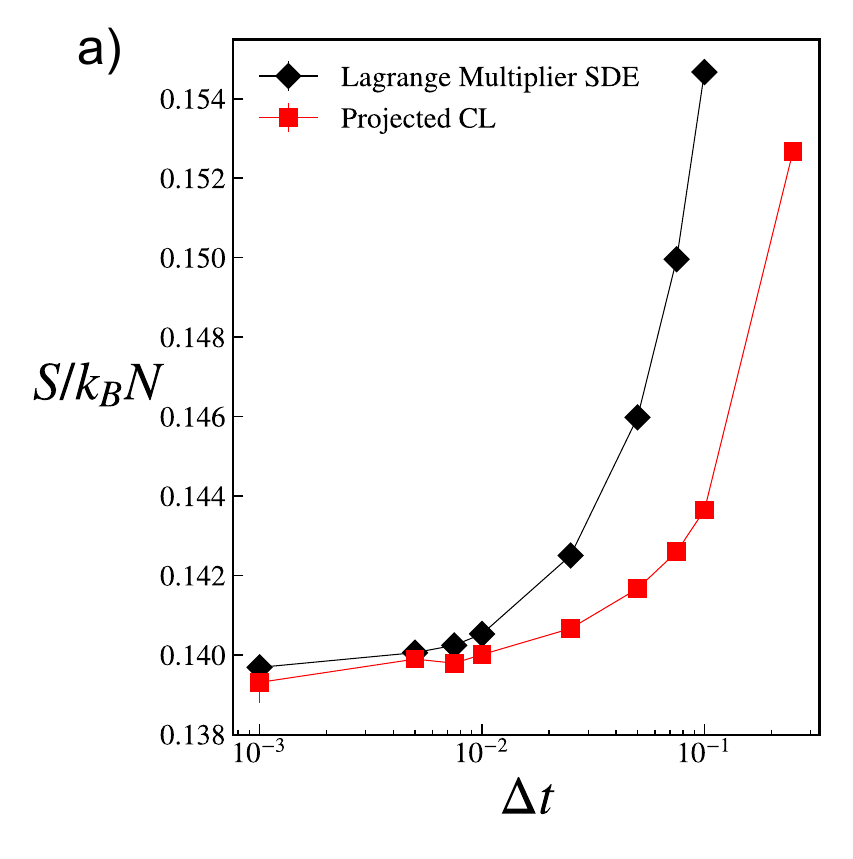}
\includegraphics[scale = 0.28]{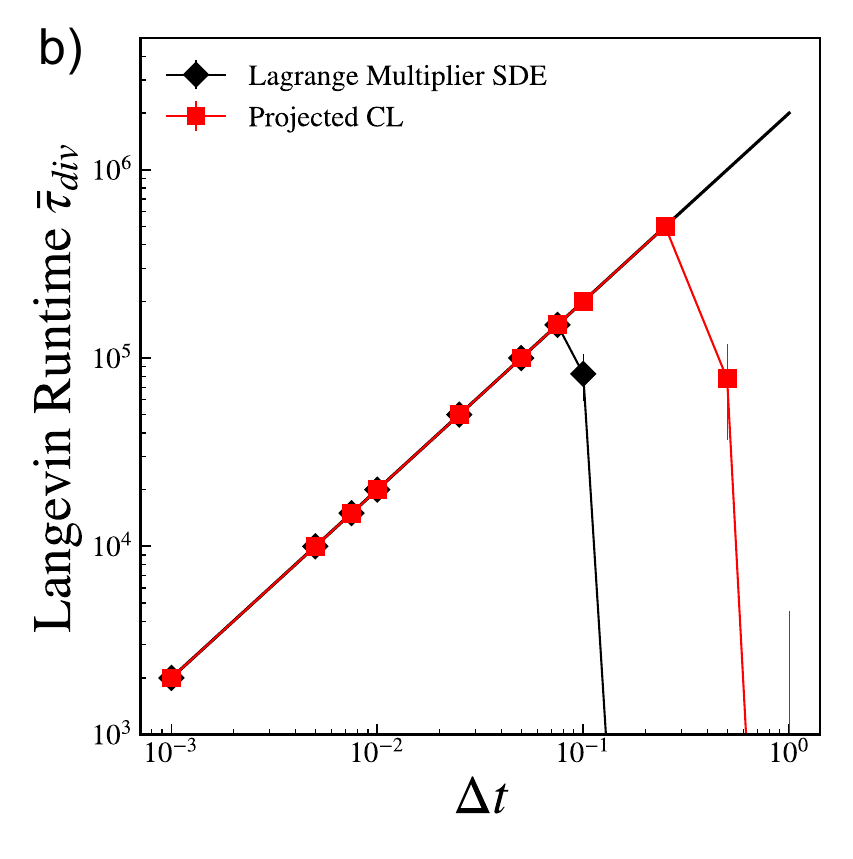}
\caption{Numerical timestep convergence and stability comparison of the projected CL (red squares) and Lagrange multiplier SDE (black diamonds) methods for a two-dimensional $N = 8000$ particle system in the microcanonical ensemble with energy $U = 16000$ Kelvin. a) Timestep convergence for the entropy per particle, and b) mean time-to-divergence, where data points lying on the solid line signal robust numerical stability. Each simulation in b) was run for 2 million CL steps, where the solid line depicts the total Langevin time accessible. All simulations were conducted using the ETD algorithm with system parameters $m = 8.0$ Da, $u_0 = 0.08$ K\r{A}$^2$, $L = 30$ \r{A}, $N_{x} = 30$ plane waves in each direction, and $N_{\tau} = 28$.}  \label{fig: MCE_numerical} 
\end{figure}

Next, we compare the timestep convergence and numerical stability of the two microcanonical ensemble methods. We study the timestep bias in Figure\ (\ref{fig: MCE_numerical}a) for calculating the entropy per particle of an interacting Bose fluid. Figure\ (\ref{fig: MCE_numerical}a) shows typical behavior for a weak first-order stochastic method, with projected CL performing significantly better in terms of accuracy at larger time steps. 

Moreover, we compare the numerical stability of the two methods by quantifying the mean time-to-divergence, shown in Figure\ (\ref{fig: MCE_numerical}b). For the Lagrange multiplier SDE approach, a mobility of $\alpha_{N} = 10^{-4}$ was used to ensure a favorable range of numerical stability. With that choice of mobility, the time-to-divergence plot shows a drop in numerical stability for $\Delta t > 0.05$ for the Lagrange multiplier SDE method. The projected method maintains robust stability throughout the range of timesteps until $\Delta t \geq 0.25$. For the larger timestep range, the numerical root-finding procedure requires many iterations to converge to a solution because the unconstrained step will produce an intermediate configuration farther from $\mathcal{M}$. Furthermore, significant time integration errors may occur for the unconstrained step at larger timesteps, leading to situations where the root-finding procedure fails to find a physical solution. In those cases, the fields may escape the physical basin and sample a divergent trajectory. It is therefore not surprising that we observe a diminished numerical stability for projected CL at large timesteps in the microcanonical ensemble. 

\section{Conclusion and Outlook} \label{conclusion}
We have detailed a general formalism and implementation of a constrained complex Langevin sampling method via projection. Our Lagrangian formalism provides a conceptually straightforward approach for incorporating constraints into a weak first-order complex Langevin algorithm without requiring modification of the noise or gradient descent scheme. As such, the generalized projected CL approach should be applicable to a wide variety of physical systems, especially those suffering from the sign problem. 

We report several advantages of the projected Langevin method: Firstly, the projected method does not require extensive hyperparameter tuning to ensure stable sampling, whereas the mobility parameter(s) in the Lagrange Multiplier SDE approach must be tuned to ensure numerical stability. Second, the projected CL method boasts superior numerical stability in a wide set of system parameters and at larger fictitious timestep discretizations. While the method does incur numerical bias at larger timesteps, such simulations provide a fast qualitative understanding of the systems' properties and equilibrium phase behavior. Lastly, we find some physical system examples where the projected method maintains its stability while the stochastic Lagrange Multiplier approach either fails from instability or requires expensive fully implicit numerical methods.

We find a natural application of projected CL to systems of ultracold BECs in the microcanonical and canonical ensembles. In this setting, projected CL emerges as a powerful simulation method that is more representative of modern cold-atom experiments and  provides access to unbiased thermodynamic properties. Moreover, projected CL provides more efficient access to thermodynamic potentials like the Helmholtz free energy or entropy, which facilitate the construction of phase diagrams. Our comparison of projected CL to the previous state-of-the-art highlights a significant improvement in numerical stability. Our study also highlights the canonical ensemble as an application where projected CL is particularly efficient.  

Although we observe success with projected CL's application to ultracold Bose gases, there remain open challenges with adapting projected CL to problems across statistical physics. A successful projected CL method requires robust root finding algorithms and strategies for root rejection and selection. In complex Langevin sampling, we cannot reject solutions with non-zero imaginary parts without incurring bias, even if we expect real-valued solutions on average. Furthermore, we anticipate that some statistical theories may present greater difficulties than others. For the example of a classical small molecule or polymer fluid \cite{fredricksonEquilibriumTheoryInhomogeneous2005} with a local incompressibility constraint, the representative auxiliary field theory has non-local character that leads to transcendental projection equations to solve at every CL timestep. While approximations can reduce the complexity of the projection equations, these approaches are likely limited in applicability to weakly fluctuating regimes where the Lagrange multiplier solution does not deviate significantly from the solution at the previous time point. 

Projected CL is anticipated to perform well for situations where the constraints have reduced complexity in form and number. In such scenarios, the required computation time and number of spurious solutions will be greatly reduced. For example, projected CL would provide robust and efficient access to multi-species cold atom systems with fixed particle numbers or spinor BECs with a global magnetization constraint. Projected CL's enhanced numerical stability would enable more careful and detailed studies of phase transitions of spinor BEC systems, where rich low-temperature phase diagrams exist \cite{stamper-kurnSpinorBoseGases2013a}. In a quantum lattice boson system such as an optical lattice system \cite{blochUltracoldQuantumGases2005} or Hubbard model \cite{fredricksonFieldTheoreticSimulationsSoft2023}, a particle number filling constraint may be imposed locally at each site. In such a case, the Lagrange multipliers can be determined in parallel because each site's filling constraint is independent. Furthermore, projected CL may enable an efficient numerical approach for quantum spin lattice models via a Schwinger boson construction, where particle number constraints enforce a faithful mapping to spin-S physics \cite{arovasFunctionalIntegralTheories1988a}. In all cases, convergence of CL trajectories must be studied to ensure correctness. 

\section*{Acknowledgements}
This work was enabled by field-theoretic simulation tools developed under support from the National Science Foundation (CMMT Program, DMR-2104255). We thank Chris Balzar, Kris Delaney, and Timothy Quah for fruitful discussions.  Use was made of computational facilities purchased with funds from the NSF (CNS-1725797) and administered by the Center for Scientific Computing (CSC). This work made use of the BioPACIFIC Materials Innovation Platform computing resources of the National Science Foundation Award No. DMR-1933487. The CSC is supported by the California NanoSystems Institute and the Materials Research Science and Engineering Center (MRSEC; NSF DMR 2308708) at UC Santa Barbara. E.C.M acknowledges support from a Mitsubishi Chemical Fellowship. 
H.D.C. acknowledges support from the National Science Foundation Award No. DMS-2410252.

\appendix

\section{Functional Derivatives and Langevin Forces} \label{appendixA}
The Langevin equations of motion require first functional derivatives of the action $S[\phi, \phi^*]$ in order to sample the corresponding stationary distribution $e^{-S}$. In the complex Langevin literature, these functional derivatives are often referred to as ``forces'' or ``drift terms''. For both statistical ensembles considered in this work, the functional derivatives of the internal energy are required for CL sampling: 
\begin{subequations}
\begin{equation}
 \begin{split}
  \frac{\delta \tilde{U}[\phi^* , \phi^{\vphantom{*}}]}{\delta \phi^*_{j} (\mathbf{r})} &= -\frac{1}{N_{\tau}} \left( \frac{\hbar^2 \nabla^2 }{2m} \right ) \phi_{j-1} (\mathbf{r}) \\ 
 &+ \frac{\phi_{j-1} (\mathbf{r})}{N_{\tau}} \int d\mathbf{r'}\ u(|\mathbf{r} - \mathbf{r'}|) \phi^*_{j} (\mathbf{r'}) \phi^{\vphantom{*}}_{j-1} (\mathbf{r'}) , \label{eq: dUdphistar}
  \end{split}
\end{equation}
\begin{equation}
 \begin{split}
  \frac{\delta \tilde{U}[\phi^* , \phi^{\vphantom{*}}]}{\delta \phi_{j} (\mathbf{r})} &= -\frac{\phi_{j+1} (\mathbf{r})}{N_{\tau}} \left( \frac{\hbar^2 \nabla^2 }{2m} \right ) \\ 
  &+ \frac{\phi^{*}_{j+1} (\mathbf{r})}{N_{\tau}} \int d\mathbf{r'}\ u(|\mathbf{r} - \mathbf{r'}|) \phi^*_{j+1} (\mathbf{r'}) \phi^{\vphantom{*}}_{j} (\mathbf{r'}) . \label{eq: dUdphi}
 \end{split}
\end{equation}
\end{subequations}
\noindent{In} practice, our pseudospectral implementation leverages FFT algorithms to evaluate the pair-interaction convolution as well as the Laplacian operators efficiently and with spectral accuracy \cite{fredricksonFieldTheoreticSimulationsSoft2023}. 

The exponential-time-differencing updates in equations\ (\ref{eq: ETD_phi}) and (\ref{eq: ETD_phistar}) use Fourier transforms of the functional derivatives to diagonalize portions of the forces. Separating the linear and non-linear portions of the functional derivatives can dramatically improve numerical stability. In the canonical ensemble, the linear force coefficients are 
\begin{equation}
  A_{\mathbf{k}, n} = 1 - e^{- 2\pi i n / N_{\tau}} \left [1 + \frac{\beta}{N_{\tau}}\frac{\hbar^2 \mathbf{k}^2}{2m} \right ]   , 
  \label{eq: A_coeff}
\end{equation}
\noindent{for} each wavevector $\mathbf{k}$ and Matsubara frequency indexed by $n$. The linear coefficient for the equation of motion for $\phi^*$ is denoted as $A^*_{\mathbf{k}, n}$, which is the exact complex-conjugate of $A_{\mathbf{k}, n}$ in this case. In the microcanonical ensemble, the contribution proportional to $\hbar^2 \mathbf{k}^2 / 2m$ is omitted from both $A_{\mathbf{k},n}$ and $A^*_{\mathbf{k},n}$. Instead, the kinetic energy terms are incorporated in the projection steps via equations\ (\ref{eq: dUdphistar}) and (\ref{eq: dUdphi}) or taken explicitly in $\mathcal{N}$ or $\mathcal{N}^*$ for the Lagrange Multiplier SDE method. 

Furthermore, we detail the nonlinear forces used in the explicit portion of the ETD time-integration algorithm, in the microcanonical ensemble as an example. For the Lagrange multiplier SDE method, the forces that incorporate the Lagrange multipliers are additionally included in these terms to be treated explicitly: 
\begin{subequations}
 \begin{equation}
 \mathcal{N}_{\mathbf{k}, n}[\phi^* , \phi^{\vphantom{*}} ; \psi_{N}, \psi_{U}] = \mathcal{F} \left [ i \psi_{U} \left ( \frac{\delta \tilde{U}}{\delta \phi^*_{j} (\mathbf{r})} \right ) + i\psi_{N} \left( \frac{\delta \tilde{N}}{\delta \phi^*_{j} (\mathbf{r})} \right ) \right ] \label{eq: N_appendix}
 \end{equation}
  \begin{equation}
  \mathcal{N}^*_{\mathbf{k}, n}[\phi^* , \phi^{\vphantom{*}} ; \psi_{N}, \psi_{U}] = \mathcal{F} \left [ i \psi_{U} \left ( \frac{\delta \tilde{U}}{\delta \phi_{j} (\mathbf{r})} \right ) + i\psi_{N} \left( \frac{\delta \tilde{N}}{\delta \phi_{j} (\mathbf{r})} \right ) \right ] \label{eq: Nstar_appendix}
 \end{equation}
\end{subequations}
\noindent{where} $\mathcal{F}$ denotes the forward $(d+1)$-dimensional Fourier transform. To clarify in the example of the microcanonical ensemble, the expressions in equations\ (\ref{eq: N_appendix}) and (\ref{eq: Nstar_appendix}) imply that the kinetic energy terms are omitted from the linear coefficients in equations\ (\ref{eq: A_coeff}). For the example of accessing the canonical ensemble with the projected method, one would replace $i\psi_{U} \to \beta$ but remove the kinetic energy contributions from equations\ (\ref{eq: N_appendix}) and (\ref{eq: Nstar_appendix}) in order to keep them in the linear coefficient equation (\ref{eq: A_coeff}). In the projected canonical ensemble, the term proportional to $i \psi_{N}$ is omitted from equations\ (\ref{eq: N_appendix}) and (\ref{eq: Nstar_appendix}) and treated in the projection step. 

Note that $\mathcal{N}^*$ is not the complex-conjugate of $\mathcal{N}$ for $N_{\tau} > 1$. Here we use the notation $\mathcal{N}^*$ to denote that it is the nonlinear force appearing in the $\phi^*$ CL equation of motion. Similar to the linear coefficients before, the $\mathcal{N}$ and $\mathcal{N}^*$ terms are omitted from the unconstrained portion of the algorithm when sampling the microcanonical ensemble, and instead the expressions in equations\ (\ref{eq: dUdphistar}) and (\ref{eq: dUdphi}) are incorporated via the projection step at each CL iteration. 

\section{Projection equations for the microcanonical ensemble} \label{appendixB}
The microcanonical ensemble presents two coupled equations to be solved simultaneously at each CL iteration for $\lambda_{N}$ and $\lambda_{U}$: 
\begin{equation}
  \begin{split}
 A^{(N)}_{\mu} \lambda^2_{N} + A^{(N)}_{\beta} \lambda^2_{U} &+ B^{(N)}_{\mu} \lambda_{N} + B^{(N)}_{\beta} \lambda_{U} \\
  &A^{(N)}_{\mu \beta} \lambda_{N} \lambda_{U} + C^{(N)} = 0 ,
   \label{eq: N_MCE}
   \end{split}
\end{equation}
\begin{equation}
 \begin{split}
  &D_{\beta} \lambda^{4}_{U} + D_{\mu} \lambda^{4}_{N} + D_{\beta \mu} \lambda^3_{U} \lambda_{N} + D_{\mu \beta} \lambda^{3}_{N} \lambda_{U} + D_{\mu^2 \beta^2} \lambda^2_{U} \lambda^2_{N}  \\
  &+ G_{\beta} \lambda^3_{U} + G_{\mu} \lambda^{3}_{N} + G_{\mu \beta} \lambda^2_{N} \lambda_{U} + G_{\beta \mu} \lambda^2_{U} \lambda_{N} + A^{(U)}_{\mu} \lambda^2_{N} \\
  &+ A^{(U)}_{\beta} \lambda^2_{U} + B^{(U)}_{\mu} \lambda_{N} + B^{(U)}_{\beta} \lambda_{U} + A^{(U)}_{\mu \beta} \lambda_{N} \lambda_{U} + C^{(U)} = 0 , 
   \label{eq: U_MCE}
   \end{split}
\end{equation}
\noindent{where} all the coefficients are calculated at the beginning of each timestep. For coefficients that share the same letter and subscript across both equations, a superscript is used to denote the equation to which it belongs, i.e. $(N)$ for the particle number constraint and $(U)$ for the energy constraint equation. The subscript(s) of each coefficient denotes whether $\lambda_{U}$ or $\lambda_{N}$ is multiplied, with combinations possible, where the $\mu$ subscript corresponds to $\lambda_{N}$ and $\beta$, $\lambda_{U}$. For example, $A^{(N)}_{\mu}$ is the quadratic coefficient appearing in the particle number constraint equation preceding $\lambda^2_{N}$. 

We provide a list of the coefficient functionals of the constraint derivatives and the intermediate CS field configurations $\bar{\phi}^* , \bar{\phi}^{\vphantom{*}}$. To condense the notation, we define a generalized kinetic energy and interaction energy functional of coherent state fields:
\begin{equation}
 K[f, g] = -\frac{1}{N_{\tau}} \sum_{j=0}^{N_{\tau} - 1} \int d\mathbf{r}\ f_{j}(\mathbf{r}) \left( \frac{\hbar^2 \nabla^2}{2m} \right )  g_{j-1} (\mathbf{r}) \label{eq: K_functional}
\end{equation}
 and
\begin{equation}
 \begin{split}
 V[f, g, h, s] = \frac{1}{2 N_{\tau}} \sum_{j=0}^{N_{\tau} - 1} & \int d\mathbf{r}\ f_{j}(\mathbf{r}) g_{j-1} (\mathbf{r}) \times \\
 & \int d\mathbf{r'} u(|\mathbf{r} - \mathbf{r'}|) h_{j} (\mathbf{r'}) s_{j-1} (\mathbf{r'}) ,\label{eq: V_functional}
 \end{split}
\end{equation}
\noindent{where} $u(|\mathbf{r} - \mathbf{r'}|)$ is the same pair-potential appearing in equation\ (\ref{eq: H}), and $f, g, h,$ and $s$ are coherent state-like functions sharing the same $d+1$-dimensionality as the CS fields $\phi$ and $\phi^*$. As such, $K[\phi^* , \phi^{\vphantom{*}}] + V[\phi^* , \phi^{\vphantom{*}}, \phi^* , \phi^{\vphantom{*}}]$ is equivalent to $\tilde{U} [\phi^* , \phi^{\vphantom{*}}]$ specified in equation\ (\ref{eq: U_tilde_functional}). With those definitions, we organize the coefficients for the microcanonical constraint equations.

Using the coherent state particle number functional $\tilde{N}[\phi^* , \phi^{\vphantom{*}}]$ defined in equation\ (\ref{eq: N_CSfunctional}), the coefficients in the particle number constraint equation (\ref{eq: N_MCE}) for the microcanonocal ensemble are
\begin{equation}
  A^{(N)}_{\mu} = \tilde{N} \left [ \frac{\delta \tilde{N} }{\delta \phi_{j} (\mathbf{r})} , \frac{\delta \tilde{N} }{\delta \phi^*_{j} (\mathbf{r}) } \right ] ,
\end{equation}
\begin{equation}
  A^{(N)}_{\beta} = \tilde{N} \left [ \frac{\delta \tilde{U} }{\delta \phi_{j} (\mathbf{r})} , \frac{\delta \tilde{U} }{\delta \phi^*_{j} (\mathbf{r}) } \right ] ,
\end{equation}
\begin{equation}
  A^{(N)}_{\mu \beta} = \tilde{N} \left [ \frac{\delta \tilde{N} }{\delta \phi_{j} (\mathbf{r})} , \frac{\delta \tilde{U} }{\delta \phi^*_{j} (\mathbf{r}) } \right ] + \tilde{N} \left [ \frac{\delta \tilde{U}}{\delta \phi_{j} (\mathbf{r})} , \frac{\delta \tilde{N} }{\delta \phi^*_{j} (\mathbf{r}) } \right ] ,
\end{equation}
\begin{equation}
  B^{(N)}_{\mu} = \tilde{N} \left [ \frac{\delta \tilde{N} }{\delta \phi_{j} (\mathbf{r})} , \bar{\phi} \right ] + \tilde{N} \left [ \bar{\phi}^* , \frac{\delta \tilde{N} }{\delta \phi^*_{j} (\mathbf{r}) } \right ] ,
\end{equation}
\begin{equation}
  B^{(N)}_{\beta} = \tilde{N} \left [ \frac{\delta \tilde{U} }{\delta \phi_{j} (\mathbf{r})} , \bar{\phi} \right ] + \tilde{N} \left [ \bar{\phi}^* , \frac{\delta \tilde{U} }{\delta \phi^*_{j} (\mathbf{r}) } \right ]  ,
\end{equation}
\begin{equation}
 C^{(N)} = \tilde{N}[\bar{\phi}^* , \bar{\phi}^{\vphantom{*}}] - N .
\end{equation}
Similarly, we provide the explicit expressions for the coefficients in the energy equation\ (\ref{eq: U_MCE}) using our functionals in equations\ (\ref{eq: K_functional}) and (\ref{eq: V_functional})):
\begin{equation}
  D_{\mu} = V \left [ \frac{\delta \tilde{N}}{\delta \phi_{j} (\mathbf{r})} , \frac{\delta \tilde{N}}{\delta \phi^*_{j} (\mathbf{r}) } , \frac{\delta \tilde{N}}{\delta \phi_{j} (\mathbf{r})} , \frac{\delta \tilde{N}}{\delta \phi^*_{j} (\mathbf{r}) } \right ] ,
\end{equation}
\begin{equation}
  D_{\beta} = V \left [ \frac{\delta \tilde{U}}{\delta \phi_{j} (\mathbf{r})} , \frac{\delta \tilde{U}}{\delta \phi^*_{j} (\mathbf{r}) } , \frac{\delta \tilde{U}}{\delta \phi_{j} (\mathbf{r})} , \frac{\delta \tilde{U}}{\delta \phi^*_{j} (\mathbf{r}) } \right ] ,
\end{equation}
\begin{equation}
 \begin{split}
  D_{\beta \mu} &= V \left [ \frac{\delta \tilde{U}}{\delta \phi_{j} (\mathbf{r})} , \frac{\delta \tilde{N}}{\delta \phi^*_{j} (\mathbf{r}) } , \frac{\delta \tilde{U}}{\delta \phi_{j} (\mathbf{r})} , \frac{\delta \tilde{U}}{\delta \phi^*_{j} (\mathbf{r}) } \right ] \\ 
  &+ V \left [ \frac{\delta \tilde{N}}{\delta \phi_{j} (\mathbf{r})} , \frac{\delta \tilde{U}}{\delta \phi^*_{j} (\mathbf{r}) } , \frac{\delta \tilde{U}}{\delta \phi_{j} (\mathbf{r})} , \frac{\delta \tilde{U}}{\delta \phi^*_{j} (\mathbf{r}) } \right ] \\
  &+ V \left [ \frac{\delta \tilde{U}}{\delta \phi_{j} (\mathbf{r})} , \frac{\delta \tilde{U}}{\delta \phi^*_{j} (\mathbf{r}) } , \frac{\delta \tilde{N}}{\delta \phi_{j} (\mathbf{r})} , \frac{\delta \tilde{U}}{\delta \phi^*_{j} (\mathbf{r}) } \right ] \\
  &+ V \left [ \frac{\delta \tilde{U}}{\delta \phi_{j} (\mathbf{r})} , \frac{\delta \tilde{U}}{\delta \phi^*_{j} (\mathbf{r}) } , \frac{\delta \tilde{U}}{\delta \phi_{j} (\mathbf{r})} , \frac{\delta \tilde{N}}{\delta \phi^*_{j} (\mathbf{r}) } \right ],
 \end{split}
\end{equation}
\begin{equation}
 \begin{split}
  D_{\mu \beta} &= V \left [ \frac{\delta \tilde{N}}{\delta \phi_{j} (\mathbf{r})} , \frac{\delta \tilde{U}}{\delta \phi^*_{j} (\mathbf{r}) } , \frac{\delta \tilde{N}}{\delta \phi_{j} (\mathbf{r})} , \frac{\delta \tilde{N}}{\delta \phi^*_{j} (\mathbf{r}) } \right ] \\ 
  &+ V \left [ \frac{\delta \tilde{U}}{\delta \phi_{j} (\mathbf{r})} , \frac{\delta \tilde{N}}{\delta \phi^*_{j} (\mathbf{r}) } , \frac{\delta \tilde{N}}{\delta \phi_{j} (\mathbf{r})} , \frac{\delta \tilde{N}}{\delta \phi^*_{j} (\mathbf{r}) } \right ] \\
  &+ V \left [ \frac{\delta \tilde{N}}{\delta \phi_{j} (\mathbf{r})} , \frac{\delta \tilde{N}}{\delta \phi^*_{j} (\mathbf{r}) } , \frac{\delta \tilde{U}}{\delta \phi_{j} (\mathbf{r})} , \frac{\delta \tilde{N}}{\delta \phi^*_{j} (\mathbf{r}) } \right ] \\
  &+ V \left [ \frac{\delta \tilde{N}}{\delta \phi_{j} (\mathbf{r})} , \frac{\delta \tilde{N}}{\delta \phi^*_{j} (\mathbf{r}) } , \frac{\delta \tilde{N}}{\delta \phi_{j} (\mathbf{r})} , \frac{\delta \tilde{U}}{\delta \phi^*_{j} (\mathbf{r}) } \right ],
 \end{split}
\end{equation}

\begin{equation}
 \begin{split}
  D_{\mu^2 \beta^2 } &= V \left [ \frac{\delta \tilde{U}}{\delta \phi_{j} (\mathbf{r})} , \frac{\delta \tilde{U}}{\delta \phi^*_{j} (\mathbf{r}) } , \frac{\delta \tilde{N}}{\delta \phi_{j} (\mathbf{r})} , \frac{\delta \tilde{N}}{\delta \phi^*_{j} (\mathbf{r}) } \right ] \\ 
  &+ V \left [ \frac{\delta \tilde{N}}{\delta \phi_{j} (\mathbf{r})} , \frac{\delta \tilde{N}}{\delta \phi^*_{j} (\mathbf{r}) } , \frac{\delta \tilde{U}}{\delta \phi_{j} (\mathbf{r})} , \frac{\delta \tilde{U}}{\delta \phi^*_{j} (\mathbf{r}) } \right ] \\
  &+ V \left [ \frac{\delta \tilde{U}}{\delta \phi_{j} (\mathbf{r})} , \frac{\delta \tilde{N}}{\delta \phi^*_{j} (\mathbf{r}) } , \frac{\delta \tilde{U}}{\delta \phi_{j} (\mathbf{r})} , \frac{\delta \tilde{N}}{\delta \phi^*_{j} (\mathbf{r}) } \right ] \\
  &+ V \left [ \frac{\delta \tilde{U}}{\delta \phi_{j} (\mathbf{r})} , \frac{\delta \tilde{N}}{\delta \phi^*_{j} (\mathbf{r}) } , \frac{\delta \tilde{N}}{\delta \phi_{j} (\mathbf{r})} , \frac{\delta \tilde{U}}{\delta \phi^*_{j} (\mathbf{r}) } \right ] \\
  &+ V \left [ \frac{\delta \tilde{N}}{\delta \phi_{j} (\mathbf{r})} , \frac{\delta \tilde{U}}{\delta \phi^*_{j} (\mathbf{r}) } , \frac{\delta \tilde{U}}{\delta \phi_{j} (\mathbf{r})} , \frac{\delta \tilde{N}}{\delta \phi^*_{j} (\mathbf{r}) } \right ] \\
  &+ V \left [ \frac{\delta \tilde{N}}{\delta \phi_{j} (\mathbf{r})} , \frac{\delta \tilde{U}}{\delta \phi^*_{j} (\mathbf{r}) } , \frac{\delta \tilde{N}}{\delta \phi_{j} (\mathbf{r})} , \frac{\delta \tilde{U}}{\delta \phi^*_{j} (\mathbf{r}) } \right ],
 \end{split}
\end{equation}
\begin{equation}
 \begin{split}
  G_{\beta} &= V \left [ \bar{\phi}^*, \frac{\delta \tilde{U}}{\delta \phi^*_{j} (\mathbf{r}) } , \frac{\delta \tilde{U}}{\delta \phi_{j} (\mathbf{r})} , \frac{\delta \tilde{U}}{\delta \phi^*_{j} (\mathbf{r}) } \right ] \\ 
  &+ V \left [ \frac{\delta \tilde{U}}{\delta \phi_{j} (\mathbf{r})} , \bar{\phi} , \frac{\delta \tilde{U}}{\delta \phi_{j} (\mathbf{r})} , \frac{\delta \tilde{U}}{\delta \phi^*_{j} (\mathbf{r}) } \right ] \\
  &+ V \left [ \frac{\delta \tilde{U}}{\delta \phi_{j} (\mathbf{r})} , \frac{\delta \tilde{U}}{\delta \phi^*_{j} (\mathbf{r}) } , \frac{\delta \tilde{U}}{\delta \phi_{j} (\mathbf{r})} , \bar{\phi} \right ] \\
  &+ V \left [ \frac{\delta \tilde{U}}{\delta \phi_{j} (\mathbf{r})} , \frac{\delta \tilde{U}}{\delta \phi^*_{j} (\mathbf{r}) } , \bar{\phi}^* , \frac{\delta \tilde{U}}{\delta \phi^*_{j} (\mathbf{r}) } \right ],
 \end{split}
\end{equation}
\begin{equation}
 \begin{split}
  G_{\mu} &= V \left [ \bar{\phi}^*, \frac{\delta \tilde{N}}{\delta \phi^*_{j} (\mathbf{r}) } , \frac{\delta \tilde{N}}{\delta \phi_{j} (\mathbf{r})} , \frac{\delta \tilde{N}}{\delta \phi^*_{j} (\mathbf{r}) } \right ] + V \left [ \frac{\delta \tilde{N}}{\delta \phi_{j} (\mathbf{r})} , \bar{\phi} , \frac{\delta \tilde{N}}{\delta \phi_{j} (\mathbf{r})} , \frac{\delta \tilde{N}}{\delta \phi^*_{j} (\mathbf{r}) } \right ] + V \left [ \frac{\delta \tilde{N}}{\delta \phi_{j} (\mathbf{r})} , \frac{\delta \tilde{N}}{\delta \phi^*_{j} (\mathbf{r}) } , \frac{\delta \tilde{N}}{\delta \phi_{j} (\mathbf{r})} , \bar{\phi} \right ] \\
  &+ V \left [ \frac{\delta \tilde{N}}{\delta \phi_{j} (\mathbf{r})} , \frac{\delta \tilde{N}}{\delta \phi^*_{j} (\mathbf{r}) } , \bar{\phi}^* , \frac{\delta \tilde{N}}{\delta \phi^*_{j} (\mathbf{r}) } \right ],
 \end{split}
\end{equation}
\begin{widetext}
    \begin{align}
  G_{\mu \beta} &= V \left [ \bar{\phi}^* , \frac{\delta \tilde{U}}{\delta \phi^*_{j} (\mathbf{r}) } , \frac{\delta \tilde{N}}{\delta \phi_{j} (\mathbf{r})} , \frac{\delta \tilde{N}}{\delta \phi^*_{j} (\mathbf{r}) } \right ] 
  + V \left [ \frac{\delta \tilde{U}}{\delta \phi_{j} (\mathbf{r})} , \bar{\phi} , \frac{\delta \tilde{N}}{\delta \phi_{j} (\mathbf{r})} , \frac{\delta \tilde{N}}{\delta \phi^*_{j} (\mathbf{r}) } \right ] + V \left [ \frac{\delta \tilde{N}}{\delta \phi_{j} (\mathbf{r})} , \bar{\phi} , \frac{\delta \tilde{U}}{\delta \phi_{j} (\mathbf{r})} , \frac{\delta \tilde{N}}{\delta \phi^*_{j} (\mathbf{r}) }  \right ] \nonumber \\
  &+ V \left [ \frac{\delta \tilde{N}}{\delta \phi_{j} (\mathbf{r})} , \bar{\phi} , \frac{\delta \tilde{N}}{\delta \phi_{j} (\mathbf{r})} , \frac{\delta \tilde{U}}{\delta \phi^*_{j} (\mathbf{r}) }  \right ] + V \left [ \bar{\phi}^* , \frac{\delta \tilde{N}}{\delta \phi^*_{j} (\mathbf{r}) } , \frac{\delta \tilde{U}}{\delta \phi_{j} (\mathbf{r})} , \frac{\delta \tilde{N}}{\delta \phi^*_{j} (\mathbf{r}) } \right ] + V \left [ \bar{\phi}^* , \frac{\delta \tilde{N}}{\delta \phi^*_{j} (\mathbf{r}) } , \frac{\delta \tilde{N}}{\delta \phi_{j} (\mathbf{r})} , \frac{\delta \tilde{U}}{\delta \phi^*_{j} (\mathbf{r}) } \right ] \nonumber \\
  &+ V \left [ \frac{\delta \tilde{N}}{\delta \phi_{j} (\mathbf{r})} , \frac{\delta \tilde{N}}{\delta \phi^*_{j} (\mathbf{r}) } , \bar{\phi}^* , \frac{\delta \tilde{U}}{\delta \phi^*_{j} (\mathbf{r}) }  \right ] + V \left [\frac{\delta \tilde{N}}{\delta \phi_{j} (\mathbf{r})} , \frac{\delta \tilde{N}}{\delta \phi^*_{j} (\mathbf{r}) } , \frac{\delta \tilde{U}}{\delta \phi_{j} (\mathbf{r})} , \bar{\phi} \right ] + V \left [ \frac{\delta \tilde{U}}{\delta \phi_{j} (\mathbf{r})} , \frac{\delta \tilde{N}}{\delta \phi^*_{j} (\mathbf{r}) } , \frac{\delta \tilde{N}}{\delta \phi_{j} (\mathbf{r})} , \bar{\phi} \right ] \nonumber \\
  &+ V \left [\frac{\delta \tilde{N}}{\delta \phi_{j} (\mathbf{r})} , \frac{\delta \tilde{U}}{\delta \phi^*_{j} (\mathbf{r}) }, \frac{\delta \tilde{N}}{\delta \phi_{j} (\mathbf{r})} , \bar{\phi} \right ] + V \left [ \frac{\delta \tilde{U}}{\delta \phi_{j} (\mathbf{r})} , \frac{\delta \tilde{N}}{\delta \phi^*_{j} (\mathbf{r}) } , \bar{\phi}^* , \frac{\delta \tilde{N}}{\delta \phi^*_{j} (\mathbf{r}) } \right ] + V \left [\frac{\delta \tilde{N}}{\delta \phi_{j} (\mathbf{r})} , \frac{\delta \tilde{U}}{\delta \phi^*_{j} (\mathbf{r}) } , \bar{\phi}^* , \frac{\delta \tilde{N}}{\delta \phi^*_{j} (\mathbf{r}) } \right ],
 \end{align}
\end{widetext}
\begin{widetext}
 \begin{align}
  G_{\beta \mu} &= V \left [ \bar{\phi}^* , \frac{\delta \tilde{N}}{\delta \phi^*_{j} (\mathbf{r}) } , \frac{\delta \tilde{U}}{\delta \phi_{j} (\mathbf{r})} , \frac{\delta \tilde{U}}{\delta \phi^*_{j} (\mathbf{r}) } \right ] + V \left [ \frac{\delta \tilde{N}}{\delta \phi_{j} (\mathbf{r})} , \bar{\phi} , \frac{\delta \tilde{U}}{\delta \phi_{j} (\mathbf{r})} , \frac{\delta \tilde{U}}{\delta \phi^*_{j} (\mathbf{r}) } \right ] + V \left [ \frac{\delta \tilde{U}}{\delta \phi_{j} (\mathbf{r})} , \bar{\phi} , \frac{\delta \tilde{U}}{\delta \phi_{j} (\mathbf{r})} , \frac{\delta \tilde{N}}{\delta \phi^*_{j} (\mathbf{r}) }  \right ] \nonumber \\ 
  &+ V \left [ \frac{\delta \tilde{U}}{\delta \phi_{j} (\mathbf{r})} , \bar{\phi} , \frac{\delta \tilde{N}}{\delta \phi_{j} (\mathbf{r})} , \frac{\delta \tilde{U}}{\delta \phi^*_{j} (\mathbf{r}) }  \right ] + V \left [ \bar{\phi}^* , \frac{\delta \tilde{U}}{\delta \phi^*_{j} (\mathbf{r}) } , \frac{\delta \tilde{U}}{\delta \phi_{j} (\mathbf{r})} , \frac{\delta \tilde{N}}{\delta \phi^*_{j} (\mathbf{r}) } \right ] + V \left [ \bar{\phi}^* , \frac{\delta \tilde{U}}{\delta \phi^*_{j} (\mathbf{r}) } , \frac{\delta \tilde{N}}{\delta \phi_{j} (\mathbf{r})} , \frac{\delta \tilde{U}}{\delta \phi^*_{j} (\mathbf{r}) } \right ] \nonumber \\
  &+ V \left [ \frac{\delta \tilde{U}}{\delta \phi_{j} (\mathbf{r})} , \frac{\delta \tilde{N}}{\delta \phi^*_{j} (\mathbf{r}) } , \bar{\phi}^* , \frac{\delta \tilde{U}}{\delta \phi^*_{j} (\mathbf{r}) }  \right ] + V \left [\frac{\delta \tilde{N}}{\delta \phi_{j} (\mathbf{r})} , \frac{\delta \tilde{U}}{\delta \phi^*_{j} (\mathbf{r}) } , \frac{\delta \tilde{U}}{\delta \phi_{j} (\mathbf{r})} , \bar{\phi} \right ] + V \left [ \frac{\delta \tilde{U}}{\delta \phi_{j} (\mathbf{r})} , \frac{\delta \tilde{N}}{\delta \phi^*_{j} (\mathbf{r}) } , \frac{\delta \tilde{U}}{\delta \phi_{j} (\mathbf{r})} , \bar{\phi} \right ] \nonumber  \\
  &+ V \left [ \frac{\delta \tilde{U}}{\delta \phi_{j} (\mathbf{r})} , \frac{\delta \tilde{U}}{\delta \phi^*_{j} (\mathbf{r}) } , \frac{\delta \tilde{N}}{\delta \phi_{j} (\mathbf{r})} , \bar{\phi} \right ] + V \left [\frac{\delta \tilde{N}}{\delta \phi_{j} (\mathbf{r})} , \frac{\delta \tilde{U}}{\delta \phi^*_{j} (\mathbf{r}) } , \bar{\phi}^* , \frac{\delta \tilde{U}}{\delta \phi^*_{j} (\mathbf{r}) } \right ] + V \left [\frac{\delta \tilde{U}}{\delta \phi_{j} (\mathbf{r})} , \frac{\delta \tilde{U}}{\delta \phi^*_{j} (\mathbf{r}) } , \bar{\phi}^* , \frac{\delta \tilde{N}}{\delta \phi^*_{j} (\mathbf{r}) } \right ],
 \end{align}
\end{widetext}
\begin{equation}
 \begin{split}
  &A^{(U)}_{\mu} = K \left [ \frac{\delta \tilde{N}}{\delta \phi_{j} (\mathbf{r})} , \frac{\delta \tilde{N}}{\delta \phi^*_{j} (\mathbf{r}) } \right ] + V \left [ \frac{\delta \tilde{N}}{\delta \phi_{j} (\mathbf{r})} , \frac{\delta \tilde{N}}{\delta \phi^*_{j} (\mathbf{r}) } , \bar{\phi}^* , \bar{\phi} \right ] \\
  &+ V \left [ \bar{\phi}^* , \bar{\phi} , \frac{\delta \tilde{N}}{\delta \phi_{j} (\mathbf{r})} , \frac{\delta \tilde{N}}{\delta \phi^*_{j} (\mathbf{r}) } \right ] + V \left [ \bar{\phi}^* , \frac{\delta \tilde{N}}{\delta \phi^*_{j} (\mathbf{r}) }, \bar{\phi}^* , \frac{\delta \tilde{N}}{\delta \phi^*_{j} (\mathbf{r}) } \right ] \\ 
  &+ V \left [ \bar{\phi}^* , \frac{\delta \tilde{N}}{\delta \phi^*_{j} (\mathbf{r}) } , \frac{\delta \tilde{N}}{\delta \phi_{j} (\mathbf{r})}, \bar{\phi}  \right ] + V \left [\frac{\delta \tilde{N}}{\delta \phi_{j} (\mathbf{r})}, \bar{\phi}, \bar{\phi}^* , \frac{\delta \tilde{N}}{\delta \phi^*_{j} (\mathbf{r}) }  \right ] \\
  &+ V \left [\frac{\delta \tilde{N}}{\delta \phi_{j} (\mathbf{r})}, \bar{\phi}, \frac{\delta \tilde{N}}{\delta \phi_{j} (\mathbf{r})}, \bar{\phi} \right ] ,
 \end{split}
\end{equation}
\begin{equation}
 \begin{split}
  &A^{(U)}_{\beta} = K \left [ \frac{\delta \tilde{U}}{\delta \phi_{j} (\mathbf{r})} , \frac{\delta \tilde{U}}{\delta \phi^*_{j} (\mathbf{r}) } \right ] + V \left [ \frac{\delta \tilde{U}}{\delta \phi_{j} (\mathbf{r})} , \frac{\delta \tilde{U}}{\delta \phi^*_{j} (\mathbf{r}) } , \bar{\phi}^* , \bar{\phi} \right ] \\
  &+ V \left [ \bar{\phi}^* , \bar{\phi} , \frac{\delta \tilde{U}}{\delta \phi_{j} (\mathbf{r})} , \frac{\delta \tilde{U}}{\delta \phi^*_{j} (\mathbf{r}) } \right ] + V \left [ \bar{\phi}^* , \frac{\delta \tilde{U}}{\delta \phi^*_{j} (\mathbf{r}) }, \bar{\phi}^* , \frac{\delta \tilde{U}}{\delta \phi^*_{j} (\mathbf{r}) } \right ] \\ 
  &+ V \left [ \bar{\phi}^* , \frac{\delta \tilde{U}}{\delta \phi^*_{j} (\mathbf{r}) } , \frac{\delta \tilde{U}}{\delta \phi_{j} (\mathbf{r})}, \bar{\phi}  \right ] + V \left [\frac{\delta \tilde{U}}{\delta \phi_{j} (\mathbf{r})}, \bar{\phi}, \bar{\phi}^* , \frac{\delta \tilde{U}}{\delta \phi^*_{j} (\mathbf{r}) }  \right ] \\
  &+ V \left [\frac{\delta \tilde{U}}{\delta \phi_{j} (\mathbf{r})}, \bar{\phi}, \frac{\delta \tilde{U}}{\delta \phi_{j} (\mathbf{r})}, \bar{\phi} \right ] ,
 \end{split}
\end{equation}
\begin{equation}
 \begin{split}
  &A^{(U)}_{\mu \beta} = K \left [ \frac{\delta \tilde{U}}{\delta \phi_{j} (\mathbf{r})} , \frac{\delta \tilde{N}}{\delta \phi^*_{j} (\mathbf{r}) } \right ] + K \left [ \frac{\delta \tilde{N}}{\delta \phi_{j} (\mathbf{r})} , \frac{\delta \tilde{U}}{\delta \phi^*_{j} (\mathbf{r}) } \right ] \\
  &+ V \left[ \frac{\delta \tilde{U}}{\delta \phi_{j} (\mathbf{r})} , \frac{\delta \tilde{N}}{\delta \phi^*_{j} (\mathbf{r}) }, \bar{\phi}^* , \bar{\phi} \right] + V \left[ \frac{\delta \tilde{N}}{\delta \phi_{j} (\mathbf{r})} , \frac{\delta \tilde{U}}{\delta \phi^*_{j} (\mathbf{r}) }, \bar{\phi}^* , \bar{\phi} \right] \\
  &+ V \left[ \bar{\phi}^* , \bar{\phi} , \frac{\delta \tilde{U}}{\delta \phi_{j} (\mathbf{r})} , \frac{\delta \tilde{N}}{\delta \phi^*_{j} (\mathbf{r}) } \right] + V \left[ \bar{\phi}^* , \bar{\phi} , \frac{\delta \tilde{N}}{\delta \phi_{j} (\mathbf{r})} , \frac{\delta \tilde{U}}{\delta \phi^*_{j} (\mathbf{r}) } \right] \\
  &+ V \left[ \bar{\phi}^* , \frac{\delta \tilde{U}}{\delta \phi^*_{j} (\mathbf{r}) }, \bar{\phi}^* , \frac{\delta \tilde{N}}{\delta \phi^*_{j} (\mathbf{r}) } \right] + V \left[ \bar{\phi}^* , \frac{\delta \tilde{U}}{\delta \phi^*_{j} (\mathbf{r}) }, \frac{\delta \tilde{N}}{\delta \phi_{j} (\mathbf{r})}, \bar{\phi} \right] \\
  &+ V \left[ \frac{\delta \tilde{U}}{\delta \phi_{j} (\mathbf{r})}, \bar{\phi}, \bar{\phi}^* , \frac{\delta \tilde{N}}{\delta \phi^*_{j} (\mathbf{r})} \right ] + V \left[ \frac{\delta \tilde{U}}{\delta \phi_{j} (\mathbf{r})}, \bar{\phi}, \frac{\delta \tilde{N}}{\delta \phi_{j} (\mathbf{r})} , \bar{\phi} \right ] \\
  &+ V \left[ \bar{\phi}^* , \frac{\delta \tilde{N}}{\delta \phi^*_{j} (\mathbf{r})} , \bar{\phi}^* , \frac{\delta \tilde{U}}{\delta \phi^*_{j} (\mathbf{r})} \right ] + V \left[ \bar{\phi}^* , \frac{\delta \tilde{N}}{\delta \phi^*_{j} (\mathbf{r})} , \frac{\delta \tilde{U}}{\delta \phi_{j} (\mathbf{r})} , \bar{\phi} \right ] \\ 
  &+ V \left[ \frac{\delta \tilde{N}}{\delta \phi_{j} (\mathbf{r})}, \bar{\phi} , \bar{\phi}^* , \frac{\delta \tilde{U}}{\delta \phi^*_{j} (\mathbf{r})} \right ] + V \left[ \frac{\delta \tilde{N}}{\delta \phi_{j} (\mathbf{r})}, \bar{\phi} , \frac{\delta \tilde{U}}{\delta \phi_{j} (\mathbf{r})} , \bar{\phi} \right ] ,
  \end{split}
\end{equation}
\begin{equation}
 \begin{split}
  &B^{(U)}_{\mu} = K \left [ \frac{\delta \tilde{N}}{\delta \phi_{j} (\mathbf{r})} , \bar{\phi} \right ] + K \left [ \bar{\phi}^{*} \frac{\delta \tilde{N}}{\delta \phi^*_{j} (\mathbf{r})} \right ] \\
  &+ V \left [ \frac{\delta \tilde{N}}{\delta \phi_{j} (\mathbf{r})}, \bar{\phi}, \bar{\phi}^* , \bar{\phi}  \right] + V \left [ \bar{\phi}^* , \bar{\phi} , \frac{\delta \tilde{N}}{\delta \phi_{j} (\mathbf{r})}, \bar{\phi} \right] \\
  &+ V \left [ \bar{\phi}^* , \frac{\delta \tilde{N}}{\delta \phi^*_{j} (\mathbf{r})}, \bar{\phi}^* , \bar{\phi} \right] + V \left [ \bar{\phi}^* , \bar{\phi} , \bar{\phi}^* , \frac{\delta \tilde{N}}{\delta \phi^*_{j} (\mathbf{r})} \right] ,
  \end{split}
\end{equation}
\begin{equation}
 \begin{split}
  &B^{(U)}_{\beta} = K \left [ \frac{\delta \tilde{U}}{\delta \phi_{j} (\mathbf{r})} , \bar{\phi} \right ] + K \left [ \bar{\phi}^{*} \frac{\delta \tilde{U}}{\delta \phi^*_{j} (\mathbf{r})} \right ] \\
  &+ V \left [ \frac{\delta \tilde{U}}{\delta \phi_{j} (\mathbf{r})}, \bar{\phi}, \bar{\phi}^* , \bar{\phi}  \right] + V \left [ \bar{\phi}^* , \bar{\phi} , \frac{\delta \tilde{U}}{\delta \phi_{j} (\mathbf{r})}, \bar{\phi} \right] \\
  &+ V \left [ \bar{\phi}^* , \frac{\delta \tilde{U}}{\delta \phi^*_{j} (\mathbf{r})}, \bar{\phi}^* , \bar{\phi} \right] + V \left [ \bar{\phi}^* , \bar{\phi} , \bar{\phi}^* , \frac{\delta \tilde{U}}{\delta \phi^*_{j} (\mathbf{r})} \right] ,
 \end{split}
\end{equation}
\begin{equation}
 C^{(U)} = \tilde{U}[\bar{\phi}^* , \bar{\phi}^{\vphantom{*}} ] - U .
\end{equation}

For evaluating the coefficients, we pay special attention to the argument ordering of the functionals defined in equations\ (\ref{eq: K_functional}) and (\ref{eq: V_functional}) in order to respect the causal properties of the theory. For the choice of contact interactions, many of the equations may be simplified using the identity $V[f, g, h, g] + V[h, g, f, g] = 2V[f, g, h, g]$ (or $2V[h,g,f,g]$) because a Fourier convolution is not required. Furthermore, this simplification would reduce the required computer time by decreasing the required number of FFTs per CL timestep by $72$.

\section{Root-finding procedure for the microcanonical ensemble} \label{appendixC}
Here we detail our algorithm and procedure for solving the energy and particle number constraints simultaneously at each Langevin iteration. We further specify our implementation of a modified Levenberg-Marquardt solver, which requires an initial guess for $\lambda_{N}$ and $\lambda_{U}$ as well as a reasonable initial coherent state field configuration. We start all microcanonical simulations with an equilibrated canonical ensemble configuration and initialize $\lambda_{N} = \langle \mu \rangle / \beta$ and $\lambda_{U} = -\beta$, where $\langle \mu \rangle$ and $\beta$ are taken from the corresponding canonical ensemble simulation. Finding initial guesses for $\lambda_{N}$, $\lambda_{U}$, and ($\phi^{\vphantom{*}}$, $\phi^*$) may be automated in practice via projected canonical ensemble simulations  due to their robust numerical stability. In such a procedure, one would vary $\beta$ until the canonical ensemble's average internal energy matches the desired internal energy for the microcanonical ensemble.
 
Importantly, the initial configuration of the coherent state fields must be reasonably close to satisfying $\mathbf{g}[\phi^* , \phi^{\vphantom{*}}] = \mathbf{0}$ for the Levenberg-Marquardt solver to work well. We found a Newton solver alone to be insufficient to find the roots in the presence of the thermal and quantum fluctuations, so we use the following variant of the Levenberg-Marquardt algorithm at each iteration ($\nu$): 
\begin{subequations}
\begin{equation}
    \left ( \underline{\underline{J}}^{(\nu)} + \gamma^{(\nu)} \text{diag}[\underline{\underline{J}}^{(\nu)}]  \right ) \bm{\delta}^{(\nu)} = -\mathbf{g}^{(\nu)} ,
\end{equation} 

\begin{equation}
    \bm{\lambda}^{(\nu + 1)} = \bm{\lambda}^{(\nu)} + \bm{\delta}^{(\nu)} ,
\end{equation}
\end{subequations}
\noindent{where} $J_{ij} \equiv \partial g_{i} / \partial \lambda_{j}$ is a Jacobian matrix element determined by differentiating equation\ (\ref{eq: N_MCE}) or (\ref{eq: U_MCE}), and $\gamma$ is a heuristic parameter used to dynamically switch between gradient descent and Newton relaxation. In our implementation, we find robust behavior from the solver with the following form for $\gamma$:
\begin{equation}
    \gamma^{(\nu)} = \gamma_{0} || \mathbf{g}^{(\nu)} ||_{2} ,
      \label{eq: gamma_LM}
\end{equation}
\noindent{where} we have chosen $\gamma_0 = 1$ and $||.||_{2}$ denotes the $L_{2}$ norm. This algorithm is a ``trust region'' approach where a Newton iteration is effectively used when close to a root ($\gamma \to 0$). On the other hand, the large $\gamma$ limit corresponds to a gradient descent approach and is more appropriate when far from a root. Fortunately for the microcanonical case, we are presented with just two equations and unknown variables, so the Jacobian and its inverse are cheap to compute at every solver iteration. As long as an initial condition for $\phi^* , \phi^{\vphantom{*}}$ is well-chosen, the Jacobian's inverse will exist and this iterative solution approach can be used. We performed this scheme iteratively to find solutions, using $||\mathbf{g}||_{2} < 10^{-14}$ as a convergence criterion. The number of iterations required to converge to a solution decreased as the timestep $\Delta t$ decreased. 


\begin{thebibliography}{61}%
\makeatletter
\providecommand \@ifxundefined [1]{%
 \@ifx{#1\undefined}
}%
\providecommand \@ifnum [1]{%
 \ifnum #1\expandafter \@firstoftwo
 \else \expandafter \@secondoftwo
 \fi
}%
\providecommand \@ifx [1]{%
 \ifx #1\expandafter \@firstoftwo
 \else \expandafter \@secondoftwo
 \fi
}%
\providecommand \natexlab [1]{#1}%
\providecommand \enquote  [1]{``#1''}%
\providecommand \bibnamefont  [1]{#1}%
\providecommand \bibfnamefont [1]{#1}%
\providecommand \citenamefont [1]{#1}%
\providecommand \href@noop [0]{\@secondoftwo}%
\providecommand \href [0]{\begingroup \@sanitize@url \@href}%
\providecommand \@href[1]{\@@startlink{#1}\@@href}%
\providecommand \@@href[1]{\endgroup#1\@@endlink}%
\providecommand \@sanitize@url [0]{\catcode `\\12\catcode `\$12\catcode
  `\&12\catcode `\#12\catcode `\^12\catcode `\_12\catcode `\%12\relax}%
\providecommand \@@startlink[1]{}%
\providecommand \@@endlink[0]{}%
\providecommand \url  [0]{\begingroup\@sanitize@url \@url }%
\providecommand \@url [1]{\endgroup\@href {#1}{\urlprefix }}%
\providecommand \urlprefix  [0]{URL }%
\providecommand \Eprint [0]{\href }%
\providecommand \doibase [0]{https://doi.org/}%
\providecommand \selectlanguage [0]{\@gobble}%
\providecommand \bibinfo  [0]{\@secondoftwo}%
\providecommand \bibfield  [0]{\@secondoftwo}%
\providecommand \translation [1]{[#1]}%
\providecommand \BibitemOpen [0]{}%
\providecommand \bibitemStop [0]{}%
\providecommand \bibitemNoStop [0]{.\EOS\space}%
\providecommand \EOS [0]{\spacefactor3000\relax}%
\providecommand \BibitemShut  [1]{\csname bibitem#1\endcsname}%
\let\auto@bib@innerbib\@empty
\bibitem [{\citenamefont {Sholl}\ and\ \citenamefont
  {Steckel}(2011)}]{shollDensityFunctionalTheory2011}%
  \BibitemOpen
  \bibfield  {author} {\bibinfo {author} {\bibfnamefont {D.~S.}\ \bibnamefont
  {Sholl}}\ and\ \bibinfo {author} {\bibfnamefont {J.~A.}\ \bibnamefont
  {Steckel}},\ }\href@noop {} {\emph {\bibinfo {title} {Density {{Functional
  Theory}}: {{A Practical Introduction}}}}}\ (\bibinfo  {publisher} {John Wiley
  \& Sons},\ \bibinfo {year} {2011})\BibitemShut {NoStop}%
\bibitem [{\citenamefont {Hess}\ \emph {et~al.}(2008)\citenamefont {Hess},
  \citenamefont {Kutzner}, \citenamefont {{van der Spoel}},\ and\ \citenamefont
  {Lindahl}}]{hessGROMACSAlgorithmsHighly2008}%
  \BibitemOpen
  \bibfield  {author} {\bibinfo {author} {\bibfnamefont {B.}~\bibnamefont
  {Hess}}, \bibinfo {author} {\bibfnamefont {C.}~\bibnamefont {Kutzner}},
  \bibinfo {author} {\bibfnamefont {D.}~\bibnamefont {{van der Spoel}}},\ and\
  \bibinfo {author} {\bibfnamefont {E.}~\bibnamefont {Lindahl}},\ }\href
  {https://doi.org/10.1021/ct700301q} {\bibfield  {journal} {\bibinfo
  {journal} {Journal of Chemical Theory and Computation}\ }\textbf {\bibinfo
  {volume} {4}},\ \bibinfo {pages} {435} (\bibinfo {year} {2008})}\BibitemShut
  {NoStop}%
\bibitem [{\citenamefont {Andrade}\ \emph {et~al.}(2021)\citenamefont
  {Andrade}, \citenamefont {Pemmaraju}, \citenamefont {Kartsev}, \citenamefont
  {Xiao}, \citenamefont {Lindenberg}, \citenamefont {Rajpurohit}, \citenamefont
  {Tan}, \citenamefont {Ogitsu},\ and\ \citenamefont
  {Correa}}]{andradeInqModernGPUAccelerated2021}%
  \BibitemOpen
  \bibfield  {author} {\bibinfo {author} {\bibfnamefont {X.}~\bibnamefont
  {Andrade}}, \bibinfo {author} {\bibfnamefont {C.~D.}\ \bibnamefont
  {Pemmaraju}}, \bibinfo {author} {\bibfnamefont {A.}~\bibnamefont {Kartsev}},
  \bibinfo {author} {\bibfnamefont {J.}~\bibnamefont {Xiao}}, \bibinfo {author}
  {\bibfnamefont {A.}~\bibnamefont {Lindenberg}}, \bibinfo {author}
  {\bibfnamefont {S.}~\bibnamefont {Rajpurohit}}, \bibinfo {author}
  {\bibfnamefont {L.~Z.}\ \bibnamefont {Tan}}, \bibinfo {author} {\bibfnamefont
  {T.}~\bibnamefont {Ogitsu}},\ and\ \bibinfo {author} {\bibfnamefont {A.~A.}\
  \bibnamefont {Correa}},\ }\href {https://doi.org/10.1021/acs.jctc.1c00562}
  {\bibfield  {journal} {\bibinfo  {journal} {Journal of Chemical Theory and
  Computation}\ }\textbf {\bibinfo {volume} {17}},\ \bibinfo {pages} {7447}
  (\bibinfo {year} {2021})}\BibitemShut {NoStop}%
\bibitem [{\citenamefont {Berendsen}\ \emph {et~al.}(1995)\citenamefont
  {Berendsen}, \citenamefont {{van der Spoel}},\ and\ \citenamefont {{van
  Drunen}}}]{berendsenGROMACSMessagepassingParallel1995}%
  \BibitemOpen
  \bibfield  {author} {\bibinfo {author} {\bibfnamefont {H.~J.~C.}\
  \bibnamefont {Berendsen}}, \bibinfo {author} {\bibfnamefont {D.}~\bibnamefont
  {{van der Spoel}}},\ and\ \bibinfo {author} {\bibfnamefont {R.}~\bibnamefont
  {{van Drunen}}},\ }\href {https://doi.org/10.1016/0010-4655(95)00042-E}
  {\bibfield  {journal} {\bibinfo  {journal} {Computer Physics Communications}\
  }\textbf {\bibinfo {volume} {91}},\ \bibinfo {pages} {43} (\bibinfo {year}
  {1995})}\BibitemShut {NoStop}%
\bibitem [{\citenamefont {Binder}(1986)}]{binderMonteCarloMethods1986}%
  \BibitemOpen
  \bibinfo {editor} {\bibfnamefont {K.}~\bibnamefont {Binder}},\ ed.,\ \href
  {https://doi.org/10.1007/978-3-642-82803-4} {\emph {\bibinfo {title} {Monte
  {{Carlo Methods}} in {{Statistical Physics}}}}},\ \bibinfo {series} {Topics
  in {{Current Physics}}}, Vol.~\bibinfo {volume} {7}\ (\bibinfo  {publisher}
  {Springer},\ \bibinfo {address} {Berlin, Heidelberg},\ \bibinfo {year}
  {1986})\BibitemShut {NoStop}%
\bibitem [{\citenamefont {Gubernatis}\ \emph {et~al.}(2016)\citenamefont
  {Gubernatis}, \citenamefont {Kawashima},\ and\ \citenamefont
  {Werner}}]{gubernatisQuantumMonteCarlo2016}%
  \BibitemOpen
  \bibfield  {author} {\bibinfo {author} {\bibfnamefont {J.}~\bibnamefont
  {Gubernatis}}, \bibinfo {author} {\bibfnamefont {N.}~\bibnamefont
  {Kawashima}},\ and\ \bibinfo {author} {\bibfnamefont {P.}~\bibnamefont
  {Werner}},\ }\href {https://doi.org/10.1017/CBO9780511902581} {\emph
  {\bibinfo {title} {Quantum {{Monte Carlo Methods}}: {{Algorithms}} for
  {{Lattice Models}}}}}\ (\bibinfo  {publisher} {Cambridge University Press},\
  \bibinfo {address} {Cambridge},\ \bibinfo {year} {2016})\BibitemShut
  {NoStop}%
\bibitem [{\citenamefont {Boninsegni}\ \emph
  {et~al.}(2006{\natexlab{a}})\citenamefont {Boninsegni}, \citenamefont
  {Prokof'ev},\ and\ \citenamefont
  {Svistunov}}]{boninsegniWormAlgorithmDiagrammatic2006}%
  \BibitemOpen
  \bibfield  {author} {\bibinfo {author} {\bibfnamefont {M.}~\bibnamefont
  {Boninsegni}}, \bibinfo {author} {\bibfnamefont {N.~V.}\ \bibnamefont
  {Prokof'ev}},\ and\ \bibinfo {author} {\bibfnamefont {B.~V.}\ \bibnamefont
  {Svistunov}},\ }\href {https://doi.org/10.1103/PhysRevE.74.036701} {\bibfield
   {journal} {\bibinfo  {journal} {Physical Review E}\ }\textbf {\bibinfo
  {volume} {74}},\ \bibinfo {pages} {036701} (\bibinfo {year}
  {2006}{\natexlab{a}})}\BibitemShut {NoStop}%
\bibitem [{\citenamefont {Zappa}\ \emph {et~al.}(2018)\citenamefont {Zappa},
  \citenamefont {{Holmes-Cerfon}},\ and\ \citenamefont
  {Goodman}}]{zappaMonteCarloManifolds2018}%
  \BibitemOpen
  \bibfield  {author} {\bibinfo {author} {\bibfnamefont {E.}~\bibnamefont
  {Zappa}}, \bibinfo {author} {\bibfnamefont {M.}~\bibnamefont
  {{Holmes-Cerfon}}},\ and\ \bibinfo {author} {\bibfnamefont {J.}~\bibnamefont
  {Goodman}},\ }\href {https://doi.org/10.1002/cpa.21783} {\bibfield  {journal}
  {\bibinfo  {journal} {Communications on Pure and Applied Mathematics}\
  }\textbf {\bibinfo {volume} {71}},\ \bibinfo {pages} {2609} (\bibinfo {year}
  {2018})}\BibitemShut {NoStop}%
\bibitem [{\citenamefont
  {Chorin}(1967)}]{chorinNumericalSolutionNavierStokes1967}%
  \BibitemOpen
  \bibfield  {author} {\bibinfo {author} {\bibfnamefont {A.~J.}\ \bibnamefont
  {Chorin}},\ }\href {https://doi.org/10.1090/S0002-9904-1967-11853-6}
  {\bibfield  {journal} {\bibinfo  {journal} {Bulletin of the American
  Mathematical Society}\ }\textbf {\bibinfo {volume} {73}},\ \bibinfo {pages}
  {928} (\bibinfo {year} {1967})}\BibitemShut {NoStop}%
\bibitem [{\citenamefont
  {T{\'e}mam}(1969)}]{temamApproximationSolutionEquations1969}%
  \BibitemOpen
  \bibfield  {author} {\bibinfo {author} {\bibfnamefont {R.}~\bibnamefont
  {T{\'e}mam}},\ }\href {https://doi.org/10.1007/BF00247696} {\bibfield
  {journal} {\bibinfo  {journal} {Archive for Rational Mechanics and Analysis}\
  }\textbf {\bibinfo {volume} {33}},\ \bibinfo {pages} {377} (\bibinfo {year}
  {1969})}\BibitemShut {NoStop}%
\bibitem [{\citenamefont {{Gr{\o}nbech-jensen}}\ and\ \citenamefont
  {Doniach}(1994)}]{gronbech-jensenLongtimeOverdampedLangevin1994}%
  \BibitemOpen
  \bibfield  {author} {\bibinfo {author} {\bibfnamefont {N.}~\bibnamefont
  {{Gr{\o}nbech-jensen}}}\ and\ \bibinfo {author} {\bibfnamefont
  {S.}~\bibnamefont {Doniach}},\ }\href {https://doi.org/10.1002/jcc.540150908}
  {\bibfield  {journal} {\bibinfo  {journal} {Journal of Computational
  Chemistry}\ }\textbf {\bibinfo {volume} {15}},\ \bibinfo {pages} {997}
  (\bibinfo {year} {1994})}\BibitemShut {NoStop}%
\bibitem [{\citenamefont {Pellegrini}\ \emph {et~al.}(1998)\citenamefont
  {Pellegrini}, \citenamefont {Larsen}, \citenamefont {Yeates},\ and\
  \citenamefont {{Gr{\o}nbech-Jensen}}}]{pellegriniNewConstrainedLangevin1998}%
  \BibitemOpen
  \bibfield  {author} {\bibinfo {author} {\bibfnamefont {M.}~\bibnamefont
  {Pellegrini}}, \bibinfo {author} {\bibfnamefont {N.~A.}\ \bibnamefont
  {Larsen}}, \bibinfo {author} {\bibfnamefont {T.~O.}\ \bibnamefont {Yeates}},\
  and\ \bibinfo {author} {\bibfnamefont {N.}~\bibnamefont
  {{Gr{\o}nbech-Jensen}}},\ }\href
  {https://doi.org/10.1016/S0378-4371(98)00375-6} {\bibfield  {journal}
  {\bibinfo  {journal} {Physica A: Statistical Mechanics and its Applications}\
  }\textbf {\bibinfo {volume} {261}},\ \bibinfo {pages} {224} (\bibinfo {year}
  {1998})}\BibitemShut {NoStop}%
\bibitem [{\citenamefont {Alfonsi}\ \emph {et~al.}(2022)\citenamefont
  {Alfonsi}, \citenamefont {Coyaud},\ and\ \citenamefont
  {Ehrlacher}}]{alfonsiConstrainedOverdampedLangevin2022}%
  \BibitemOpen
  \bibfield  {author} {\bibinfo {author} {\bibfnamefont {A.}~\bibnamefont
  {Alfonsi}}, \bibinfo {author} {\bibfnamefont {R.}~\bibnamefont {Coyaud}},\
  and\ \bibinfo {author} {\bibfnamefont {V.}~\bibnamefont {Ehrlacher}},\ }\href
  {https://doi.org/10.1142/S0218202522500105} {\bibfield  {journal} {\bibinfo
  {journal} {Mathematical Models and Methods in Applied Sciences}\ }\textbf
  {\bibinfo {volume} {32}},\ \bibinfo {pages} {403} (\bibinfo {year}
  {2022})}\BibitemShut {NoStop}%
\bibitem [{\citenamefont {Kallemov}\ and\ \citenamefont
  {Miller}(2011)}]{kallemovSecondOrderStrongMethod2011}%
  \BibitemOpen
  \bibfield  {author} {\bibinfo {author} {\bibfnamefont {B.}~\bibnamefont
  {Kallemov}}\ and\ \bibinfo {author} {\bibfnamefont {G.~H.}\ \bibnamefont
  {Miller}},\ }\href {https://doi.org/10.1137/100785600} {\bibfield  {journal}
  {\bibinfo  {journal} {SIAM Journal on Scientific Computing}\ }\textbf
  {\bibinfo {volume} {33}},\ \bibinfo {pages} {653} (\bibinfo {year}
  {2011})}\BibitemShut {NoStop}%
\bibitem [{\citenamefont {Parisi}(1983)}]{parisiComplexProbabilities1983}%
  \BibitemOpen
  \bibfield  {author} {\bibinfo {author} {\bibfnamefont {G.}~\bibnamefont
  {Parisi}},\ }\href {https://doi.org/10.1016/0370-2693(83)90525-7} {\bibfield
  {journal} {\bibinfo  {journal} {Physics Letters B}\ }\textbf {\bibinfo
  {volume} {131}},\ \bibinfo {pages} {393} (\bibinfo {year}
  {1983})}\BibitemShut {NoStop}%
\bibitem [{\citenamefont {Klauder}(1983)}]{klauderLangevinApproachFermion1983}%
  \BibitemOpen
  \bibfield  {author} {\bibinfo {author} {\bibfnamefont {J.~R.}\ \bibnamefont
  {Klauder}},\ }\href {https://doi.org/10.1088/0305-4470/16/10/001} {\bibfield
  {journal} {\bibinfo  {journal} {Journal of Physics A: Mathematical and
  General}\ }\textbf {\bibinfo {volume} {16}},\ \bibinfo {pages} {L317}
  (\bibinfo {year} {1983})}\BibitemShut {NoStop}%
\bibitem [{\citenamefont {Berger}\ \emph {et~al.}(2021)\citenamefont {Berger},
  \citenamefont {Rammelm{\"u}ller}, \citenamefont {Loheac}, \citenamefont
  {Ehmann}, \citenamefont {Braun},\ and\ \citenamefont
  {Drut}}]{bergerComplexLangevinOther2021a}%
  \BibitemOpen
  \bibfield  {author} {\bibinfo {author} {\bibfnamefont {C.~E.}\ \bibnamefont
  {Berger}}, \bibinfo {author} {\bibfnamefont {L.}~\bibnamefont
  {Rammelm{\"u}ller}}, \bibinfo {author} {\bibfnamefont {A.~C.}\ \bibnamefont
  {Loheac}}, \bibinfo {author} {\bibfnamefont {F.}~\bibnamefont {Ehmann}},
  \bibinfo {author} {\bibfnamefont {J.}~\bibnamefont {Braun}},\ and\ \bibinfo
  {author} {\bibfnamefont {J.~E.}\ \bibnamefont {Drut}},\ }\href
  {https://doi.org/10.1016/j.physrep.2020.09.002} {\bibfield  {journal}
  {\bibinfo  {journal} {Physics Reports}\ }\bibinfo {series} {Complex
  {{Langevin}} and Other Approaches to the Sign Problem in Quantum Many-Body
  Physics},\ \textbf {\bibinfo {volume} {892}},\ \bibinfo {pages} {1} (\bibinfo
  {year} {2021})}\BibitemShut {NoStop}%
\bibitem [{\citenamefont {Loh}\ \emph {et~al.}(1990)\citenamefont {Loh},
  \citenamefont {Gubernatis}, \citenamefont {Scalettar}, \citenamefont {White},
  \citenamefont {Scalapino},\ and\ \citenamefont
  {Sugar}}]{lohSignProblemNumerical1990}%
  \BibitemOpen
  \bibfield  {author} {\bibinfo {author} {\bibfnamefont {E.~Y.}\ \bibnamefont
  {Loh}}, \bibinfo {author} {\bibfnamefont {J.~E.}\ \bibnamefont {Gubernatis}},
  \bibinfo {author} {\bibfnamefont {R.~T.}\ \bibnamefont {Scalettar}}, \bibinfo
  {author} {\bibfnamefont {S.~R.}\ \bibnamefont {White}}, \bibinfo {author}
  {\bibfnamefont {D.~J.}\ \bibnamefont {Scalapino}},\ and\ \bibinfo {author}
  {\bibfnamefont {R.~L.}\ \bibnamefont {Sugar}},\ }\href
  {https://doi.org/10.1103/PhysRevB.41.9301} {\bibfield  {journal} {\bibinfo
  {journal} {Physical Review B}\ }\textbf {\bibinfo {volume} {41}},\ \bibinfo
  {pages} {9301} (\bibinfo {year} {1990})}\BibitemShut {NoStop}%
\bibitem [{\citenamefont {Dornheim}(2019)}]{dornheimFermionSignProblem2019}%
  \BibitemOpen
  \bibfield  {author} {\bibinfo {author} {\bibfnamefont {T.}~\bibnamefont
  {Dornheim}},\ }\href {https://doi.org/10.1103/PhysRevE.100.023307} {\bibfield
   {journal} {\bibinfo  {journal} {Physical Review E}\ }\textbf {\bibinfo
  {volume} {100}},\ \bibinfo {pages} {023307} (\bibinfo {year}
  {2019})}\BibitemShut {NoStop}%
\bibitem [{\citenamefont {Negele}\ and\ \citenamefont
  {Orland}(1988)}]{negeleQuantumManyparticleSystems1988}%
  \BibitemOpen
  \bibfield  {author} {\bibinfo {author} {\bibfnamefont {J.~W.}\ \bibnamefont
  {Negele}}\ and\ \bibinfo {author} {\bibfnamefont {H.}~\bibnamefont
  {Orland}},\ }\href@noop {} {\emph {\bibinfo {title} {Quantum {{Many-particle
  Systems}}}}}\ (\bibinfo  {publisher} {Basic Books},\ \bibinfo {year}
  {1988})\BibitemShut {NoStop}%
\bibitem [{\citenamefont {Troyer}\ and\ \citenamefont
  {Wiese}(2005)}]{troyerComputationalComplexityFundamental2005}%
  \BibitemOpen
  \bibfield  {author} {\bibinfo {author} {\bibfnamefont {M.}~\bibnamefont
  {Troyer}}\ and\ \bibinfo {author} {\bibfnamefont {U.-J.}\ \bibnamefont
  {Wiese}},\ }\href {https://doi.org/10.1103/PhysRevLett.94.170201} {\bibfield
  {journal} {\bibinfo  {journal} {Physical Review Letters}\ }\textbf {\bibinfo
  {volume} {94}},\ \bibinfo {pages} {170201} (\bibinfo {year}
  {2005})}\BibitemShut {NoStop}%
\bibitem [{\citenamefont
  {Fredrickson}(2005)}]{fredricksonEquilibriumTheoryInhomogeneous2005}%
  \BibitemOpen
  \bibfield  {author} {\bibinfo {author} {\bibfnamefont {G.}~\bibnamefont
  {Fredrickson}},\ }\href
  {https://doi.org/10.1093/acprof:oso/9780198567295.001.0001} {\emph {\bibinfo
  {title} {The {{Equilibrium Theory}} of {{Inhomogeneous Polymers}}}}},\
  International {{Series}} of {{Monographs}} on {{Physics}}\ (\bibinfo
  {publisher} {Oxford University Press},\ \bibinfo {address} {Oxford},\
  \bibinfo {year} {2005})\BibitemShut {NoStop}%
\bibitem [{\citenamefont {Beardsley}\ \emph {et~al.}(2019)\citenamefont
  {Beardsley}, \citenamefont {Spencer},\ and\ \citenamefont
  {Matsen}}]{beardsleyComputationallyEfficientFieldTheoretic2019}%
  \BibitemOpen
  \bibfield  {author} {\bibinfo {author} {\bibfnamefont {T.~M.}\ \bibnamefont
  {Beardsley}}, \bibinfo {author} {\bibfnamefont {R.~K.~W.}\ \bibnamefont
  {Spencer}},\ and\ \bibinfo {author} {\bibfnamefont {M.~W.}\ \bibnamefont
  {Matsen}},\ }\href {https://doi.org/10.1021/acs.macromol.9b01904} {\bibfield
  {journal} {\bibinfo  {journal} {Macromolecules}\ }\textbf {\bibinfo {volume}
  {52}},\ \bibinfo {pages} {8840} (\bibinfo {year} {2019})}\BibitemShut
  {NoStop}%
\bibitem [{\citenamefont {Aarts}\ \emph {et~al.}(2010)\citenamefont {Aarts},
  \citenamefont {James}, \citenamefont {Seiler},\ and\ \citenamefont
  {Stamatescu}}]{aartsAdaptiveStepsizeInstabilities2010}%
  \BibitemOpen
  \bibfield  {author} {\bibinfo {author} {\bibfnamefont {G.}~\bibnamefont
  {Aarts}}, \bibinfo {author} {\bibfnamefont {F.~A.}\ \bibnamefont {James}},
  \bibinfo {author} {\bibfnamefont {E.}~\bibnamefont {Seiler}},\ and\ \bibinfo
  {author} {\bibfnamefont {I.-O.}\ \bibnamefont {Stamatescu}},\ }\href
  {https://doi.org/10.1016/j.physletb.2010.03.012} {\bibfield  {journal}
  {\bibinfo  {journal} {Physics Letters B}\ }\textbf {\bibinfo {volume}
  {687}},\ \bibinfo {pages} {154} (\bibinfo {year} {2010})},\ \Eprint
  {https://arxiv.org/abs/0912.0617} {arXiv:0912.0617} \BibitemShut {NoStop}%
\bibitem [{\citenamefont {Aarts}\ and\ \citenamefont
  {James}(2012)}]{aartsComplexLangevinDynamics2012}%
  \BibitemOpen
  \bibfield  {author} {\bibinfo {author} {\bibfnamefont {G.}~\bibnamefont
  {Aarts}}\ and\ \bibinfo {author} {\bibfnamefont {F.~A.}\ \bibnamefont
  {James}},\ }\href {https://doi.org/10.1007/JHEP01(2012)118} {\bibfield
  {journal} {\bibinfo  {journal} {Journal of High Energy Physics}\ }\textbf
  {\bibinfo {volume} {2012}},\ \bibinfo {pages} {118} (\bibinfo {year}
  {2012})}\BibitemShut {NoStop}%
\bibitem [{\citenamefont {Bloch}\ \emph {et~al.}(2018)\citenamefont {Bloch},
  \citenamefont {Glesaaen}, \citenamefont {Verbaarschot},\ and\ \citenamefont
  {Zafeiropoulos}}]{blochComplexLangevinSimulation2018}%
  \BibitemOpen
  \bibfield  {author} {\bibinfo {author} {\bibfnamefont {J.}~\bibnamefont
  {Bloch}}, \bibinfo {author} {\bibfnamefont {J.}~\bibnamefont {Glesaaen}},
  \bibinfo {author} {\bibfnamefont {J.~J.~M.}\ \bibnamefont {Verbaarschot}},\
  and\ \bibinfo {author} {\bibfnamefont {S.}~\bibnamefont {Zafeiropoulos}},\
  }\href {https://doi.org/10.1007/JHEP03(2018)015} {\bibfield  {journal}
  {\bibinfo  {journal} {Journal of High Energy Physics}\ }\textbf {\bibinfo
  {volume} {2018}},\ \bibinfo {pages} {15} (\bibinfo {year}
  {2018})}\BibitemShut {NoStop}%
\bibitem [{\citenamefont {Nagata}\ \emph {et~al.}(2018)\citenamefont {Nagata},
  \citenamefont {Nishimura},\ and\ \citenamefont
  {Shimasaki}}]{nagataComplexLangevinCalculations2018}%
  \BibitemOpen
  \bibfield  {author} {\bibinfo {author} {\bibfnamefont {K.}~\bibnamefont
  {Nagata}}, \bibinfo {author} {\bibfnamefont {J.}~\bibnamefont {Nishimura}},\
  and\ \bibinfo {author} {\bibfnamefont {S.}~\bibnamefont {Shimasaki}},\ }\href
  {https://doi.org/10.1103/PhysRevD.98.114513} {\bibfield  {journal} {\bibinfo
  {journal} {Physical Review D}\ }\textbf {\bibinfo {volume} {98}},\ \bibinfo
  {pages} {114513} (\bibinfo {year} {2018})}\BibitemShut {NoStop}%
\bibitem [{\citenamefont {Nagata}\ \emph {et~al.}(2016)\citenamefont {Nagata},
  \citenamefont {Nishimura},\ and\ \citenamefont
  {Shimasaki}}]{nagataJustificationComplexLangevin2016}%
  \BibitemOpen
  \bibfield  {author} {\bibinfo {author} {\bibfnamefont {K.}~\bibnamefont
  {Nagata}}, \bibinfo {author} {\bibfnamefont {J.}~\bibnamefont {Nishimura}},\
  and\ \bibinfo {author} {\bibfnamefont {S.}~\bibnamefont {Shimasaki}},\ }\href
  {https://doi.org/10.1093/ptep/ptv173} {\bibfield  {journal} {\bibinfo
  {journal} {Progress of Theoretical and Experimental Physics}\ }\textbf
  {\bibinfo {volume} {2016}},\ \bibinfo {pages} {013B01} (\bibinfo {year}
  {2016})},\ \Eprint {https://arxiv.org/abs/1508.02377} {arXiv:1508.02377
  [cond-mat, physics:hep-lat, physics:hep-th]} \BibitemShut {NoStop}%
\bibitem [{\citenamefont {Makino}\ \emph {et~al.}(2015)\citenamefont {Makino},
  \citenamefont {Suzuki},\ and\ \citenamefont
  {Takeda}}]{makinoComplexLangevinMethod2015}%
  \BibitemOpen
  \bibfield  {author} {\bibinfo {author} {\bibfnamefont {H.}~\bibnamefont
  {Makino}}, \bibinfo {author} {\bibfnamefont {H.}~\bibnamefont {Suzuki}},\
  and\ \bibinfo {author} {\bibfnamefont {D.}~\bibnamefont {Takeda}},\ }\href
  {https://doi.org/10.1103/PhysRevD.92.085020} {\bibfield  {journal} {\bibinfo
  {journal} {Physical Review D}\ }\textbf {\bibinfo {volume} {92}},\ \bibinfo
  {pages} {085020} (\bibinfo {year} {2015})}\BibitemShut {NoStop}%
\bibitem [{\citenamefont {Mollgaard}\ and\ \citenamefont
  {Splittorff}(2013)}]{mollgaardComplexLangevinDynamics2013}%
  \BibitemOpen
  \bibfield  {author} {\bibinfo {author} {\bibfnamefont {A.}~\bibnamefont
  {Mollgaard}}\ and\ \bibinfo {author} {\bibfnamefont {K.}~\bibnamefont
  {Splittorff}},\ }\href {https://doi.org/10.1103/PhysRevD.88.116007}
  {\bibfield  {journal} {\bibinfo  {journal} {Physical Review D}\ }\textbf
  {\bibinfo {volume} {88}},\ \bibinfo {pages} {116007} (\bibinfo {year}
  {2013})}\BibitemShut {NoStop}%
\bibitem [{\citenamefont {Hirasawa}\ \emph {et~al.}(2020)\citenamefont
  {Hirasawa}, \citenamefont {Matsumoto}, \citenamefont {Nishimura},\ and\
  \citenamefont {Yosprakob}}]{hirasawaComplexLangevinAnalysis2020a}%
  \BibitemOpen
  \bibfield  {author} {\bibinfo {author} {\bibfnamefont {M.}~\bibnamefont
  {Hirasawa}}, \bibinfo {author} {\bibfnamefont {A.}~\bibnamefont {Matsumoto}},
  \bibinfo {author} {\bibfnamefont {J.}~\bibnamefont {Nishimura}},\ and\
  \bibinfo {author} {\bibfnamefont {A.}~\bibnamefont {Yosprakob}},\ }\href
  {https://doi.org/10.1007/JHEP09(2020)023} {\bibfield  {journal} {\bibinfo
  {journal} {Journal of High Energy Physics}\ }\textbf {\bibinfo {volume}
  {2020}},\ \bibinfo {pages} {23} (\bibinfo {year} {2020})}\BibitemShut
  {NoStop}%
\bibitem [{\citenamefont
  {Greensite}(2014)}]{greensiteComparisonComplexLangevin2014}%
  \BibitemOpen
  \bibfield  {author} {\bibinfo {author} {\bibfnamefont {J.}~\bibnamefont
  {Greensite}},\ }\href {https://doi.org/10.1103/PhysRevD.90.114507} {\bibfield
   {journal} {\bibinfo  {journal} {Physical Review D}\ }\textbf {\bibinfo
  {volume} {90}},\ \bibinfo {pages} {114507} (\bibinfo {year}
  {2014})}\BibitemShut {NoStop}%
\bibitem [{\citenamefont {Fredrickson}\ \emph {et~al.}(2023)\citenamefont
  {Fredrickson}, \citenamefont {Delaney}, \citenamefont {Fredrickson},\ and\
  \citenamefont {Delaney}}]{fredricksonFieldTheoreticSimulationsSoft2023}%
  \BibitemOpen
  \bibfield  {author} {\bibinfo {author} {\bibfnamefont {G.~H.}\ \bibnamefont
  {Fredrickson}}, \bibinfo {author} {\bibfnamefont {K.~T.}\ \bibnamefont
  {Delaney}}, \bibinfo {author} {\bibfnamefont {G.~H.}\ \bibnamefont
  {Fredrickson}},\ and\ \bibinfo {author} {\bibfnamefont {K.~T.}\ \bibnamefont
  {Delaney}},\ }\href@noop {} {\emph {\bibinfo {title} {Field-{{Theoretic
  Simulations}} in {{Soft Matter}} and {{Quantum Fluids}}}}},\ International
  {{Series}} of {{Monographs}} on {{Physics}}\ (\bibinfo  {publisher} {Oxford
  University Press},\ \bibinfo {address} {Oxford, New York},\ \bibinfo {year}
  {2023})\BibitemShut {NoStop}%
\bibitem [{\citenamefont {Simmons}\ \emph {et~al.}(2023)\citenamefont
  {Simmons}, \citenamefont {Sajjad}, \citenamefont {Keithley}, \citenamefont
  {Mas}, \citenamefont {Tanlimco}, \citenamefont {{Nolasco-Martinez}},
  \citenamefont {Bai}, \citenamefont {Fredrickson},\ and\ \citenamefont
  {Weld}}]{simmonsThermodynamicEngineQuantum2023}%
  \BibitemOpen
  \bibfield  {author} {\bibinfo {author} {\bibfnamefont {E.~Q.}\ \bibnamefont
  {Simmons}}, \bibinfo {author} {\bibfnamefont {R.}~\bibnamefont {Sajjad}},
  \bibinfo {author} {\bibfnamefont {K.}~\bibnamefont {Keithley}}, \bibinfo
  {author} {\bibfnamefont {H.}~\bibnamefont {Mas}}, \bibinfo {author}
  {\bibfnamefont {J.~L.}\ \bibnamefont {Tanlimco}}, \bibinfo {author}
  {\bibfnamefont {E.}~\bibnamefont {{Nolasco-Martinez}}}, \bibinfo {author}
  {\bibfnamefont {Y.}~\bibnamefont {Bai}}, \bibinfo {author} {\bibfnamefont
  {G.~H.}\ \bibnamefont {Fredrickson}},\ and\ \bibinfo {author} {\bibfnamefont
  {D.~M.}\ \bibnamefont {Weld}},\ }\href
  {https://doi.org/10.1103/PhysRevResearch.5.L042009} {\bibfield  {journal}
  {\bibinfo  {journal} {Physical Review Research}\ }\textbf {\bibinfo {volume}
  {5}},\ \bibinfo {pages} {L042009} (\bibinfo {year} {2023})}\BibitemShut
  {NoStop}%
\bibitem [{\citenamefont {Attanasio}\ and\ \citenamefont
  {Drut}(2020)}]{attanasioThermodynamicsSpinorbitcoupledBosons2020}%
  \BibitemOpen
  \bibfield  {author} {\bibinfo {author} {\bibfnamefont {F.}~\bibnamefont
  {Attanasio}}\ and\ \bibinfo {author} {\bibfnamefont {J.~E.}\ \bibnamefont
  {Drut}},\ }\href {https://doi.org/10.1103/PhysRevA.101.033617} {\bibfield
  {journal} {\bibinfo  {journal} {Physical Review A}\ }\textbf {\bibinfo
  {volume} {101}},\ \bibinfo {pages} {033617} (\bibinfo {year}
  {2020})}\BibitemShut {NoStop}%
\bibitem [{\citenamefont {Heinen}\ and\ \citenamefont
  {Gasenzer}(2022)}]{heinenComplexLangevinApproach2022}%
  \BibitemOpen
  \bibfield  {author} {\bibinfo {author} {\bibfnamefont {P.}~\bibnamefont
  {Heinen}}\ and\ \bibinfo {author} {\bibfnamefont {T.}~\bibnamefont
  {Gasenzer}},\ }\href {https://doi.org/10.1103/PhysRevA.106.063308} {\bibfield
   {journal} {\bibinfo  {journal} {Physical Review A}\ }\textbf {\bibinfo
  {volume} {106}},\ \bibinfo {pages} {063308} (\bibinfo {year}
  {2022})}\BibitemShut {NoStop}%
\bibitem [{\citenamefont {Gross}\ and\ \citenamefont
  {Bakr}(2021)}]{grossQuantumGasMicroscopy2021}%
  \BibitemOpen
  \bibfield  {author} {\bibinfo {author} {\bibfnamefont {C.}~\bibnamefont
  {Gross}}\ and\ \bibinfo {author} {\bibfnamefont {W.~S.}\ \bibnamefont
  {Bakr}},\ }\href {https://doi.org/10.1038/s41567-021-01370-5} {\bibfield
  {journal} {\bibinfo  {journal} {Nature Physics}\ }\textbf {\bibinfo {volume}
  {17}},\ \bibinfo {pages} {1316} (\bibinfo {year} {2021})}\BibitemShut
  {NoStop}%
\bibitem [{\citenamefont {Haller}\ \emph {et~al.}(2015)\citenamefont {Haller},
  \citenamefont {Hudson}, \citenamefont {Kelly}, \citenamefont {Cotta},
  \citenamefont {Peaudecerf}, \citenamefont {Bruce},\ and\ \citenamefont
  {Kuhr}}]{hallerSingleatomImagingFermions2015}%
  \BibitemOpen
  \bibfield  {author} {\bibinfo {author} {\bibfnamefont {E.}~\bibnamefont
  {Haller}}, \bibinfo {author} {\bibfnamefont {J.}~\bibnamefont {Hudson}},
  \bibinfo {author} {\bibfnamefont {A.}~\bibnamefont {Kelly}}, \bibinfo
  {author} {\bibfnamefont {D.~A.}\ \bibnamefont {Cotta}}, \bibinfo {author}
  {\bibfnamefont {B.}~\bibnamefont {Peaudecerf}}, \bibinfo {author}
  {\bibfnamefont {G.~D.}\ \bibnamefont {Bruce}},\ and\ \bibinfo {author}
  {\bibfnamefont {S.}~\bibnamefont {Kuhr}},\ }\href
  {https://doi.org/10.1038/nphys3403} {\bibfield  {journal} {\bibinfo
  {journal} {Nature Physics}\ }\textbf {\bibinfo {volume} {11}},\ \bibinfo
  {pages} {738} (\bibinfo {year} {2015})}\BibitemShut {NoStop}%
\bibitem [{\citenamefont {Wenz}\ \emph {et~al.}(2013)\citenamefont {Wenz},
  \citenamefont {Z{\"u}rn}, \citenamefont {Murmann}, \citenamefont {Brouzos},
  \citenamefont {Lompe},\ and\ \citenamefont
  {Jochim}}]{wenzFewManyObserving2013}%
  \BibitemOpen
  \bibfield  {author} {\bibinfo {author} {\bibfnamefont {A.~N.}\ \bibnamefont
  {Wenz}}, \bibinfo {author} {\bibfnamefont {G.}~\bibnamefont {Z{\"u}rn}},
  \bibinfo {author} {\bibfnamefont {S.}~\bibnamefont {Murmann}}, \bibinfo
  {author} {\bibfnamefont {I.}~\bibnamefont {Brouzos}}, \bibinfo {author}
  {\bibfnamefont {T.}~\bibnamefont {Lompe}},\ and\ \bibinfo {author}
  {\bibfnamefont {S.}~\bibnamefont {Jochim}},\ }\href
  {https://doi.org/10.1126/science.1240516} {\bibfield  {journal} {\bibinfo
  {journal} {Science}\ }\textbf {\bibinfo {volume} {342}},\ \bibinfo {pages}
  {457} (\bibinfo {year} {2013})}\BibitemShut {NoStop}%
\bibitem [{\citenamefont
  {Bedingham}(2003)}]{bedinghamBoseEinsteinCondensationCanonical2003}%
  \BibitemOpen
  \bibfield  {author} {\bibinfo {author} {\bibfnamefont {D.~J.}\ \bibnamefont
  {Bedingham}},\ }\href {https://doi.org/10.1103/PhysRevD.68.105007} {\bibfield
   {journal} {\bibinfo  {journal} {Physical Review D}\ }\textbf {\bibinfo
  {volume} {68}},\ \bibinfo {pages} {105007} (\bibinfo {year}
  {2003})}\BibitemShut {NoStop}%
\bibitem [{\citenamefont {Barghathi}\ \emph {et~al.}(2020)\citenamefont
  {Barghathi}, \citenamefont {Yu},\ and\ \citenamefont
  {Del~Maestro}}]{barghathiTheoryNoninteractingFermions2020}%
  \BibitemOpen
  \bibfield  {author} {\bibinfo {author} {\bibfnamefont {H.}~\bibnamefont
  {Barghathi}}, \bibinfo {author} {\bibfnamefont {J.}~\bibnamefont {Yu}},\ and\
  \bibinfo {author} {\bibfnamefont {A.}~\bibnamefont {Del~Maestro}},\ }\href
  {https://doi.org/10.1103/PhysRevResearch.2.043206} {\bibfield  {journal}
  {\bibinfo  {journal} {Physical Review Research}\ }\textbf {\bibinfo {volume}
  {2}},\ \bibinfo {pages} {043206} (\bibinfo {year} {2020})}\BibitemShut
  {NoStop}%
\bibitem [{\citenamefont {Boninsegni}\ \emph
  {et~al.}(2006{\natexlab{b}})\citenamefont {Boninsegni}, \citenamefont
  {Prokof'ev},\ and\ \citenamefont
  {Svistunov}}]{boninsegniWormAlgorithmContinuousSpace2006}%
  \BibitemOpen
  \bibfield  {author} {\bibinfo {author} {\bibfnamefont {M.}~\bibnamefont
  {Boninsegni}}, \bibinfo {author} {\bibfnamefont {N.}~\bibnamefont
  {Prokof'ev}},\ and\ \bibinfo {author} {\bibfnamefont {B.}~\bibnamefont
  {Svistunov}},\ }\href {https://doi.org/10.1103/PhysRevLett.96.070601}
  {\bibfield  {journal} {\bibinfo  {journal} {Physical Review Letters}\
  }\textbf {\bibinfo {volume} {96}},\ \bibinfo {pages} {070601} (\bibinfo
  {year} {2006}{\natexlab{b}})}\BibitemShut {NoStop}%
\bibitem [{\citenamefont {McGarrigle}\ \emph {et~al.}(2023)\citenamefont
  {McGarrigle}, \citenamefont {Delaney}, \citenamefont {Balents},\ and\
  \citenamefont {Fredrickson}}]{mcgarrigleEmergenceSpinMicroemulsion2023}%
  \BibitemOpen
  \bibfield  {author} {\bibinfo {author} {\bibfnamefont {E.~C.}\ \bibnamefont
  {McGarrigle}}, \bibinfo {author} {\bibfnamefont {K.~T.}\ \bibnamefont
  {Delaney}}, \bibinfo {author} {\bibfnamefont {L.}~\bibnamefont {Balents}},\
  and\ \bibinfo {author} {\bibfnamefont {G.~H.}\ \bibnamefont {Fredrickson}},\
  }\href {https://doi.org/10.1103/PhysRevLett.131.173403} {\bibfield  {journal}
  {\bibinfo  {journal} {Physical Review Letters}\ }\textbf {\bibinfo {volume}
  {131}},\ \bibinfo {pages} {173403} (\bibinfo {year} {2023})}\BibitemShut
  {NoStop}%
\bibitem [{\citenamefont {Delaney}\ \emph {et~al.}(2020)\citenamefont
  {Delaney}, \citenamefont {Orland},\ and\ \citenamefont
  {Fredrickson}}]{delaneyNumericalSimulationFiniteTemperature2020}%
  \BibitemOpen
  \bibfield  {author} {\bibinfo {author} {\bibfnamefont {K.~T.}\ \bibnamefont
  {Delaney}}, \bibinfo {author} {\bibfnamefont {H.}~\bibnamefont {Orland}},\
  and\ \bibinfo {author} {\bibfnamefont {G.~H.}\ \bibnamefont {Fredrickson}},\
  }\href {https://doi.org/10.1103/PhysRevLett.124.070601} {\bibfield  {journal}
  {\bibinfo  {journal} {Physical Review Letters}\ }\textbf {\bibinfo {volume}
  {124}},\ \bibinfo {pages} {070601} (\bibinfo {year} {2020})}\BibitemShut
  {NoStop}%
\bibitem [{\citenamefont {Fetter}\ and\ \citenamefont
  {Walecka}(2012)}]{fetterQuantumTheoryManyParticle2012}%
  \BibitemOpen
  \bibfield  {author} {\bibinfo {author} {\bibfnamefont {A.~L.}\ \bibnamefont
  {Fetter}}\ and\ \bibinfo {author} {\bibfnamefont {J.~D.}\ \bibnamefont
  {Walecka}},\ }\href@noop {} {\emph {\bibinfo {title} {Quantum {{Theory}} of
  {{Many-Particle Systems}}}}}\ (\bibinfo  {publisher} {Courier Corporation},\
  \bibinfo {year} {2012})\BibitemShut {NoStop}%
\bibitem [{\citenamefont {Pitaevskii}\ and\ \citenamefont
  {Stringari}(2016)}]{pitaevskiiBoseEinsteinCondensationSuperfluidity2016}%
  \BibitemOpen
  \bibfield  {author} {\bibinfo {author} {\bibfnamefont {L.}~\bibnamefont
  {Pitaevskii}}\ and\ \bibinfo {author} {\bibfnamefont {S.}~\bibnamefont
  {Stringari}},\ }\href@noop {} {\emph {\bibinfo {title} {Bose-{{Einstein
  Condensation}} and {{Superfluidity}}}}}\ (\bibinfo  {publisher} {Oxford
  University Press},\ \bibinfo {year} {2016})\BibitemShut {NoStop}%
\bibitem [{\citenamefont {Ceperley}(1995)}]{ceperleyPathIntegralsTheory1995a}%
  \BibitemOpen
  \bibfield  {author} {\bibinfo {author} {\bibfnamefont {D.~M.}\ \bibnamefont
  {Ceperley}},\ }\href {https://doi.org/10.1103/RevModPhys.67.279} {\bibfield
  {journal} {\bibinfo  {journal} {Reviews of Modern Physics}\ }\textbf
  {\bibinfo {volume} {67}},\ \bibinfo {pages} {279} (\bibinfo {year}
  {1995})}\BibitemShut {NoStop}%
\bibitem [{\citenamefont {Shen}\ \emph {et~al.}(2023)\citenamefont {Shen},
  \citenamefont {Barghathi}, \citenamefont {Yu}, \citenamefont {Del~Maestro},\
  and\ \citenamefont {Rubenstein}}]{shenStableRecursiveAuxiliary2023}%
  \BibitemOpen
  \bibfield  {author} {\bibinfo {author} {\bibfnamefont {T.}~\bibnamefont
  {Shen}}, \bibinfo {author} {\bibfnamefont {H.}~\bibnamefont {Barghathi}},
  \bibinfo {author} {\bibfnamefont {J.}~\bibnamefont {Yu}}, \bibinfo {author}
  {\bibfnamefont {A.}~\bibnamefont {Del~Maestro}},\ and\ \bibinfo {author}
  {\bibfnamefont {B.~M.}\ \bibnamefont {Rubenstein}},\ }\href
  {https://doi.org/10.1103/PhysRevE.107.055302} {\bibfield  {journal} {\bibinfo
   {journal} {Physical Review E}\ }\textbf {\bibinfo {volume} {107}},\ \bibinfo
  {pages} {055302} (\bibinfo {year} {2023})}\BibitemShut {NoStop}%
\bibitem [{\citenamefont {Prokof'ev}\ \emph {et~al.}(1998)\citenamefont
  {Prokof'ev}, \citenamefont {Svistunov},\ and\ \citenamefont
  {Tupitsyn}}]{prokofevWormAlgorithmQuantum1998}%
  \BibitemOpen
  \bibfield  {author} {\bibinfo {author} {\bibfnamefont {N.~V.}\ \bibnamefont
  {Prokof'ev}}, \bibinfo {author} {\bibfnamefont {B.~V.}\ \bibnamefont
  {Svistunov}},\ and\ \bibinfo {author} {\bibfnamefont {I.~S.}\ \bibnamefont
  {Tupitsyn}},\ }\href {https://doi.org/10.1016/S0375-9601(97)00957-2}
  {\bibfield  {journal} {\bibinfo  {journal} {Physics Letters A}\ }\textbf
  {\bibinfo {volume} {238}},\ \bibinfo {pages} {253} (\bibinfo {year}
  {1998})}\BibitemShut {NoStop}%
\bibitem [{\citenamefont {Villet}\ and\ \citenamefont
  {Fredrickson}(2014)}]{villetEfficientFieldtheoreticSimulation2014a}%
  \BibitemOpen
  \bibfield  {author} {\bibinfo {author} {\bibfnamefont {M.~C.}\ \bibnamefont
  {Villet}}\ and\ \bibinfo {author} {\bibfnamefont {G.~H.}\ \bibnamefont
  {Fredrickson}},\ }\href {https://doi.org/10.1063/1.4902886} {\bibfield
  {journal} {\bibinfo  {journal} {The Journal of Chemical Physics}\ }\textbf
  {\bibinfo {volume} {141}},\ \bibinfo {pages} {224115} (\bibinfo {year}
  {2014})}\BibitemShut {NoStop}%
\bibitem [{\citenamefont
  {McQuarrie}(2000)}]{mcquarrieStatisticalMechanics2000}%
  \BibitemOpen
  \bibfield  {author} {\bibinfo {author} {\bibfnamefont {D.~A.}\ \bibnamefont
  {McQuarrie}},\ }\href@noop {} {\emph {\bibinfo {title} {Statistical
  {{Mechanics}}}}}\ (\bibinfo  {publisher} {University Science Books},\
  \bibinfo {year} {2000})\BibitemShut {NoStop}%
\bibitem [{\citenamefont {Fredrickson}\ and\ \citenamefont
  {Delaney}(2022)}]{fredricksonDirectFreeEnergy2022}%
  \BibitemOpen
  \bibfield  {author} {\bibinfo {author} {\bibfnamefont {G.~H.}\ \bibnamefont
  {Fredrickson}}\ and\ \bibinfo {author} {\bibfnamefont {K.~T.}\ \bibnamefont
  {Delaney}},\ }\href {https://doi.org/10.1073/pnas.2201804119} {\bibfield
  {journal} {\bibinfo  {journal} {Proceedings of the National Academy of
  Sciences}\ }\textbf {\bibinfo {volume} {119}},\ \bibinfo {pages}
  {e2201804119} (\bibinfo {year} {2022})}\BibitemShut {NoStop}%
\bibitem [{\citenamefont {Leli{\`e}vre}\ \emph {et~al.}(2012)\citenamefont
  {Leli{\`e}vre}, \citenamefont {Rousset},\ and\ \citenamefont
  {Stoltz}}]{lelievreLangevinDynamicsConstraints2012a}%
  \BibitemOpen
  \bibfield  {author} {\bibinfo {author} {\bibfnamefont {T.}~\bibnamefont
  {Leli{\`e}vre}}, \bibinfo {author} {\bibfnamefont {M.}~\bibnamefont
  {Rousset}},\ and\ \bibinfo {author} {\bibfnamefont {G.}~\bibnamefont
  {Stoltz}},\ }\href {https://doi.org/10.1090/S0025-5718-2012-02594-4}
  {\bibfield  {journal} {\bibinfo  {journal} {Mathematics of Computation}\
  }\textbf {\bibinfo {volume} {81}},\ \bibinfo {pages} {2071} (\bibinfo {year}
  {2012})}\BibitemShut {NoStop}%
\bibitem [{\citenamefont {Lelivre}\ \emph {et~al.}(2010)\citenamefont
  {Lelivre}, \citenamefont {Stoltz},\ and\ \citenamefont
  {Rousset}}]{lelivreFreeEnergyComputations2010}%
  \BibitemOpen
  \bibfield  {author} {\bibinfo {author} {\bibfnamefont {T.}~\bibnamefont
  {Lelivre}}, \bibinfo {author} {\bibfnamefont {G.}~\bibnamefont {Stoltz}},\
  and\ \bibinfo {author} {\bibfnamefont {M.}~\bibnamefont {Rousset}},\
  }\href@noop {} {\emph {\bibinfo {title} {Free {{Energy Computations}}: {{A
  Mathematical Perspective}}}}}\ (\bibinfo  {publisher} {World Scientific},\
  \bibinfo {year} {2010})\BibitemShut {NoStop}%
\bibitem [{\citenamefont {Leli{\`e}vre}\ \emph {et~al.}(2019)\citenamefont
  {Leli{\`e}vre}, \citenamefont {Rousset},\ and\ \citenamefont
  {Stoltz}}]{lelievreHybridMonteCarlo2019}%
  \BibitemOpen
  \bibfield  {author} {\bibinfo {author} {\bibfnamefont {T.}~\bibnamefont
  {Leli{\`e}vre}}, \bibinfo {author} {\bibfnamefont {M.}~\bibnamefont
  {Rousset}},\ and\ \bibinfo {author} {\bibfnamefont {G.}~\bibnamefont
  {Stoltz}},\ }\href {https://doi.org/10.48550/arXiv.1807.02356} {\bibinfo
  {title} {Hybrid {{Monte Carlo}} methods for sampling probability measures on
  submanifolds}} (\bibinfo {year} {2019}),\ \Eprint
  {https://arxiv.org/abs/1807.02356} {arXiv:1807.02356 [cs, math]} \BibitemShut
  {NoStop}%
\bibitem [{\citenamefont {Man}\ \emph {et~al.}(2014)\citenamefont {Man},
  \citenamefont {Delaney}, \citenamefont {Villet}, \citenamefont {Orland},\
  and\ \citenamefont {Fredrickson}}]{manCoherentStatesFormulation2014}%
  \BibitemOpen
  \bibfield  {author} {\bibinfo {author} {\bibfnamefont {X.}~\bibnamefont
  {Man}}, \bibinfo {author} {\bibfnamefont {K.~T.}\ \bibnamefont {Delaney}},
  \bibinfo {author} {\bibfnamefont {M.~C.}\ \bibnamefont {Villet}}, \bibinfo
  {author} {\bibfnamefont {H.}~\bibnamefont {Orland}},\ and\ \bibinfo {author}
  {\bibfnamefont {G.~H.}\ \bibnamefont {Fredrickson}},\ }\href
  {https://doi.org/10.1063/1.4860978} {\bibfield  {journal} {\bibinfo
  {journal} {The Journal of Chemical Physics}\ }\textbf {\bibinfo {volume}
  {140}},\ \bibinfo {pages} {024905} (\bibinfo {year} {2014})}\BibitemShut
  {NoStop}%
\bibitem [{\citenamefont {Kanzow}\ \emph {et~al.}(2004)\citenamefont {Kanzow},
  \citenamefont {Yamashita},\ and\ \citenamefont
  {Fukushima}}]{kanzowLevenbergMarquardtMethods2004}%
  \BibitemOpen
  \bibfield  {author} {\bibinfo {author} {\bibfnamefont {C.}~\bibnamefont
  {Kanzow}}, \bibinfo {author} {\bibfnamefont {N.}~\bibnamefont {Yamashita}},\
  and\ \bibinfo {author} {\bibfnamefont {M.}~\bibnamefont {Fukushima}},\ }\href
  {https://doi.org/10.1016/j.cam.2004.02.013} {\bibfield  {journal} {\bibinfo
  {journal} {Journal of Computational and Applied Mathematics}\ }\textbf
  {\bibinfo {volume} {172}},\ \bibinfo {pages} {375} (\bibinfo {year}
  {2004})}\BibitemShut {NoStop}%
\bibitem [{\citenamefont {Vigil}\ \emph {et~al.}(2021)\citenamefont {Vigil},
  \citenamefont {Delaney},\ and\ \citenamefont
  {Fredrickson}}]{vigilQuantitativeComparisonFieldUpdate2021}%
  \BibitemOpen
  \bibfield  {author} {\bibinfo {author} {\bibfnamefont {D.~L.}\ \bibnamefont
  {Vigil}}, \bibinfo {author} {\bibfnamefont {K.~T.}\ \bibnamefont {Delaney}},\
  and\ \bibinfo {author} {\bibfnamefont {G.~H.}\ \bibnamefont {Fredrickson}},\
  }\href {https://doi.org/10.1021/acs.macromol.1c01804} {\bibfield  {journal}
  {\bibinfo  {journal} {Macromolecules}\ }\textbf {\bibinfo {volume} {54}},\
  \bibinfo {pages} {9804} (\bibinfo {year} {2021})}\BibitemShut {NoStop}%
\bibitem [{\citenamefont {{Stamper-Kurn}}\ and\ \citenamefont
  {Ueda}(2013)}]{stamper-kurnSpinorBoseGases2013a}%
  \BibitemOpen
  \bibfield  {author} {\bibinfo {author} {\bibfnamefont {D.~M.}\ \bibnamefont
  {{Stamper-Kurn}}}\ and\ \bibinfo {author} {\bibfnamefont {M.}~\bibnamefont
  {Ueda}},\ }\href {https://doi.org/10.1103/RevModPhys.85.1191} {\bibfield
  {journal} {\bibinfo  {journal} {Reviews of Modern Physics}\ }\textbf
  {\bibinfo {volume} {85}},\ \bibinfo {pages} {1191} (\bibinfo {year}
  {2013})}\BibitemShut {NoStop}%
\bibitem [{\citenamefont {Bloch}(2005)}]{blochUltracoldQuantumGases2005}%
  \BibitemOpen
  \bibfield  {author} {\bibinfo {author} {\bibfnamefont {I.}~\bibnamefont
  {Bloch}},\ }\href {https://doi.org/10.1038/nphys138} {\bibfield  {journal}
  {\bibinfo  {journal} {Nature Physics}\ }\textbf {\bibinfo {volume} {1}},\
  \bibinfo {pages} {23} (\bibinfo {year} {2005})}\BibitemShut {NoStop}%
\bibitem [{\citenamefont {Arovas}\ and\ \citenamefont
  {Auerbach}(1988)}]{arovasFunctionalIntegralTheories1988a}%
  \BibitemOpen
  \bibfield  {author} {\bibinfo {author} {\bibfnamefont {D.~P.}\ \bibnamefont
  {Arovas}}\ and\ \bibinfo {author} {\bibfnamefont {A.}~\bibnamefont
  {Auerbach}},\ }\href {https://doi.org/10.1103/PhysRevB.38.316} {\bibfield
  {journal} {\bibinfo  {journal} {Physical Review B}\ }\textbf {\bibinfo
  {volume} {38}},\ \bibinfo {pages} {316} (\bibinfo {year} {1988})}\BibitemShut
  {NoStop}%
\end{thebibliography}
%

\end{document}